\documentclass[a4paper,12pt,twoside,openright]{report}
\usepackage[latin1]{inputenc}
\pdfoutput=1
\usepackage[bookmarks=false]{hyperref}

\usepackage[english]{babel}
\usepackage{graphicx}
\usepackage{longtable}
\usepackage{amssymb,amsmath}
\usepackage{amssymb}
\usepackage{longtable}
\usepackage{latexsym}
\usepackage{epsfig}
\usepackage{amsbsy}
\usepackage{fancyhdr}
\usepackage{array}
\usepackage{calc}
\let\origdoublepage\cleardoublepage
\newcommand{\clearemptydoublepage}{%
  \clearpage
  {\pagestyle{empty}\origdoublepage}%
}
\let\cleardoublepage\clearemptydoublepage

\begin{document}

\thispagestyle{empty}

\begin{center}
    \begin{figure}
        \includegraphics[width=2.5cm]{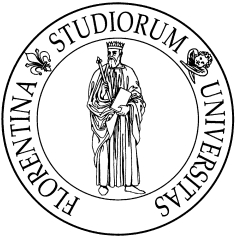}
        \hspace{8cm}
        \includegraphics[width=2.5cm]{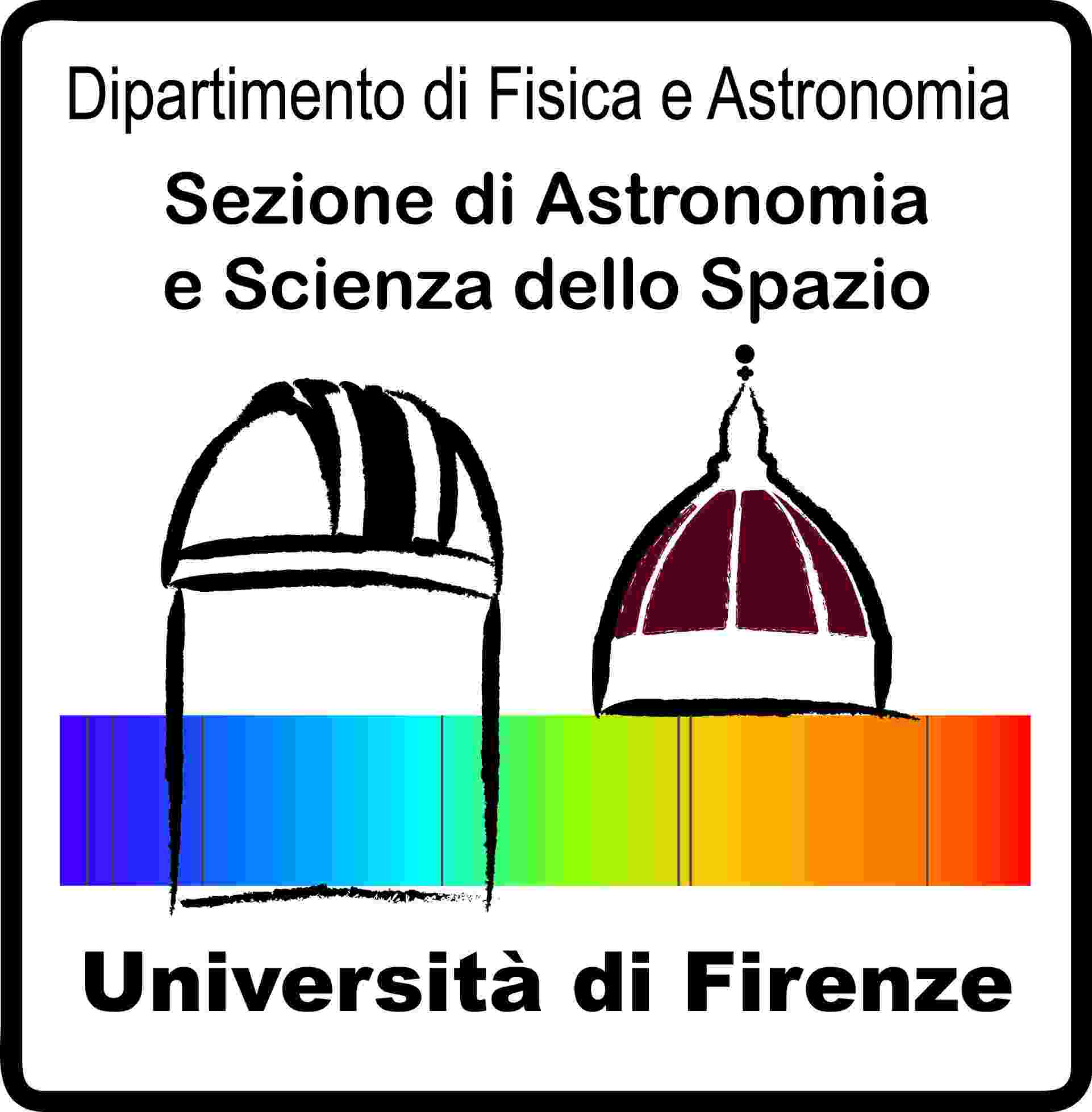}
        \label{fig:uni_logo.jpg}
    \end{figure}

        \begin{large}
               \mbox{UNIVERSIT\`{A} DEGLI STUDI DI FIRENZE} \\
               \vspace{0.3cm}
                Dipartimento di Fisica e Astronomia \\
               \vspace{0.3cm}
                Scuola di Dottorato in Astronomia \\
               \vspace{0.3cm}
                Ciclo XXIV - FIS05\\
        \end{large}

        \vspace{2cm}

       \begin{huge}
        \textbf{Design of the wavefront sensor unit of ARGOS, the LBT laser guide star system}\linebreak[4]
            \textbf{}\linebreak[4]

        \end{huge}

        \vspace{3cm}

\end{center}

        \begin{tabular} {l@{    }l}
        \vspace{0.5cm}
                \textit{Candidato:} &  Marco Bonaglia \\
        \vspace{0.5cm}
                \textit{Tutore:} & Prof. Alberto Righini \\
        \vspace{0.5cm}
                \textit{Cotutore:} & Dott. Simone Esposito \\
        \end{tabular}
\begin{center}

\end{center}

\newpage
\thispagestyle{empty}
\mbox{}
%


\newpage
\thispagestyle{empty}
\voffset=4cm
\begin{flushright}
    \emph{A common use for a glass plate is as a beam splitter,
    \\tilted at an angle of $45^{\circ}$ [$\ldots$]
    \\Since this can severely degrade the image,
    \\such plate beam splitters are not recommended
    \\in convergent or divergent beams.}
    \\W. J. Smith, Modern Optical Engineering.
\end{flushright}

\newpage
\thispagestyle{empty}
\mbox{}

\newpage
\thispagestyle{empty}
\voffset=0cm
\pagenumbering{roman}
\tableofcontents

\newpage
\thispagestyle{empty}
\pagenumbering{arabic}

%
%
%
%

\chapter{Introduction}
\label{cap:intro}

\begin{figure}
\begin{center}
  \includegraphics[width=8cm]{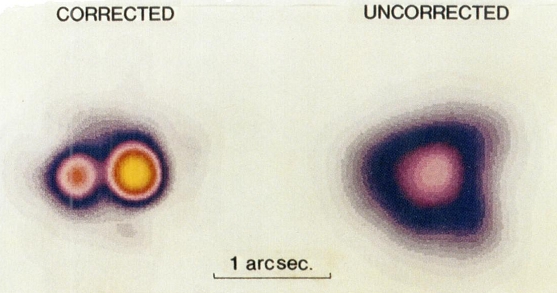}
\end{center}
\vspace{-0.5cm}
\caption{\footnotesize{Images of the group of stars $\gamma_2$ And at K band taken the night of the 22nd of October 1989 at Observatoire de Haute Provence. Left: image taken with AO correction performed with the COME-ON system. The reference star used for wavefront sensing is $\gamma_1$ And at a $9.6arcsec$ distance from the system. Right: seeing limited image of the system, $r_0\simeq12cm$. Images taken from \cite{1990_Rousset_Come-on}.}}
\label{fig:Come-on}
\end{figure}

The growth of astronomical knowledge is a direct consequence of the increase in the number and size of telescopes and their instrumentation efficiency. The large collecting areas of modern telescopes allow to detect faint, and hence distant, astronomical objects. However ground based telescopes suffer of a limited angular resolution because of the aberrations introduced by the turbulent Earth atmosphere \cite{1953_Edlen_Air_refr_index}. The main effect of atmosphere is to spread the light of a point-like object over a finite area of the telescope image plane, making astronomical images taken with long exposure times to appear blurred. The angular dimension of the blurred image is often called the \emph{seeing} value of atmosphere \cite{1981_Roddier_Seeing_PrOpt}. In seeing limited observations the resolution of a ground based telescope with an aperture of diameter $D$ is reduced by a factor $D/r_0$ with respect to its diffraction limited resolution. The $r_0$ parameter is commonly used to characterize the strength of turbulence aberrations\footnote{$r_0$ is defined as the wavefront area over which the atmospheric aberrations reach the value of $1rad$ rms.} and it is entitled to D. L. Fried, who first give a definition of the effects of wavefront distortions on angular resolution \cite{1965Fried_Statist_WF_distortion}. The seeing value $\beta$ varies linearly with the radiation wavelength $\lambda$ and with the inverse of $r_0$. Good astronomical sites have seeing values, evaluated at $0.5\mu m$, ranging in the $0.5-1arcsec$ range, corresponding to $r_0$ of $10-20cm$. Since also $r_0$ varies with wavelength as $\lambda^{6/5}$ \cite{1978_Boyd_r0_vs_wave}, the loss of performance experienced by an $8m$ class ground based telescope observing at infrared wavelengths amounts to a factor $D/r_0\simeq20$.
\\In 1953 H. W. Babcock first proposed a technique to measure the wavefront distortions introduced by atmosphere and to compensate for them inserting a deformable element in the telescope optical path \cite{1953_Babcock}. However due to the fast time scales of evolution of the atmospheric aberrations many technical difficulties were related to the possibility to effectively measure and compensate for them in real time. The first system able to demonstrate on sky the feasibility of Adaptive Optics (AO) was produced only in the 1990 \cite{1990_Rousset_Come-on} giving the first diffraction limited images with a ground based $1.5m$ diameter telescope at $2.3\mu m$ wavelength, see figure \ref{fig:Come-on}.

\section{The Large Binocular Telescope}
\label{sec:LBT}
\begin{figure}
\begin{center}
  \includegraphics[width=13.5cm]{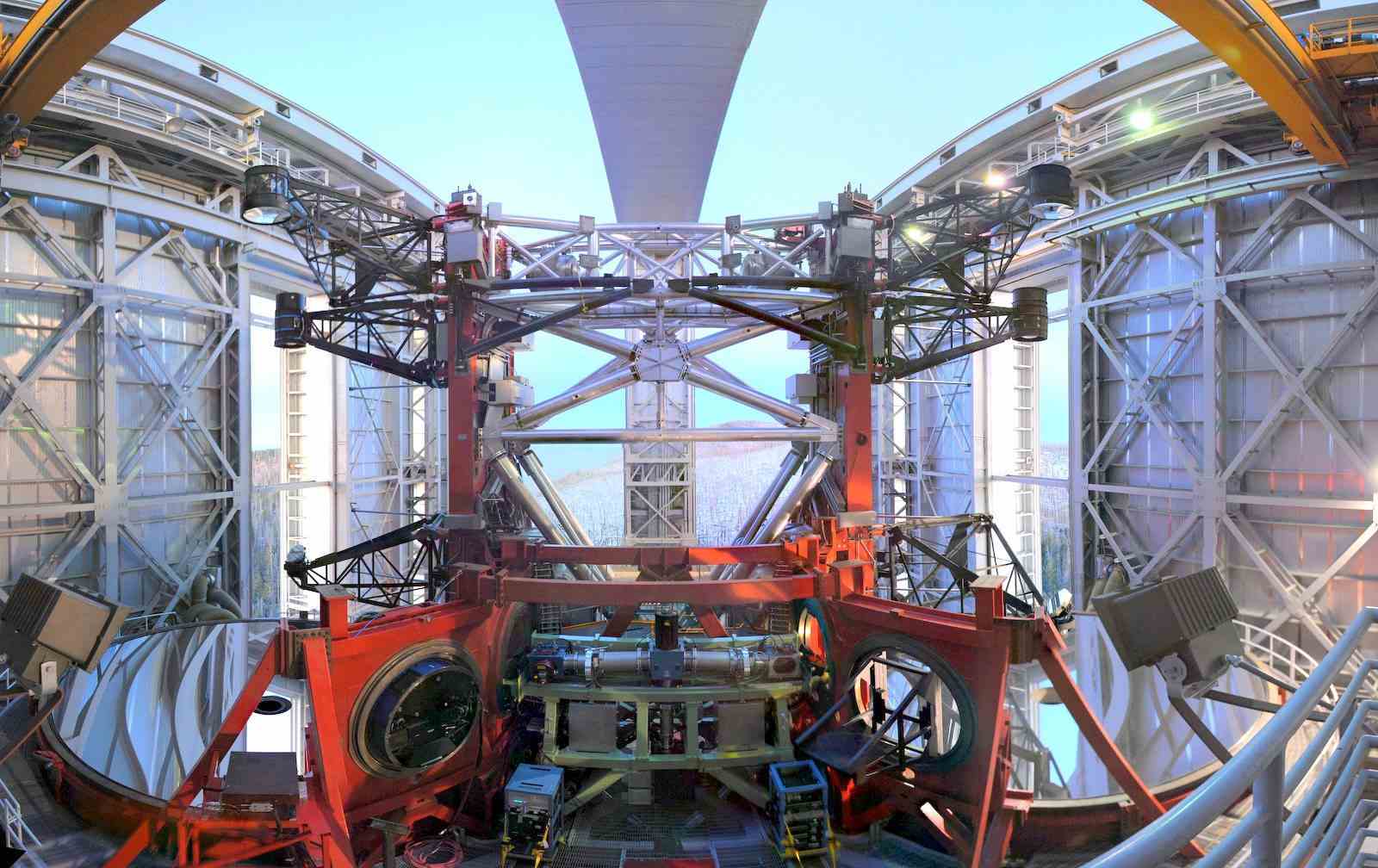}
\end{center}
\vspace{-0.5cm}
\caption{\footnotesize{Picture of the LBT taken inside the telescope dome. Red-painted parts are mounted on a rotating platform and can be moved in elevation on hydrostatic bearings to allow the telescope pointing and tracking. The black-painted swing arms can deploy over the primary mirrors the ASM and the rotating tertiary or a prime focus camera. Instruments served by AO are installed on the platform in between the 2 primary mirrors. Image courtesy of L. Busoni, OAA.}}
\label{fig:LBT_pic}
\end{figure}

The Large Binocular Telescope (LBT) has been the first $8m$ class telescope optimized for infrared observations to include an AO control of seeing by design \cite{1994_Hill_LBT_SPIE}. The mechanical structure of LBT is shown in figure \ref{fig:LBT_pic}, it provides a single alt-azimuth mount for two $8.4m$ diameter primary mirrors. Using fast focal ratios ($f_{1.142}$) for the primary mirrors \cite{2003_Martin_LBT+GMT-mirrors} allowed to locate two elliptical secondary mirrors of $0.91m$ diameter in a Gregorian configuration. The secondaries produce $f_{15}$ beams and they are also used as adaptive optics correctors \cite{2010_Riccardi_LBT_ASM}. A single mirror is made of a $1.6mm$ thin shell of Zerodur \cite{2007_Brusa_thin_shell_SPIE} and it is deformed at kHz frequencies by 672 voice coil actuators \cite{2008_DelVecchio_Voice_coil_act}.
\\The main advantages of having the secondary mirror that works also as AO corrector are several, among these the most important is that all the telescope focal stations can benefit of AO correction. This prevent to replicate the DM at all the focal stations hence reducing the number of optics placed before the instruments. At LBT the different focal stations are accessible rotating the flat tertiary mirror. In this way it is possible to feed either 2 single beam instruments or to combine the 2 $f_{15}$ beams in a single interferometric instrument to achieve the equivalent resolution of a $22.8m$ diameter telescope \cite{2000_Hill_LBT_SPIE}. The three instruments that are placed at the bent Gregorian stations of LBT and benefit of the AO correction provided by the adaptive secondary mirrors (ASM) are:
\begin{itemize}
  \item LUCI: a single beam multi-object spectrograph (MOS) and imaging camera, working at near infrared wavelengths. This instrument is replicated for each eye of the telescope \cite{2000_Mandel_LUCIFER_SPIE}.
  \item LBTI: an interferometric instrument combining mid infrared wavelengths in a nulling imaging camera \cite{2003_Hinz_LBTI_SPIE}.
  \item LINC-NIRVANA: a Fizeau interferometer operating in the near infrared \cite{2003_Herbst_LINC-NIRVANA}.
\end{itemize}
LBT is also provided with other instruments that work in seeing limited conditions: 2 wide field prime focus cameras (LBC) \cite{2004_Ragazzoni_LBC_SPIE}, 2 visible MOS and imaging cameras (MODS) \cite{2000_Byard_MODS_SPIE} mounted at the direct Gregorian focus under the primary cells, and a fiber fed high-resolution echelle spectrograph (PEPSI) \cite{2003_Strassmeier_PEPSI_SPIE}.

\section{LUCI}
\label{sec:LUCI}
LUCI is one of the principal instruments of the LBT. It combines the MOS and imaging capabilities over a wide field of view ($4\times4arcmin$) in the $1.0-2.5\mu m$ range. In MOS operation it can deploy almost 20 slits over a single mask giving a large multiplexing capability. This, combined with the multiwavelength coverage, makes LUCI one of the top level instruments for spectroscopic follow-ups and surveys of high-redshift galaxies properties, such as metallicities and kinematics of star-forming galaxies, stellar absorption properties of mass-selected samples, line emitters.
\begin{figure}
\begin{center}
  \includegraphics[width=6.8cm]{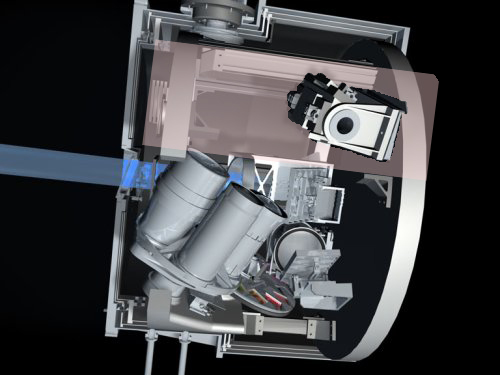}
  \includegraphics[width=6.2cm]{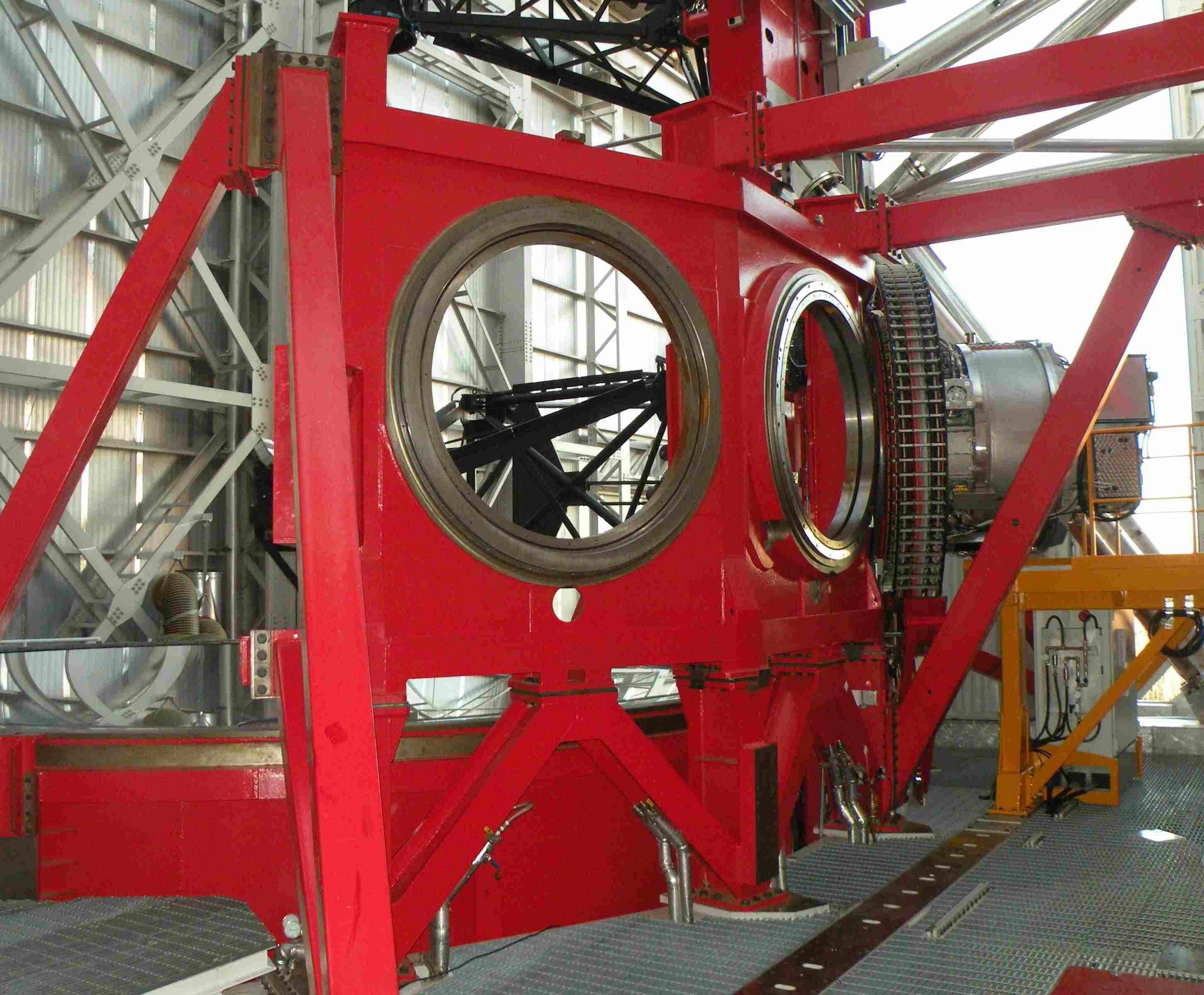}
\end{center}
\vspace{-0.5cm}
\caption{\footnotesize{Left: sectioned model of the LUCI MOS and imaging camera. Telescope $f_{15}$ light arrives from left in figure. The gray shaded area at the top highlights the robotic arm that positions the spectroscopy masks on the instrument input focal plane. The optic above this area is the exchangeable mirror-grating unit and the 3 large optic mounted on the rotation stage are the 3 cameras necessary to adapt the instrument resolution to the observing mode. Right: picture of the instrumentation platform of the LBT showing on the extreme right the first unit of LUCI installed. Image courtesy of LBTO.}}
\label{fig:LUCI}
\end{figure}
\\Figure \ref{fig:LUCI} shows a section of the instrument cryostat, cooled with liquid nitrogen during operations. In MOS operating mode a robotic arm allows to select different masks and to place them on the instrument input focal plane, placed just after the tilted entrance window on the left in figure \ref{fig:LUCI}. LUCI is provided with different sets of broad (J, H, K, H+K) and narrow (Brackett-$\gamma$, FeII, H2, HeI, J low and high, OH-hole) band filters. The instrument resolution can be adapted to different seeing conditions placing different camera optics in front of the detector. Three different scales are available by design: 2 seeing limited modes, having respectively $0.25$ and $0.12arcsec\;px^{-1}$ scale and FoV of $4\times4arcmin$ and a diffraction limited mode, having $0.015arcsec\;px^{-1}$ scale and FoV reduced to $30\times30arcsec$. The diffraction limited operation of LUCI is dedicated to the study of details of high-redshift galaxies, distinguishing for example the properties of the bulk of stellar mass from the rest-frame morphologies.
\begin{table}
\caption{\footnotesize{Summary of the operating modes of LUCI and relative properties on the instrument image plane.}}
\vspace{-0.25cm}
\begin{center}
    \begin{tabular}{|l|c|c|}
    \hline
    \textbf{Mode} & \textbf{Parameter} & \textbf{Value} \\
    \hline
    \hline
    Seeing limited & Scale  &   $0.25arcsec\;px^{-1}$ \\
                    & FoV  &   $4\times4arcmin$ \\
                    & Resolution & $500\div5000$ \\
    \hline
    Reduced seeing & Scale  &   $0.12arcsec\;px^{-1}$ \\
                    & FoV  &   $4\times4arcmin$ \\
                    & Resolution & $1000\div10000$ \\
    \hline
    Diffraction limited & Scale  &   $0.015arcsec\;px^{-1}$ \\
                    & FoV  &   $0.5\times0.5arcmin$ \\
                    & Resolution & $2000\div20000$ \\
    \hline
    \end{tabular}
\end{center}
\label{tab:LUCI_modes}
\end{table}

\section{First Light AO system}
\label{sec:FLAO}
\begin{figure}
\begin{center}
  \includegraphics[width=8cm]{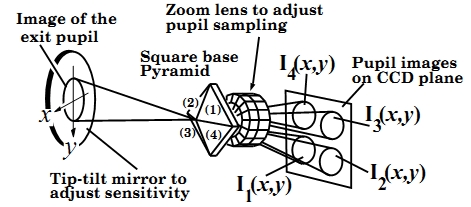}
\end{center}
\vspace{-0.5cm}
\caption{\footnotesize{Schematic layout of a Pyramid WFS. Image taken from \cite{2000_SPIE_Esposito_SP-Cloop-Perform}.}}
\label{fig:PWFS_scheme}
\end{figure}

LUCI will be the first of the LBT instruments to be provided with an AO system able to compensate for the atmospheric turbulence aberrations. This is the First Light AO (FLAO) system \cite{2010_Esposito_FLAO_ApOpt} that uses the visible light from an astronomical object close to the LUCI science target to measure the wavefront distortions introduced by the atmosphere. The device used to sense the wavefront is a Pyramid wavefront sensor (PWFS) \cite{1996_JMOp_Ragazzoni_osc-prism-WFS}.
\\The PWFS is based on the same concept of the Focault knife edge test \cite{1859_Foucault_knife_test}: a sharp edge element (a transmissive glass square-based pyramid) is placed on the focal plane where a point-like source is imaged by the telescope. In presence of aberrations the image formed after the sensor focal plane is not fully illuminated. A camera lens placed after the transmissive pyramid generates 4 images of the telescope pupil on the active area of a CCD detector. The difference between the illuminations of the 4 pupil images recorded by the PWFS detector is proportional to the wavefront local derivative. Figure \ref{fig:PWFS_scheme} shows a schematic layout of the working principle of the PWFS. On the PWFS measurements the FLAO system reconstructs the turbulence aberrations in the direction of the guide star (GS) and it compensates for them in real time applying an opposite shape to the ASM.
\\The PWFS of the FLAO system is located inside the acquisition, guiding, and wavefront sensing (AGW) \cite{2004_Storm_AGW_SPIE} unit installed at the instrument focal station, just in front of LUCI to minimize the differential flexures between the WFS and the instrument. Depending on the magnitude of the available GS it is possible to select on the PWFS a variable sampling of the telescope pupil binning the CCD at readout. The PWFS is designed to have a maximum sampling of $30\times30$ subapertures, equivalent to sample the telescope pupil at a $28cm$ resolution that corresponds to the $r_0$ value at $1\mu m$ in average seeing conditions ($\beta\simeq0.7arcsec$).
\begin{figure}
\begin{center}
  \includegraphics[width=13.5cm]{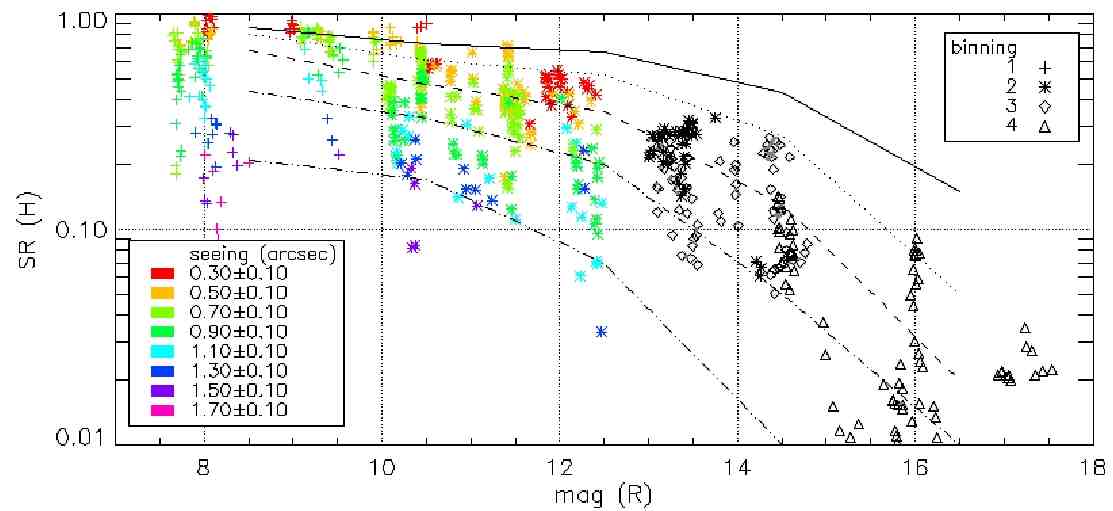}
\end{center}
\vspace{-0.5cm}
\caption{\footnotesize{Summary of the FLAO performance expressed in terms of SR at $1.6\mu m$. The data represent almost 500 different AO closed loops performed under different seeing conditions. Image taken from \cite{2010_Esposito_FLAO_SPIEAO}.}}
\label{fig:FLAO_SR}
\end{figure}
\\The first unit of FLAO has been commissioned at LBT during 2010. The system performance have been measured during the commissioning using an infrared test camera (IRTC) \cite{2008_Foppiani_IRTC_SPIE}. Figure \ref{fig:FLAO_SR} summarizes the level of correction reached by the AO system in terms of achieved Strehl ratio\footnote{Strehl ratio (SR) is a measure for the optical quality of imaging systems. It is defined as the ratio of the observed peak intensity generated by a point-like object on the image plane compared to the theoretical maximum peak intensity of an imaging system working at the diffraction limit.} at $1.6\mu m$ \cite{2010_Esposito_FLAO_SPIEAO}. The maximum value obtained in case of an high flux ($m_r=8.0$) GS exceeds $85\%$ when 495 modes have been corrected at full pupil sampling. The system demonstrated to be able to provide a partial AO correction up to GS $m_r=17.5$, reaching $SR\simeq5\%$ correcting for 36 modes with a pupil sampling of $7\times7$ subapertures. The different colors in figure \ref{fig:FLAO_SR} shows data collected under different seeing conditions ($\beta=0.6-1.5arcsec$) \cite{2010_Esposito_FLAO_SPIE}. They are well in agreement with the results obtained during the laboratory acceptance test of the system \cite{2010_QuirosPacheco_FLAO_char}.
\begin{figure}
\begin{center}
  \includegraphics[height=6.5cm]{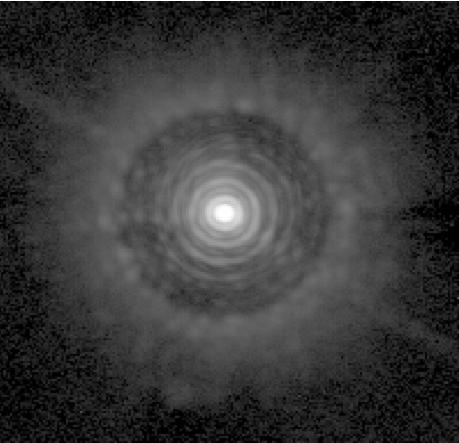}
  \includegraphics[height=6.5cm]{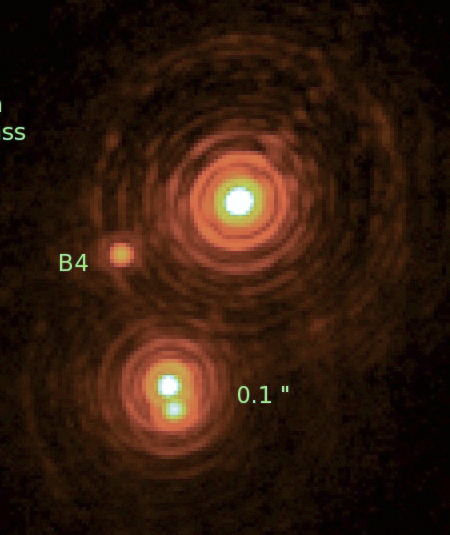}
\end{center}
\vspace{-0.5cm}
\caption{\footnotesize{Images from the commissioning run of the first unit of the FLAO system. Left: example of the diffraction limit achieved by FLAO pointing HD175658 (R=6.5, H=2.5), the image has been taken in H band using the IRTC \cite{2010_Esposito_FLAO_SPIE}. Right: image of the Orion Trapezium $\theta$ Ori cluster taken in H band with PISCES (image courtesy of L. Close). More details on these images can be found in the text.}}
\label{fig:FLAO_imas}
\end{figure}
\\The image on the left of figure \ref{fig:FLAO_imas} is an example of the diffraction limit resolution that FLAO achieved: it is a composition of two $10s$ integration time images taken with the IRTC pointing at pointing HD175658 (R=6.5, H=2.5), 10 diffraction rings are visible. In this case the measured H band SR was almost $80\%$. The seeing value was $0.9 arcsec$ and 400 modes have been corrected to obtain this image.
\\The image on the right of figure \ref{fig:FLAO_imas} has been taken instead with PISCES \cite{2001_McCarthy_PISCES_PASP} an infrared imaging camera with diffraction limited sampling. It shows the inner part of the Orion Trapezium $\theta$ Ori cluster. The spatial resolution of this image is $\sim50mas$ evaluated at $1.64\mu m$. It is visible that the 2 stars placed at $\sim100mas$ are well resolved by the system.
\begin{figure}
\begin{center}
  \includegraphics[width=13cm]{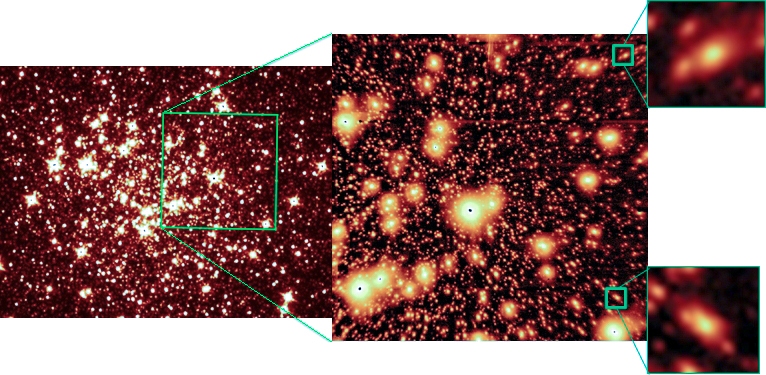}
\end{center}
\vspace{-0.5cm}
\caption{\footnotesize{Images of the globular cluster M92. Left: HST/WFC3 image in F110W ($21 min$ of exposure). Center: LBT AO K band image ($3min$ of exposure). The area of the big green square in the left image is about $20\times20arcsec$. The bright object close to the center of the central image is the reference star used to drive the AO WFS. Right: the 2 small squares shows the elongation of the PSF in the direction of the GS. This effect is due to angular anisoplanatism.}}
\label{fig:FLAO_M92}
\end{figure}
\\Figure \ref{fig:FLAO_M92} shows two images of the globular cluster M92. The left one has been taken with WFC3 of HST, integrating 21 minutes in the F110W filter. The green box highlights the area shown also in the central image. This is an AO corrected K band image taken with PISCES at LBT. The integration time was 3 minutes and the image side measures $20arcsec$ on-sky. The limiting magnitude detected in the 2 images are respectively 20.5 for HST and 23 for the LBT. Within the green area of the HST image 890 stars are detected while in the LBT they amount to 3300.
\\The GS used to perform the wavefront sensing in the central image of figure \ref{fig:FLAO_M92} is the bright object close to the center. Looking at stars at an increasing distance from the center a slight elongation in the radial direction appears (see the 2 zooms on the right of figure \ref{fig:FLAO_M92}). This effect is caused by the \emph{angular anisoplanatism} \cite{1982Fried-Anisoplanatism_in_AO} and it is the main limitation of AO systems that use a single GS, called Single Conjugate AO systems (SCAO).

\subsection{Angular anisoplanatism}
\label{ssec:ang_anisopl}
\begin{figure}
\begin{center}
  \includegraphics[width=4.5cm]{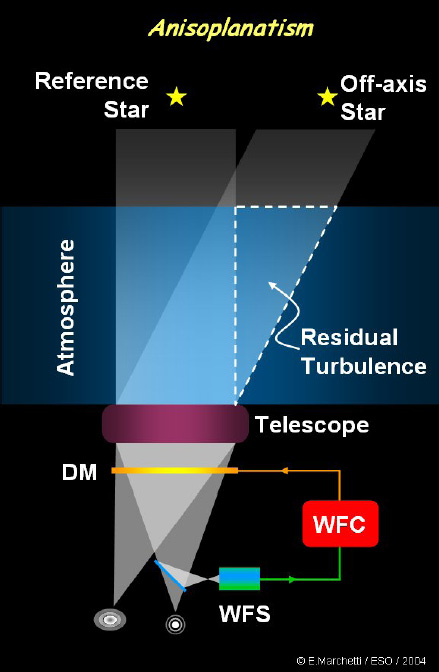}
\end{center}
\vspace{-0.5cm}
\caption{\footnotesize{Scheme of the effects of angular anisoplanatism on AO correction. An increasing angular separation between the AO GS and the science target causes the wavefronts from the 2 sources to become uncorrelated because the part of atmosphere they have passed through is different. Image courtesy of E. Marchetti (ESO).}}
\label{fig:ang_aniso_scheme}
\end{figure}

SCAO systems measure the wavefront aberrations integrated on the overall cylinder of turbulence in the direction of the GS. So an increasing angular distance between the GS and the scientific object will cause the wavefronts coming from the 2 direction to be uncorrelated and it will hence decrease the AO correction achievable. This effect is exemplified in figure \ref{fig:ang_aniso_scheme}. The angular separation at which the wavefronts from the GS and the science object will be uncorrelated is called the \emph{isoplanatic angle}. Typically $\theta_0$ is less $30arcsec$ at infrared wavelengths.

\section{Wide field AO correction}
\label{sec:glao}
\begin{figure}
\begin{center}
  \includegraphics[width=6cm]{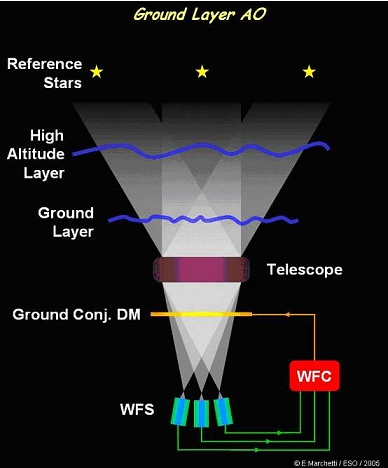}
\end{center}
\vspace{-0.5cm}
\caption{\footnotesize{Scheme of the working principle of GLAO systems. Several GS placed at the edge of the science FoV are used to perform a tomographic measure of the atmospheric turbulence. The ground layer structure is extracted averaging the several WFS measurements and it is compensated by a single DM optically conjugated close to the telescope aperture. Image courtesy of E. Marchetti (ESO).}}
\label{fig:GLAO_scheme}
\end{figure}

To increase the FoV over which the AO correction is effective it is necessary to perform a tomographic measure of the turbulence using several WFS measuring the wavefront distortions in the direction of several GS \cite{1975_Dicke_MCAO}. To fully compensate for the the atmospheric turbulence in 3 dimensions it is necessary to introduce many DMs in the optical path. Typically each DM is optically conjugated to a certain distance from the telescope pupil where the strongest turbulence layers are located, these are Multi Conjugate AO systems \cite{1988_Beckers_MCAO}.
\\In case of a single DM is available the turbulence compensation can only be partial. The typical choice in this case is to conjugate the single DM to an altitude of several hundreds meters above the ground where most of the turbulence lays, so these are called Ground Layer AO systems. GLAO systems extract the structure of the ground layer turbulence averaging the measurements of the several WFS pointing at GS placed at the edges of the instrument FoV \cite{2004_Tokovinin_GLAO_GS_position}. In this way the lower layers, where light patterns of the different GS overlap, are enhanced while the higher layers, where the light patterns are uncorrelated, contribute with a residual error to the GLAO correction performed \cite{2004_Nicolle_GLAO_SPIE}. A scheme of the working principle of GLAO system is shown in figure \ref{fig:GLAO_scheme}. Due to the large contribution of the residual high layer turbulence the AO correction provided by GLAO systems is only partial but is however sufficient to give a substantial reduction of the atmospheric seeing giving an uniform and stable PSF over a wide FoV \cite{2002_Rigaut_GLAO}.

\section{Laser guide star AO}
\label{sec:LGS_AO}
AO systems need to find a bright star close to the science target, or many of them just a little bit away form the object in MCAO systems, to perform the wavefront sensing. The lack of a bright GS will limit the performance of the AO system: in fact looking for a fainter GS increases the noise contribution to the wavefront measurement while extending the search radius of the  Natural GS (NGS) increases the angular anisoplanatism contribution. So the limiting magnitude of the GS that can be used defines the \emph{sky coverage} of the AO system.
\\A solution to the limited sky coverage of AO systems based on NGS is to artificially generate a bright GS close to the science object \cite{1985_Foy_LGS}. The light of the artificial stars is provided by projecting a laser beam into the atmosphere. The physics used to generate the reference source in the atmosphere divides the Laser Guide Stars (LGS) in 2 types: Sodium and Rayleigh LGS.
\\Sodium LGS \cite{2003_Gavel_Lick_LGS_SPIE} \cite{2006_Wizinowich_LGS_KECK_SPIE} exploit the absorption and re-emission of laser light tuned at $589.6nm$ from Sodium atoms in the Earth's mesosphere to generate sources at $\sim100km$ altitude.
\\Rayleigh LSG \cite{2008_Martin_GLAS} base on the elastic scattering by atmospheric particles smaller than light wavelength that diffuses photons of a pulsed laser beam, usually focussed at $10\div20km$ altitude. A deeper analysis of the solutions necessary to implement an AO system based on Rayleigh LGS is given in section \ref{ssec:Ray_LGS}.
\\Both types of LGS have their pros and cons and one type is preferred to the other one depending on each specific AO application.
\\A common limitation of both types of LGS is due to the so called \emph{cone effect}. The finite height of the artificial reference sources effectively limit the capacity of the AO system to sense the higher layers of atmosphere. A more detailed description of the cone effect and its consequences on the performance of the AO system will be given in section \ref{ssec:LGS_lim}.
\\Reaching higher altitudes the Sodium LGS are less affected by cone effect than Rayleigh LGS, so AO systems designed to produce high level of correction must relay on Sodium LGS. The GLAO case however is different: these systems are designed to compensate only for the lower layers of atmosphere and hence keeping the LGS at lower altitude could help in reducing the residual wavefront error due to the bad sampled higher layers.
\\The choice of implementing a GLAO system on a $8m$ class telescope using Rayleigh LGS is driven also by technical requirements. A big issue in this case is to keep simple and reliable the system with low maintenance costs. Since nowadays lasers used to generate Sodium LGS are still experimental fully custom made, the possibility to use commercially available solid-state lasers to generate Rayleigh LGS make them preferable for GLAO applications (see section \ref{sec:LGS_GLAO}).

\subsection{Limits of LGS AO}
\label{ssec:LGS_lim}
\begin{figure}
\begin{center}
  \includegraphics[width=4.5cm]{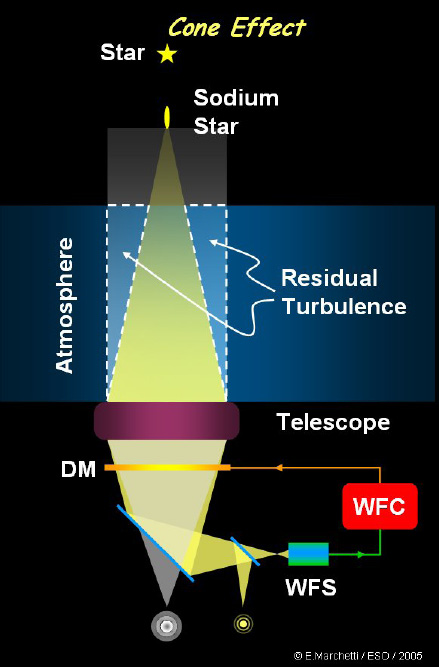}
\end{center}
\vspace{-0.5cm}
\caption{\footnotesize{Scheme of the cone effect due to the use of a Sodium LGS as reference star for AO correction. The finite distance at which is generated the LGS causes a mismatch between the volume of atmosphere sensed with the LGS and the volume that has to be corrected for the science target. Image courtesy of E. Marchetti, ESO.}}
\label{fig:focus_aniso}
\end{figure}

The main issues related to the use of artificial GS in AO systems are related to:
\begin{itemize}
  \item the increase in complexity, since the telescope has to be provided with a high power laser generation system and launch optics,
  \item the loss of performance because the finite height of the LGS since they both need atmosphere to be generated.
\end{itemize}
Typically the second point is referred as \emph{focus anisoplanatism} \cite{1995_Fried_focus_aniso}, or cone effect. Focussing the laser beam at an altitude $H$ means that the atmospheric aberrations introduced by the turbulence placed above $H$ are not sensed by the WFS. In addition since the LGS is focussed at $H$ the part of atmosphere sensed by the WFS at an altitude $h<H$ is reduced by a factor $(1-h/H)$, so the outer part of the science wavefront is not sensed at all while aberrations experience by the inner part are scaled differently in the science and LGS wavefronts. The scheme in figure \ref{fig:focus_aniso} give a 2D representation of the cone effect.
\\The wavefront variance between the measured LGS wavefront and the wavefront coming from a scientific object is dependent on the vertical distribution of turbulence. In LGS-AO applications additional contributions to the residual wavefront error come from:
\begin{itemize}
  \item the finite depth on sky of the LGS (due to range gating or Sodium layer thickness) generates an elongated spot in case of a SH type WFS is used.
  \item Vibrations in the launch optics, telescope pointing instabilities and differential tilt experienced in the launch and return paths of the laser will move the LGS position on sky. This effect, often called LGS jitter, introduces spurious tip-tilt error in the LGS-WFS measurements and it can be compensated providing each LGS-WFS with an independent stabilizing mirror.
  \item The double propagation of the laser light in the atmosphere means that the LGS beam is deflected twice, on its way upwards and downwards, whereas science beam experiences only one deflection. Since the propagation time is much smaller than the time scale of turbulence fluctuations the two paths tends to coincide in the particular case in which the LGS is launched by the primary mirror of the telescope. In general this means that the atmosphere tip-tilt error is not properly sensed by the LGS. So to determine the wavefront tilt of science objects, a dedicated NGS based WFS must be used.
\end{itemize}

\subsection{Rayleigh LGS}
\label{ssec:Ray_LGS}
\begin{figure}
\begin{center}
  \includegraphics[width=10cm]{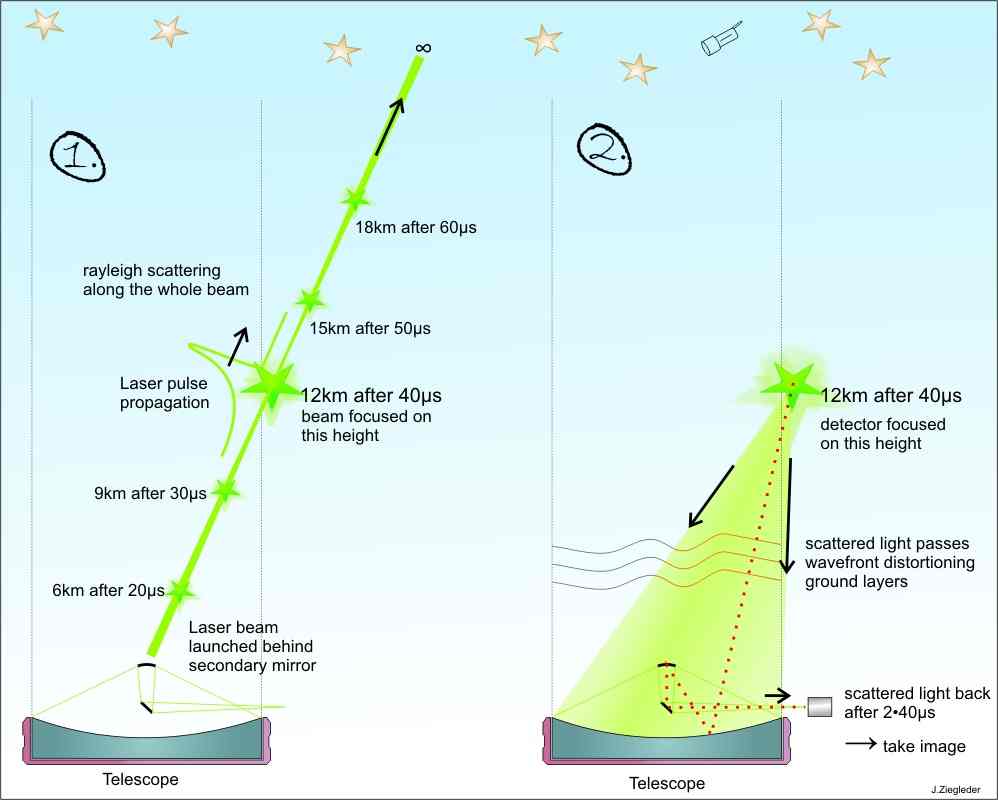}
\end{center}
\vspace{-0.5cm}
\caption{\footnotesize{Left: scheme of the upward propagation of a laser pulse in the Earth atmosphere with increasing time. Right: scheme of the backward propagation of the light scattered at $12km$ altitude toward the telescope and the WFS unit. Images courtesy of J. Ziegleder (MPE).}}
\label{fig:ARGOS_rayleigh}
\end{figure}

The Rayleigh LGS are generated by the laser photons that are backscattered from the air molecules. This elastic scattering is due to the displacement of the electronic cloud surrounding gaseous molecules, or atoms, that are perturbed by the incoming electromagnetic field \cite{2009_Handbook_of_Optics_Bass}. The phenomenon is associated with optical scattering where the wavelengths of light are larger than the physical size of the scatterers.
\\The Rayleigh scattering cross section varies with wavelength as $\lambda^{-4}$ so using lasers at shorter wavelengths helps to increase the efficiency of the Rayleigh LGS. At the same time the Rayleigh cross section varies with the density of air particles as $N^{-2}$. Since atmosphere particle density strongly decrease with the altitude to generate a bright guide star for AO with the modern $10-20W$ class solid-state laser the beam must be focussed at altitudes lower than approximately $20km$.
\\Since a large amount of light is backscattered by lower atmosphere to reduce the light pollution effects in the LGS-WFS, that lower the signal-to-noise ratio achievable by the WFS, it is preferable to use pulsed lasers to generate Rayleigh LGS and to couple them a temporal range gate system. This system allows to gate a small vertical section from which the backscattered photons reach the WFS detector. Many technologies are employed to create the range gate \cite{2002_Thompson_range_gating_PASP}, mainly divided in mechanical (like shutters) and electro-optical (as the Pockels cell).
\\The timing input of the range gate system sets the distance from the launch aperture and the altitude elongation of the LGS. This timing input is triggered according to the laser pulses width, that are defined by the FWHM of the pulse energy. To generate Rayleigh LGS are commonly used solid-state Nd:YAG lasers, although other materials, such as Yb:YAG have been used. A Nd:YAG has a pulse width of approximately $100ns$ FWHM, while Yb:YAG has a longer pulse width of $400ns$ FWHM that corresponds approximately to range gating of $30m$ and $120m$ respectively.
\\Figure \ref{fig:ARGOS_rayleigh} shows a scheme of the propagation of a laser pulse in the atmosphere and its position at increasing time. While the main pulse continues to propagate upward with decreasing energy, a portion of the backscattered light is collected by the telescope aperture and it is directed toward the WFS, focussed in correspondence of the center of the gated range.

\section{LGS-GLAO facilities}
\label{sec:LGS_GLAO}
Despite the more complications related to generate and use artificial GS in wide field AO systems, the possibility to perform the AO correction over almost the full sky is much more attractive. In the past years many LGS based wide field AO system have been proven on sky, among which the ones that have tested the performance of GLAO correction are: GLAS at WHT \cite{2006_Rutten_GLAS_WHT}, the GLAO system of the MMT \cite{2009_Baranec_MMT_GLAO_ApJ} and the SAM project at SOAR \cite{2003_Tokovinin_SOAR_LGS_AO}.
\\Also the LBT is building a LGS-GLAO facility. Such a system is designed to increase the scientific return and efficiency of the LUCI imaging and spectroscopic modes. A detailed description of the LBT LGS-GLAO facility will be given in chapter \ref{cap:argos} where also will describe the expected performance of the system.

\subsection{GLAS}
\label{ssec:GLAS}
\begin{figure}
\begin{center}
  \includegraphics[height=6cm]{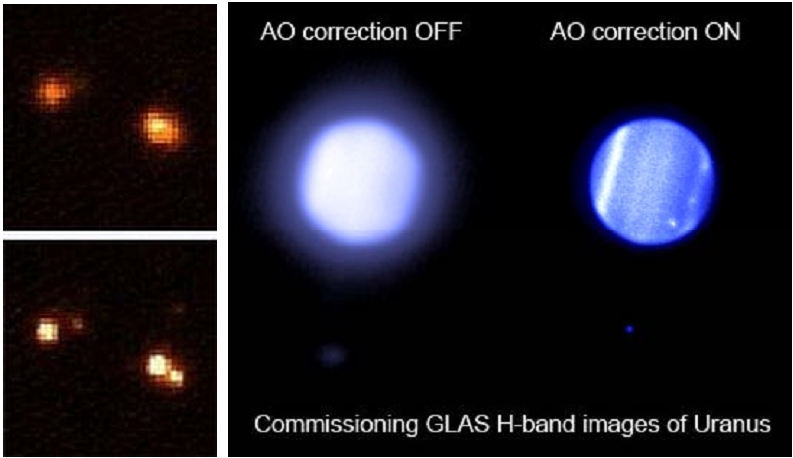}
\end{center}
\vspace{-0.5cm}
\caption{\footnotesize{Two astronomical images taken from the commissioning of GLAS. Left: comparison of stars in M15 in seeing limited conditions (top) and with the GLAO correction provided by GLAS (bottom). Rright: Uranus with and without AO.}}
\label{fig:GLAS}
\end{figure}

GLAS uses a single Rayleigh beacon to create an artificial star at an altitude of $15 km$. The light source is a commercial frequency doubled Yb:YAG disc laser, emitting an output power of $20 W$ at a wavelength of $515 nm$. The laser is launched on sky by a launch telescope designed as a folded Galilean refractor. The launch telescope aperture is $350 mm$ and it is fully mounted behind the secondary mirror of  the telescope. GLAS does not implement fast steering optics to stabilize the uplink path of the beam.
\\On the downlink side the LGS is sensed by a Shack-Hartmann array based on a CCD395. The range gating of the LGS is realized with Pockels cells with crossed polarizations to be independent of the incoming light's polarization \cite{2008_Martin_GLAS}. The AO correction is performed with a segmented DM having 8 segments across the telescope $4.2m$ diameter.
\\The scientific drivers for GLAS were mainly to increase the sky coverage of the AO instrument NAOMI to nearly 100\% \cite{2005_Benn_NAOMI_AO@WHT}, coupled with point-and-shoot capabilities. The instruments foreseen to be used with GLAS are first OASIS \cite{2006_Morris_OASIS_LGS_WHT}, a visual spectrograph employing an integral field unit with $0.2arcsec$ lenslets and $20arcsec$ FoV, and then INGRID \cite{2003_Packham_INGRID_WHT_MNRAS}, a $40arcsec$ FoV NIR camera with $0.04arcsec\;px^{-1}$ scale.
\\GLAS demonstrated to be able to provide an improved image over a moderately wide field of view. The gain in FWHM is a factor of two or better. Figure \ref{fig:GLAS} shows two examples of the performance of the GLAO correction provided by GLAS. On the left 2 images of stars in M15 shows the difference in light concentration in seeing limited conditions and with GLAO correction. On right image the details of the upper atmosphere of Uranus are clearly visible in the AO corrected image.

\subsection{The MMT GLAO system}
\label{ssec:MMT_GLAO}
\begin{figure}
\begin{center}
  \includegraphics[height=6cm]{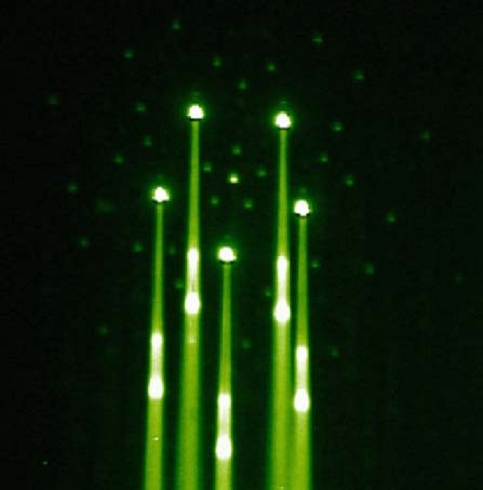}
  \includegraphics[height=6cm]{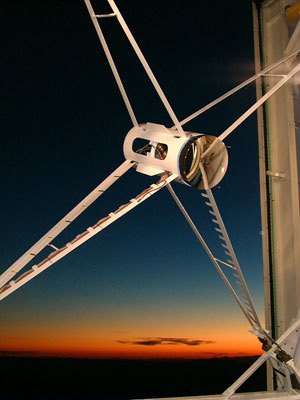}
\end{center}
\vspace{-0.5cm}
\caption{\footnotesize{Left: asterism of the 5 Rayleigh LGS projected on sky at MMT. Right: picture of the ASM mounted on the spider arm at MMT. Image courtesy of MMTO.}}
\label{fig:MMT_GLAO}
\end{figure}

\begin{figure}
\begin{center}
  \includegraphics[width=13.5cm]{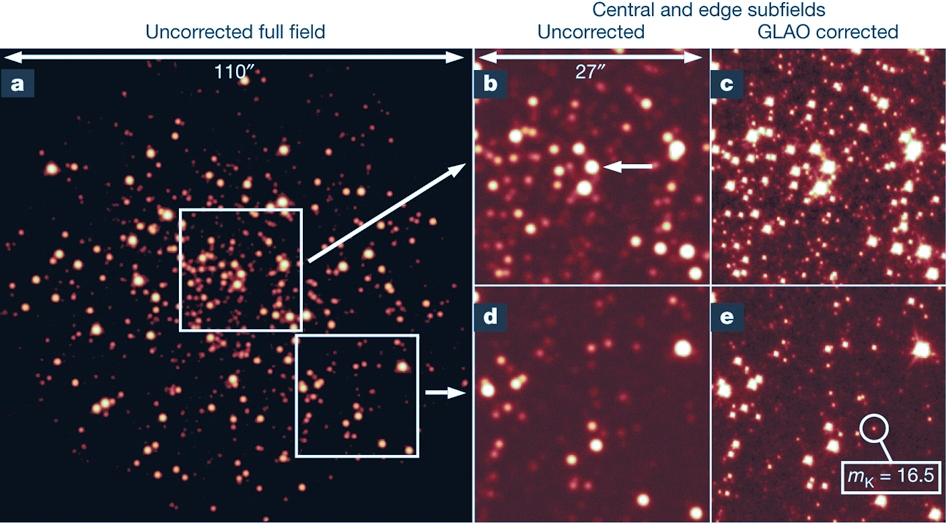}
\end{center}
\vspace{-0.5cm}
\caption{\footnotesize{Image of the globular cluster M3 observed at MMT at K band. Left: seeing limited image of a $110arcsec$ diameter area of the cluster core ($\beta\simeq0.7arcsec$. Right: zoomed subareas of $27\times27arcsec$ in seeing limited conditions (center) and with GLAO correction (right). The white arrow highlights the star used for tip-tilt sensing. More details about images can be found in the text. Images taken from \cite{2010_Hart_MMT_GLAO_Nature}.}}
\label{fig:MMT_Nature}
\end{figure}

The MMT combines the output of two frequency doubled commercial Nd:YAG laser heads to generate one single beam with $24 W$ power. The laser beam is relayed in free air to a pupil box at the upper ring of the telescope, where a holographic phase plate is used to produce a five stars asterism out of the single incoming beam. Once divided the five beams are put through to the folded refractive launch telescope, mounted behind the secondary mirror. A picture of the 5 separated beams propagating through atmosphere is shown on left in figure \ref{fig:MMT_GLAO}.
\\On the downlink side of the system MMT employs unique techniques. A mechanical resonator is used to refocus the laser beacon while the pulse is traveling upwards through the atmosphere. This allows to integrate the scattered light over a longer altitude  interval, collecting more photons, and reaching higher altitudes with the same laser power. The resonator itself is an aluminum cylinder with an eigenfrequency of approximately $5kHz$. A single WFS, a gated CCD manufactured by LLNL, is used to measure all five beacons. This unique feature saves the need for external optical gating mechanisms like a Pockels cell. The AO correction is performed at MMT using the adaptive secondary build on the same technology of the LBT ASM and deformed by 335 voice coil actuators \cite{2003_Brusa_MMT_ASM_SPIE}. A picture of the MMT ASM is shown on right in figure \ref{fig:MMT_GLAO}.
\\Figure \ref{fig:MMT_Nature} shows the core of the globular cluster M3 observed at MMT in K band \cite{2010_Hart_MMT_GLAO_Nature}. The image on left shows an area of $110arcsec$ diameter taken in seeing limited conditions ($\beta\simeq0.7arcsec$). The two white squares highlight subareas of $27\times27arcsec$ in correspondence of the center and at the edge of the cluster. The 2 subareas are zoomed on the right where the images taken in seeing limited conditions are compared to the GLAO corrected ones. It is noticeable the improvement in limiting magnitude for the GLAO images, a gain of $2mag$ is achieved. The star used for tip-tilt sensing is marked with the white arrow in subarea \emph{b}. The PSF shape is very uniform across the areas imaged with GLAO correction, the standard deviation of the measured PSF FWHM is $9mas$.

\subsection{SAM}
\label{ssec:SAM}
\begin{figure}
\begin{center}
  \includegraphics[height=6cm]{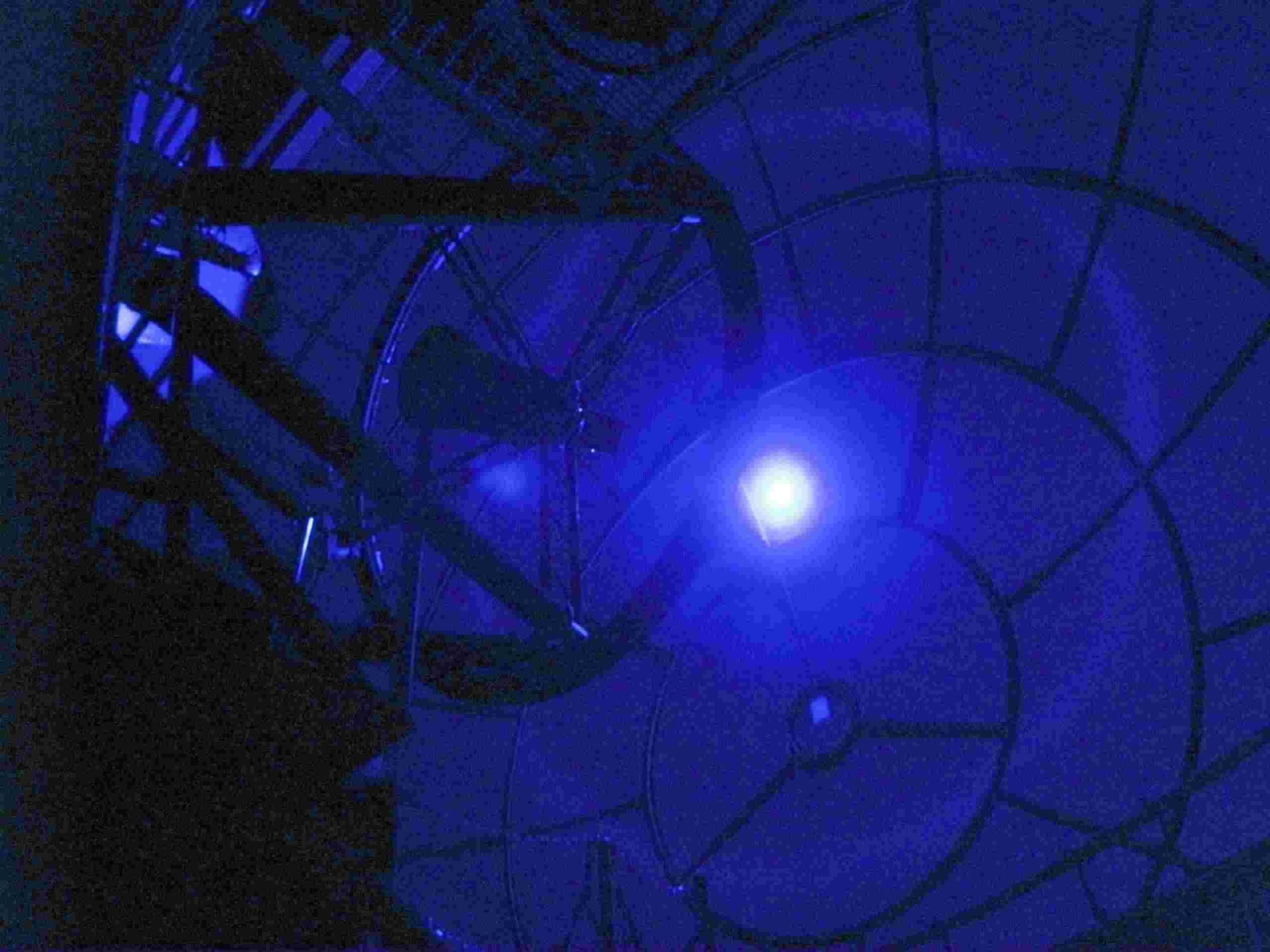}
\end{center}
\vspace{-0.5cm}
\caption{\footnotesize{Picture taken inside the SOAR dome when the SAM UV laser was shined during the AO system commissioning. Image courtesy of SOAR.}}
\label{fig:SAM}
\end{figure}

\begin{figure}
\begin{center}
  \includegraphics[width=13.5cm]{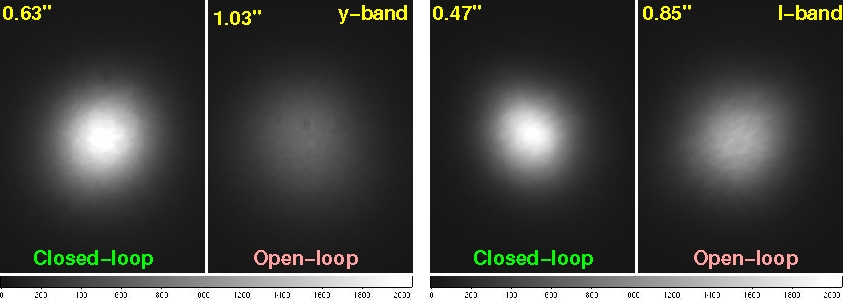}
\end{center}
\vspace{-0.5cm}
\caption{\footnotesize{Images of a bright star at y and I bands taken during the commissioning of the SAM. The improvement in image quality obtained closing the AO loop on the LGS is shown in terms of PSF FWHM reduction. Image courtesy of SOAR.}}
\label{fig:SAM_CL}
\end{figure}

SAM is the SOAR Adaptive Module. It uses an UV laser with $8W$ output power at $355nm$ to generate a Rayleigh LGS. The main advantage of the UV beacon is that the increased efficiency of the Rayleigh scattering at shorter wavelengths allows the use of lower power lasers. The use of this short wavelength imposes consequences for the employed optics; transmission and chromaticity characteristics were chosen accordingly. The launch telescope works with an aperture of $350 mm$. The beam path is enclosed up to the launch telescope that is mounted behind the secondary. Figure \ref{fig:SAM} shows the SAM UV laser shined inside the telescope dome during the system commissioning.
\\In the SAM module the LGS WFS is housed, together with re-imaging optics and a bimorph DM, in a sealed aluminum enclosure mounted on one of the Nasmyth foci of the telescope. The system enclosure rotates to compensate field rotation. The Shack-Hartman wavefront sensing uses a CCD39 as detector. The gating is carried out by a single Pockels cell. Fiber-coupled avalanche photodiodes (APD) are be used for the tip-tilt measurement. On the instrument side the SOAR will be provided with a visible and infrared imaging camera with $3arcmin$ FOV. Figure \ref{fig:SAM_CL} shows improvement in image quality obtained during the SAM commissioning. The AO loop was closed on the LGS and a bright star was images at y and I bands measuring a reduction of the PFS FWHM of a factor 2.

%

%
%
%
%

\chapter{ARGOS: a laser guide star AO system for the LBT}
\label{cap:argos}
\begin{figure}
\begin{center}
\includegraphics[width=6.5cm]{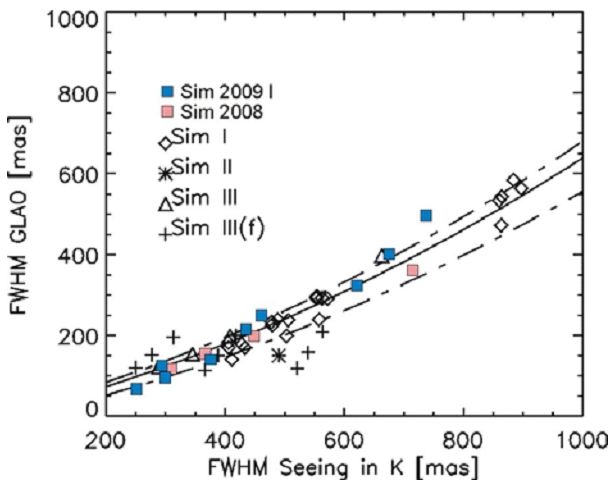}
\includegraphics[width=6.5cm]{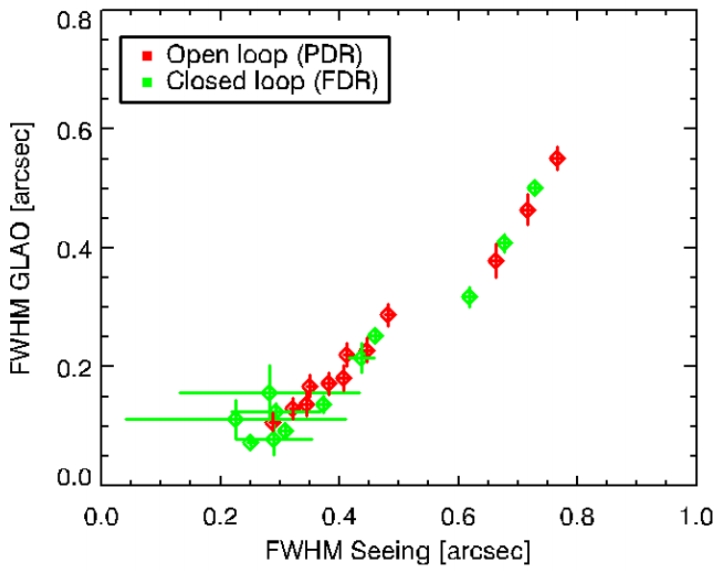}
\end{center}
\vspace{-0.5cm}
\caption{\footnotesize{Left: summary of the ARGOS performance numbers as have been calculated within the various project phases. It is visible that ARGOS can bring an improvement of a factor 2 in FWHM and accordingly in the energy coupled to the MOS slits. Right: comparison between closed-loop and open-loop simulations. Red points show the performance as has been retrieved with an open loop model of the AO. The green dots denote the full closed loop simulation points. The data is shown for the K-band.}}
\label{fig:argos_fwhm}
\end{figure}

The ARGOS project started in 2007 to provide the LBT with a wide field AO system assisted by laser guide stars to ensure an almost full sky coverage. At first ARGOS will implement a ground layer correction to increase the scientific return and efficiency of the available LBT instrumentation, LUCI in particular. The main science requirements on which the system has been designed are:
\begin{enumerate}
  \item The resolution should be improved by a factor 2 for $75\%$ of observable nights, and should reach $0.25arcsec$ on a significant fraction of nights.
  \item An optical tip-tilt sensor should be implemented and it has to be able to perform tip-tilt correction on stars of $m_r\leq18.5 mag$.
  \item Using the LGS-GLAO system should not increase, with respect to the NGS based AO system, the acquisition time by more than $10 min$, and the observing time by more than $30s$ for each dither point during a series of exposures.
  \item A quantitative expression for how the PSF varies across the full field should be provided by the GLAO system with an accuracy (in terms of FWHM uncertainty) of better than $10\%$.
\end{enumerate}
To fulfil these requirements the ARGOS system will implement:
\begin{enumerate}
  \item Multiple high power lasers to achieve a high photon flux on the WFS. In conjunction with a high WFS framerate the ground layer can be corrected with LBT adaptive secondary mirror.
  \item Using Avalanche Photo Diodes (APD) as tip-tilt sensor the image motion due to telescope vibrations and on-sky laser jitter can be reduced to $100mas$ using an $18.5 mag$ star.
  \item The selected strategy to keep overheads as small as possible rely on: implement industrial style robust laser systems, automate the LGS position tracking, implement LGS patrol cameras and automatic acquisition procedures, implement a daytime deployable calibration unit.
  \item The ARGOS software will be provided with an automatic tool to perform the PSF reconstruction on wavefront and seeing measurements.
\end{enumerate}
Section \ref{sec:sys_design} details the design of ARGOS and it gives a brief introduction of the concept of the wavefront sensing unit that I contributed to design and develop since the beginning of my PhD work in 2009.
\\Section \ref{sec:performance} describes a study that I carried out at the beginning of my PhD work to refine the closed-loop analysis of the ARGOS performance, using numerical simulations. The performance of the wide field correction yield by ARGOS had been evaluated by different groups participating to the project during the design studies. These groups performed numerical simulations assuming different approaches evaluating the performance in terms of gain in PSF FWHM between the seeing limited and the GLAO assisted case \cite{2008_SPIE_Rabien_LBT_LGS}. A comparison of the results obtained by the different groups are represented in the left plot in figure \ref{fig:argos_fwhm}. The right plot of figure \ref{fig:argos_fwhm} instead compares the results of open loop and closed loop simulations \cite{2010_Rabien_SysDesign_FDR} carried out for the \emph{Final Design Review} of ARGOS.
\\My contribution to the ARGOS performance estimation went in the direction of refining the results obtained during the design phases, producing end-to-end simulations aimed to be as close as possible to the real system. Section \ref{ssec:FDR_perf} presents the simulation code main features and it analyzes the more sensible parameters. The results I obtained in the study are summarized and commented in section \ref{ssec:sim_res}.

\section{System design}
\label{sec:sys_design}
\begin{figure}
\begin{center}
\includegraphics[width=10cm]{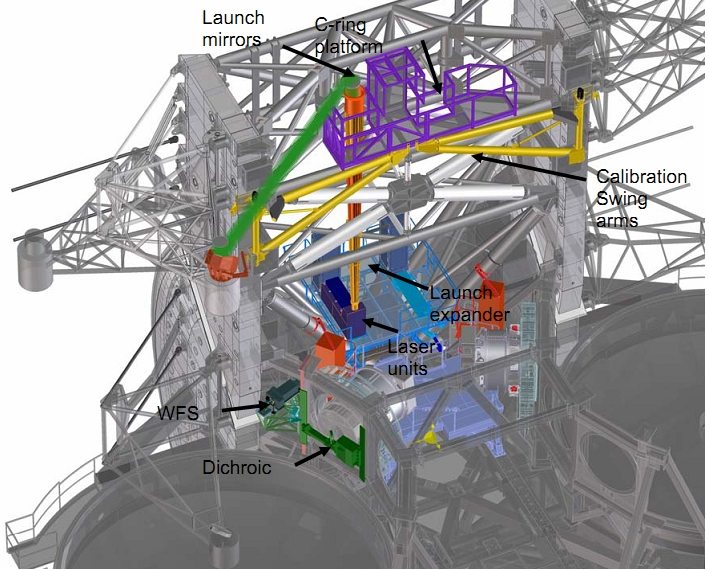}
\end{center}
\vspace{-0.5cm}
\caption{\footnotesize{3D model showing the location of the components of ARGOS (in color) on the LBT structure (in grey). The left eye of the telescope is fully populated while few elements on the right eye are missing.}}
\label{fig:argos_overview}
\end{figure}

\begin{figure}
\begin{center}
\includegraphics[width=7cm]{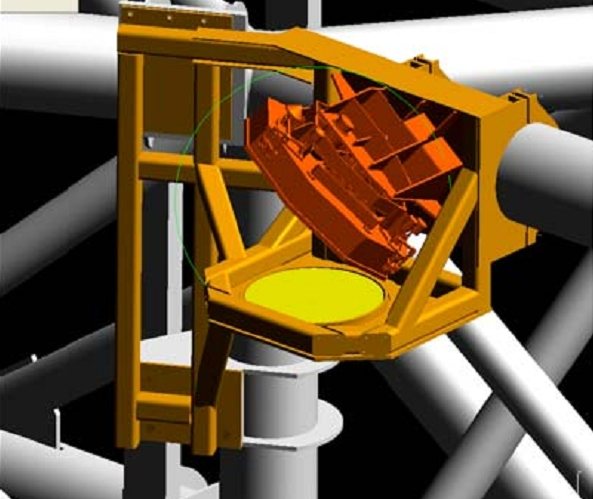}
\includegraphics[width=6cm]{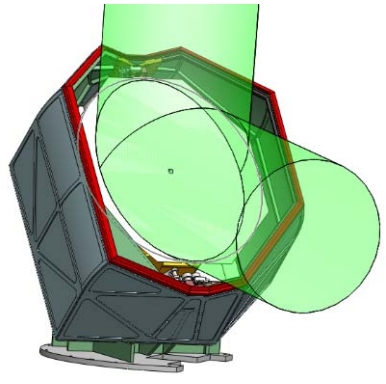}
\end{center}
\vspace{-0.5cm}
\caption{\footnotesize{Detail of the launch system. Left: the large expander lens (yellow) and the first large fold flat (red) are clamped to the upper part of the LBT windbraces. Right: the second fold mirror with dust cover open. This unit will be mounted on top of the M2 hub. From that location on the laser beams are travelling through the amtosphere.}}
\label{fig:ARGOS_launch}
\end{figure}

\begin{figure}
\begin{center}
\includegraphics[width=12cm]{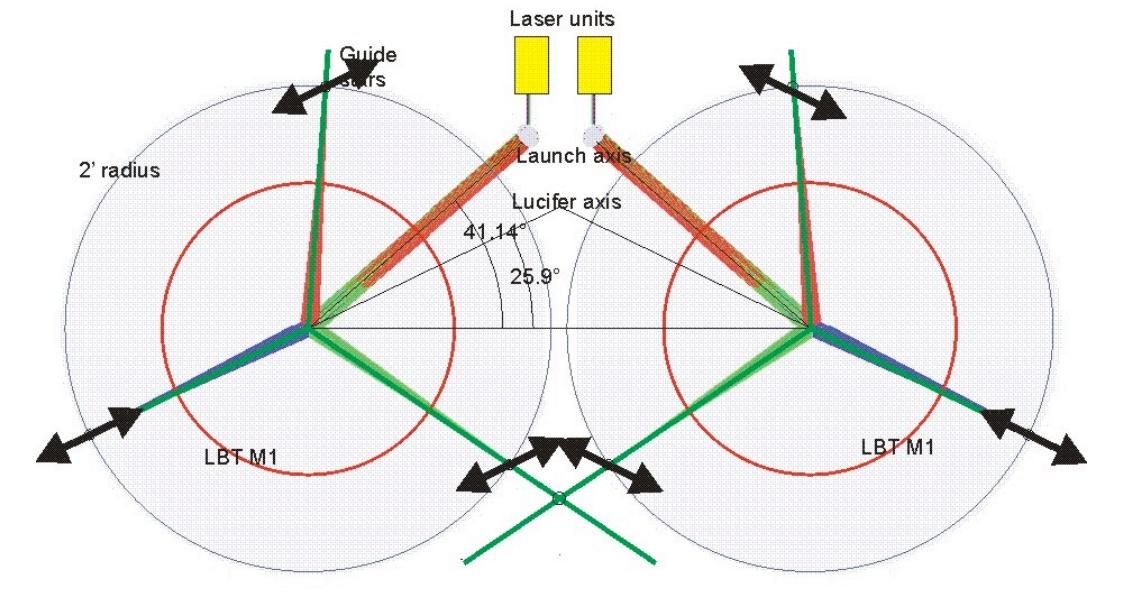}
\end{center}
\vspace{-0.5cm}
\caption{\footnotesize{Top view of the arrangement of the launch beams and of the LGS constellation. Polarization planes of the LGS on sky are denoted with black arrows. The polarization alignment has been selected to be parallel to the LUCI optical axis. The two $8.4m$ diameter primary mirrors of LBT are drawn in red.}}
\label{fig:LGS_asterism}
\end{figure}

\begin{figure}
\begin{center}
\includegraphics[width=6cm]{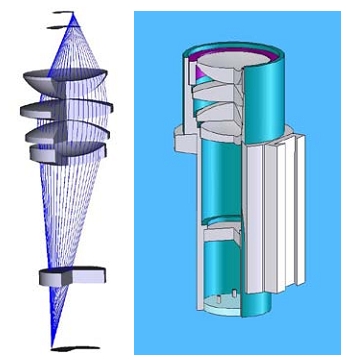}
\includegraphics[width=7.15cm]{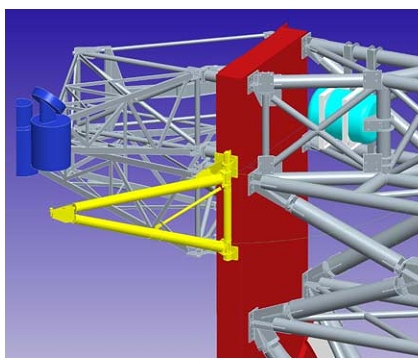}
\end{center}
\vspace{-0.5cm}
\caption{\footnotesize{Optics system within the lens barrel of the calibration unit. The off axis fibres mimic the laser guide stars on sky with the help of a computer generated hologram adding the required phase aberration. The calibration unit optic is placed in the prime focus position by a carbon fiber swing arm mounted on the telescope C-ring (drawn in yellow on right).}}
\label{fig:argos_cal_unit}
\end{figure}

ARGOS projects multiple Rayleigh LGS to perform a 3D measurement of the atmospheric turbulence. The ground layer structure is extracted averaging the measurements of 3 Shack-Hartmann type WFS. The AO correction is performed using the LBT ASM.
\begin{figure}
\begin{center}
\includegraphics[width=12cm]{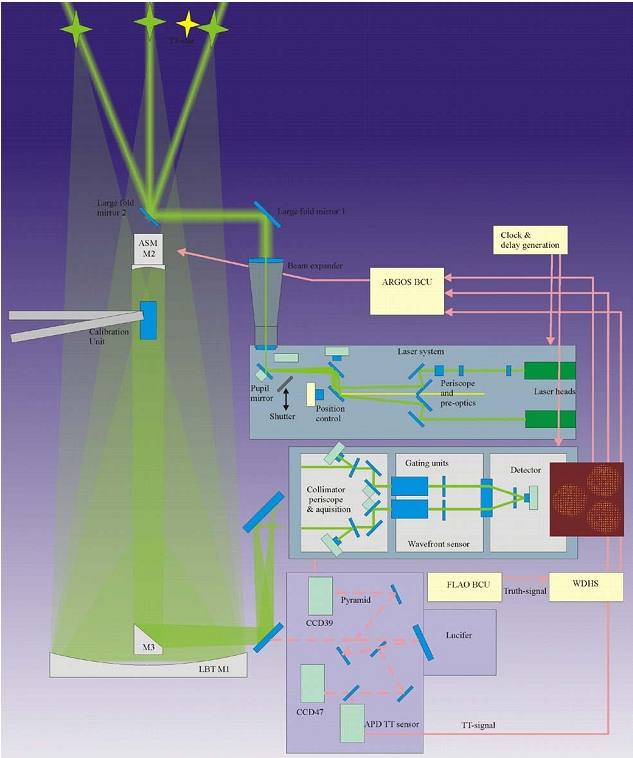}
\end{center}
\vspace{-0.5cm}
\caption{\footnotesize{Scheme of the ARGOS system. A laser system contains all units to generate the laser beams. A launch beam expander and 2 large fold flats direct the laser beams to sky. A dichroic beam splitter separates the green light out and directs it to the ARGOS LGS wavefront sensor. The infrared scientific light is transmitted toward LUCI while visible light is used inside the AGW unit to measure the atmospheric tip-tilt and for truth-sensing. The data collected by the LGS WFS are transferred to the real time computer where the ground layer turbulence is reconstructed averaging the measurements. Finally the BCU on LBT ASM calculate the required mirror shape to compensate for the GL atmospheric distortions. A calibration unit connected to a swing arm can be placed just below the ASM focus to calibrate the LGS WFS and align it with LUCI. Image courtesy of S. Rabien (MPE).}}
\label{fig:sys_scheme}
\end{figure}
\\Figure \ref{fig:sys_scheme} shows a scheme of the devices that compose the ARGOS system at LBT. The laser systems are mounted with a support structure to the windbraces between the C-ring extensions (see figure \ref{fig:argos_overview} to have a detail of the systems allocations at LBT). The unit contain three lasers each emitting, upon a $10kHz$ trigger command, pulses of synchronized $532nm$ light $40ns$ wide. The optics inside the laser system pre-expand the beams to a $6mm$ width and adjust the required polarization. The pointing direction is controlled by a pupil mirror, in common to the 3 beams, placed at the low end of the launch telescope \cite{2010_Kenneganti_ARGOS_LAS_SPIE}.
\\The launch system consists of a refractive beam expander with long focal length being built into the LBT structure. The small entrance lenses are fixed to the laser system and provided with the ability to move along the optical axis to compensate for thermal expansions. On the other end of the expander a large aspheric exit lens of $400mm$ diameter is mounted to the top of the LBT and focusses the 3 laser beams at $12km$ distance. The expanded beam is then folded towards the ASM hub, where a second large fold flat directs it to sky. Figure \ref{fig:ARGOS_launch} shows on left the beam expander exit and the first flat mirror mounted at the top of the LBT windbracing, on right it shows the second flat mirror mounted on top of M2.
\\Figure \ref{fig:LGS_asterism} shows the geometry assumed by the reference sources on-sky. For each source the direction of polarization is also shown. After $80.06\mu s$ from being emitted photons scattered at a $12km$ distance arrive back again at the telescope.
\\The large diameter of the LBT ASM and the fact that is optically conjugated to a $\sim100m$ altitude over M1 makes it an optimal AO corrector for ARGOS. The ASM produces an $f_{16.6}$ beam with the laser light and it focuses objects placed at $12km$ distance $\sim1.4m$ after the $f_{15}$ focal plane of the scientific objects. In front of the LUCI rotator structure the laser light is separated from the scientific light by a dichroic window and it is directed aside the LUCI focal station by a large fold mirror coupled to the dichroic \cite{2008_SPIE_Busoni_ARGOS_WFS}. The dichroic material, shape and coating has been studied in detail to minimize the image quality degradation and light loss due to transmission through the plate. This study is described in detail in chapter \ref{cap:dich}.
\begin{figure}
\begin{center}
\includegraphics[width=12cm]{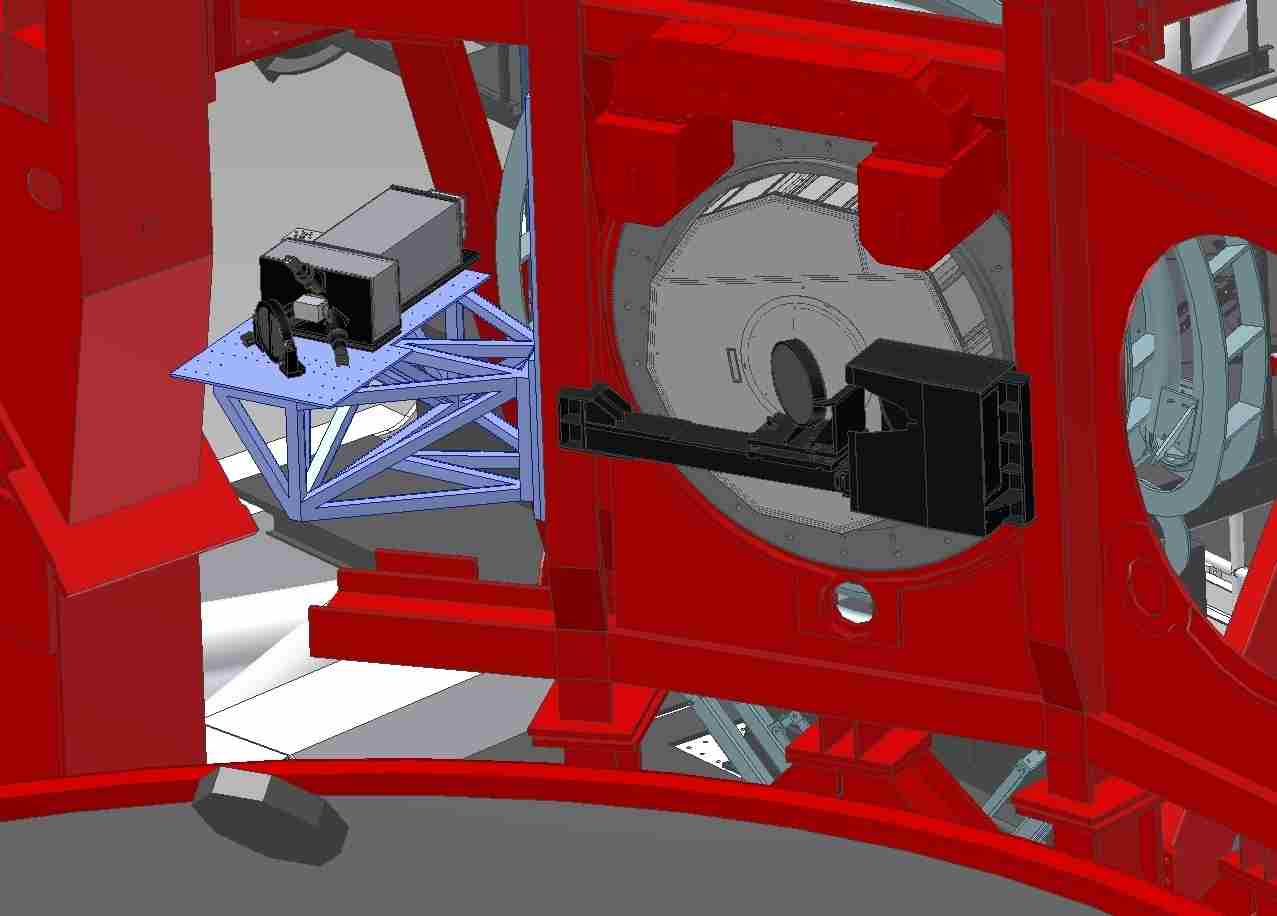}
\end{center}
\vspace{-0.5cm}
\caption{\footnotesize{3D model showing the ARGOS wavefront sensing system. The Rayleigh WFS takes place aside the LUCI focal station, clamped on a dedicated platform. When ARGOS system is in use the dichroic window is placed in the telescope optical path by a motorized chart, to reflect the laser light toward the WFS and transmit the scientific one to the instrument.}}
\label{fig:wfs_overview}
\end{figure}
\begin{figure}
\begin{center}
\includegraphics[width=6.5cm]{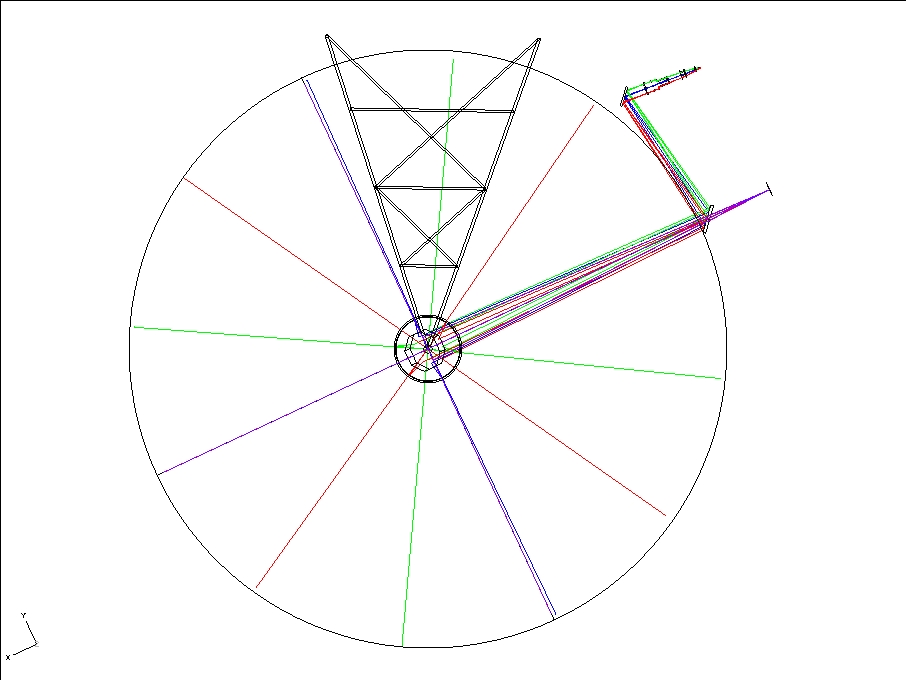}
\includegraphics[width=6.5cm]{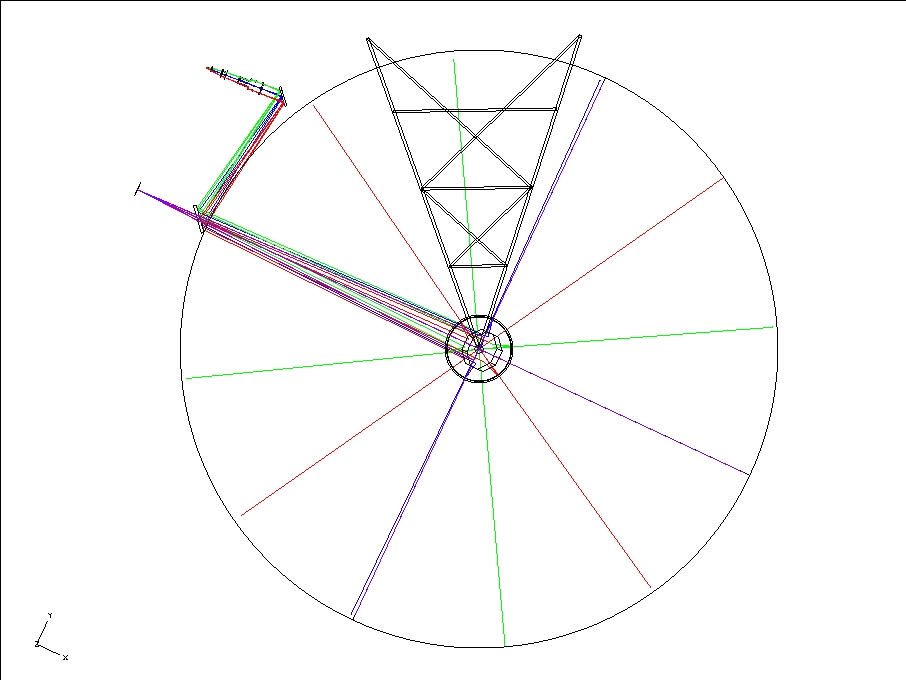}
\end{center}
\vspace{-0.5cm}
\caption{\footnotesize{Optical layout of the 2 eyes of LBT presenting the position of the dichroic and WFS optics as seen from the top of the telescope. The violet rays represent the $f_{15}$ beam of an on-axis object transmitted by the dichroic toward LUCI. The LGS beams instead are separated by the dichroic and directed aside the LUCI focal station by a flat mirror. The structure of the spider arm supporting the ASM is also shown.}}
\label{fig:argos_wfs_position}
\end{figure}
\\Figure \ref{fig:argos_wfs_position} shows a schematic view of the LBT optics from the top with the ARGOS WFS in place. The WFS for the Rayleigh beacons is placed in correspondence of the $f_{16.6}$ focal plane \cite{2010_Busoni_ARGOS_WFS_SPIE}. Three large field cameras patrol this focal plane and allow to locate the LGS on a $1arcmin$ diameter field and recenter them into the $4.7arcsec$ FoV of the WFS. Inside the WFS the beams are first collimated and then stabilized for the uplink vibration induced jitter by a pupil conjugated piezo driven mirror. Pockels cells gate the laser light and transmit towards the WFS only photons backscattered in a $300m$ range centered at the $12km$ distance. The light out of this limited volume then falls through a lenslet array onto the detector creating 3 Shack-Hartman type WFS on the same chip. In the WFS the telescope pupil is sampled with 15 subapertures across the diameter and each subaperture is imaged on a $8\times8px$ area on the chip. The detector itself is a fast large frame PnCCD \cite{2010_Xivry_pnCCD_SPIE} with $256 \times 248$ pixels of $48\mu m$ width. A more detailed description of the WFS is given in chapter \ref{cap:wfs}.
\begin{figure}
\begin{center}
\includegraphics[width=5cm]{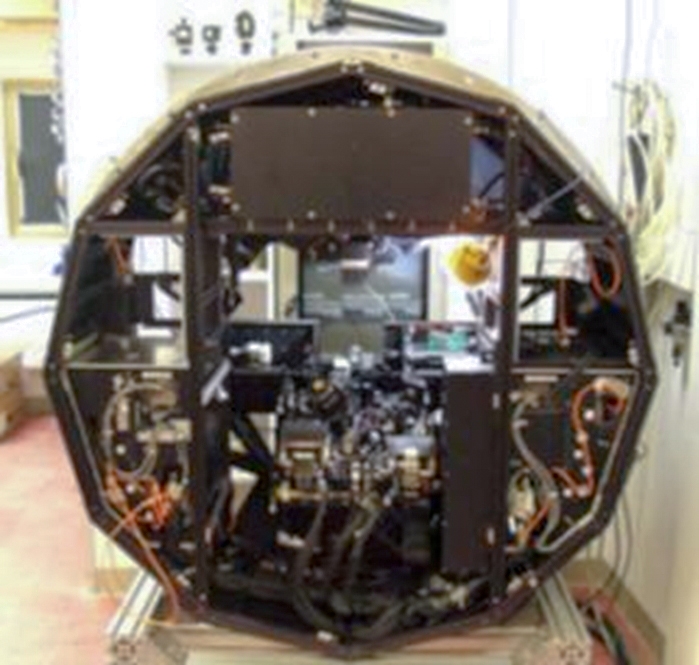}
\includegraphics[width=8cm]{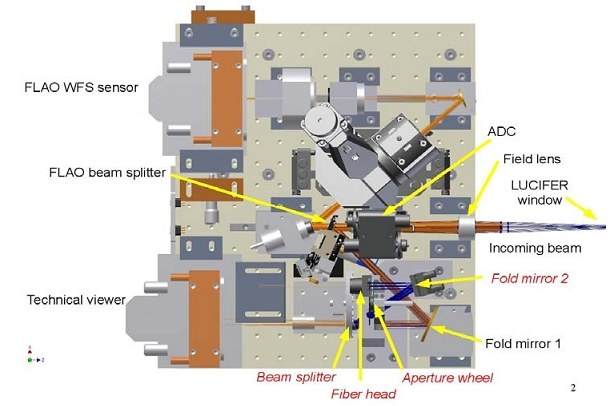}
\end{center}
\vspace{-0.5cm}
\caption{\footnotesize{Left: picture of the AGW unit before it will be installed on the right eye of the LBT during the FLAO commissioning. Right: 3D model showing the FLAO W-board and the foreseen upgrade of the technical viewer arm to install the ARGOS TT sensor.}}
\label{fig:argos_tts}
\end{figure}
\\Since the upward propagation of the laser beacons washes out the atmospheric tip-tilt contribution these low order aberrations are sensed independently by a dedicated Tip-Tilt Sensor (TTS). This sensor is placed inside the AGW unit of the FLAO system, mounted at the rotator structure just in front of the LUCI instrument. A picture of the AGW unit before the installation at LBT is shown in figure \ref{fig:argos_tts}. The TTS takes place on the technical arm of the W-board of the FLAO WFS (see the 3D model in figure \ref{fig:argos_tts}) and it is fed by the $600-1000nm$ light of a NGS picked up within the $1arcmin$ radius of the W-unit FoV. The TT sensor implements a quadrant detection of the NGS with a 4 lenses array with a FoV of $2.3arcsec$ feeding optical fibres. At the end of those fibres commercial Avalanche Photo Diode (APD) units are placed. The use of APDs is justified by the choice to have an high-efficiency photo-counting device for the TT sensing that must be able to work with very faint NGS ($m_r\simeq18.5$), since it is just the ability to find a proper tip-tilt star (TT-NGS) that limits the sky-coverage of an AO system based on LGS. The tip-tilt signals are evaluated by a quad-cell algorithm on the counts recorded by the APDs.
\\The other arm of the W-board, hosting the Pyramid WFS, will be used for Truth Sensing (TS) in ARGOS. This task is necessary since a large portion of the light paths of the Rayleigh WFS and of the instrument are independent and they can be affected by differential variations due to mechanical flexures of the telescope structure during observation. The TS is also sensible to a drift of the LGS altitude on-sky that will be sensed as a focus term by the Rayleigh WFS and that will affect the telescope collimation. Truth sensing is performed integrating few percent of the TT-NGS light ($\sim10\%$) with the Pyramid sensor and off-loading at cadence of $10-20s$ the measured non common path aberrations on the slopes measured by the Rayleigh WFS.
\begin{figure}
\begin{center}
\includegraphics[width=10cm]{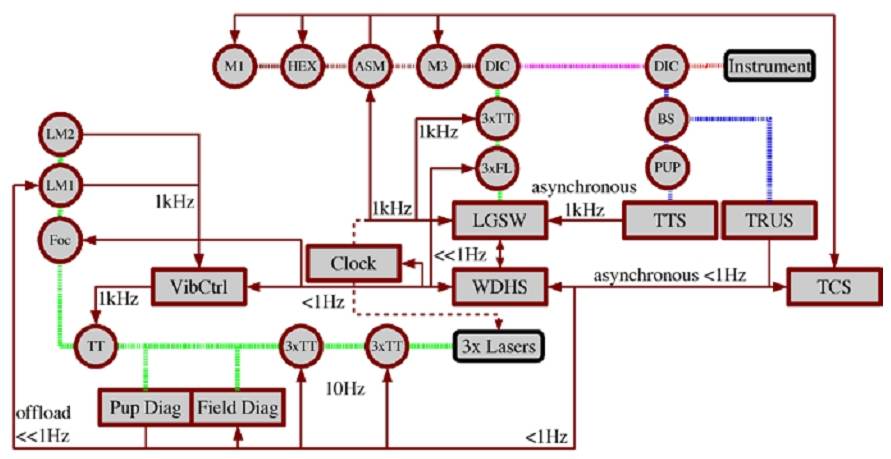}
\end{center}
\vspace{-0.5cm}
\caption{\footnotesize{The scheme shows the control architecture for ARGOS. Circles are optical elements, while rectangles are control units or hardware. The optical path is drawn in ultra fine dashed lines with different colors highlighting the various light paths. The control interaction instead are drawn with solid lines and arrows. The clock is synchronizing the WFS shutters and the lasers emitters, the connection is drawn in dashed red line.}}
\label{fig:argos_loops}
\end{figure}
\\The tip-tilt signals are sent to a Basic Computational Unit (BCU) where are merged with the higher order signals evaluated by a centroid algorithm on the WFS frames. The resulting slope vectors are then sent via optical link to the ASM BCU where the reconstruction is performed and applied as correction, through the 672 actuators, to the thin shell. A complete overview of the control loops that ARGOS will implement at LBT (such as vibration control, synchronization between lasers and shutters) and their typical operating frequencies are schematized in figure \ref{fig:argos_loops}.
\\To calibrate the AO system is necessary to record the interaction matrix between the ASM and the WFS. This is done using artificial light sources placed by a deployable calibration unit at the prime focus \cite{2010_Schwab_ARGOS_CalUnit_SPIE}. An hologram incorporated in the calibration unit optics allow to reproduce the aberrations of the LGS due to their off-axis position. Figure \ref{fig:argos_cal_unit} shows a detail of the optics of the calibration unit and of the swing arm used to position and keep it at the telescope prime focus.

\section{Study of ARGOS performance}
\label{sec:performance}
We already mentioned in the first part of this chapter the results obtained by the extensive campaign of characterization and summarized by the plots in figure \ref{fig:argos_fwhm}. This first set of calculations was run to identify the best solutions for the various ARGOS subsystems during the evolution of the system design: such as number and configuration of LGS on-sky, the limiting magnitude and detector system for the TT sensor, the robustness of the system under different seeing conditions. Then this analysis has been refined introducing the AO loop control and wavefront reconstruction to check for the contribution of these parameters to the residual error. The need to perform end-to-end simulations goes in the direction of perform an analysis of the system performance that will be as complete as possible, taking into account all the features of the system and the real design parameters.

\subsection{The simulation code}
\label{ssec:FDR_perf}
\begin{figure}
\begin{center}
\includegraphics[width=13.5cm]{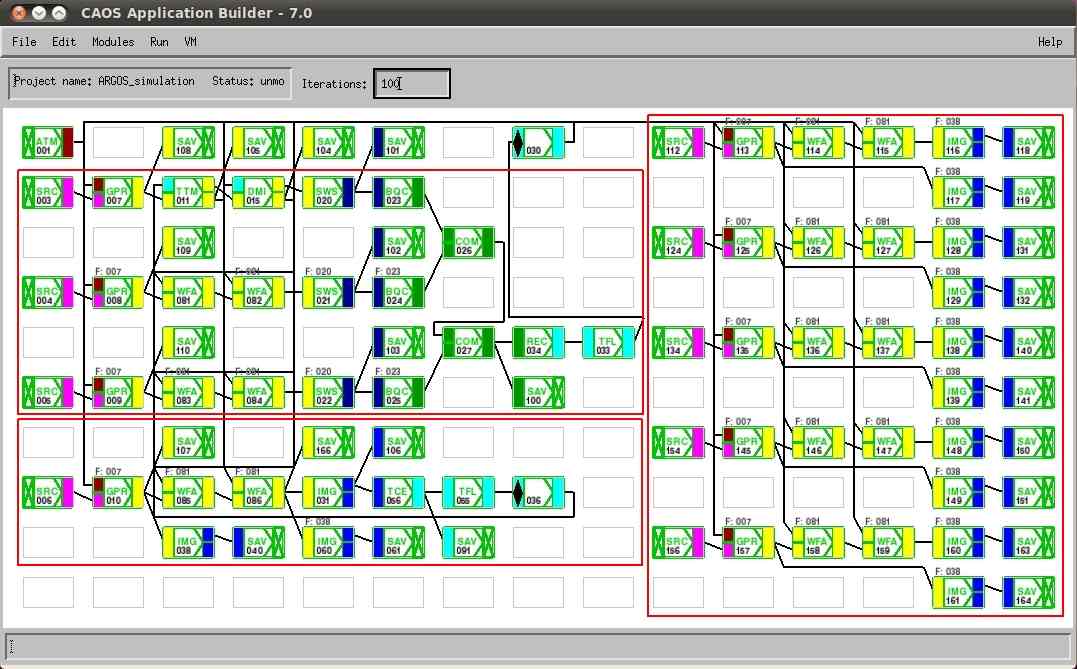}
\end{center}
\vspace{-0.5cm}
\caption{\footnotesize{Screenshot of the CAOS Application Builder showing the modules that compose the ARGOS simulator. The 3 red rectangles are the sets of modules simulating the high order LGS WFS, the tip-tilt NGS WFS and the constellation of reference stars to evaluate the system performance.}}
\label{fig:CAOS_overview}
\end{figure}

The end-to-end simulations of ARGOS we describe in this section have been performed using the Code for Adaptive Optics Systems (CAOS) \cite{2001_Carbillet_CAOS_ASPC} a software widely used in AO simulations. Figure \ref{fig:CAOS_overview} shows the setup of the CAOS Application Builder \cite{2001_Fini_CAOS_AppBuilder_ASPC} where the ARGOS system has been represented.
The main tasks performed by the ARGOS simulator, highlighted with 3 red rectangles in figure \ref{fig:CAOS_overview}, are:
\begin{enumerate}
  \item Sample the atmosphere through 3 extended sources placed at a finite distance from the telescope with 3 Shack-Hartman type WFS to measure the higher order aberrations introduced by the atmospheric turbulence.
  \item Sample the atmosphere through a point-like source at infinity with a quad-cell algorithm to measure the atmospheric tip-tilt.
  \item To image a constellation of five point-like sources to evaluate the AO system performance in different directions of the FoV.
\end{enumerate}
The following sections are dedicated to describe in detail the parameters we used to represent the ARGOS system with CAOS, while the results of the analysis are discussed in section \ref{ssec:sim_res}.

\subsubsection{Atmosphere model}
\label{sssec:atmo}
CAOS simulates the aberration introduced by the atmospheric turbulence through a set of \emph{phase screens}, generated using the \emph{Fourier's method} \cite{1976_McGlamery_PhScreen_SPIE} and weighted by a specific $C_n^2$ profile to account for the specific environmental condition of an astronomical site.
\\As in previous ARGOS simulations in this study we made use of four different $C_n^2$ profiles retrieved in a measurement campaign carried out at Mt. Graham \cite{2008_Stoesz_LBTseeing_SPIE}. These profiles allow to simulate for the statistical distribution of the seeing at LBT and they are resumed in table \ref{tab:phase_screens}.
\begin{table}
\caption{\footnotesize{Weights ($C^2_n$) and wind speeds ($v_{wind}$) associated to the 6 layers on which has been divided the atmosphere. The altitude parameter expresses the distance of the layer from the telescope pupil.}}
\vspace{-0.25cm}
\begin{center}
    \begin{tabular}{|l|c|c|c|c|c|}
    \hline
    \multicolumn{2}{|l|}{\textbf{Profile}} & \textbf{Bad} & \textbf{75\%} & \textbf{50\%} & \textbf{25\%} \\
    \hline
    \hline
    \multicolumn{2}{|l|}{Seeing [arcsec]} & 1.20 & 0.75 & 0.66 & 0.57 \\
    \hline
    \multicolumn{2}{|l|}{$r_0 \;[cm]$} & 8.7 & 13.5 & 15.6 & 18.2 \\
    \hline
    \hline
    \textbf{Altitude [m]} & $\mathbf{v_{wind} [m\;s^{-1}]}$ & \multicolumn{4}{c|}{$\mathbf{C^2_n}$ \textbf{weights}} \\
    \hline
    \hline
    125     & 9     & 0.459 & 0.478 & 0.491 & 0.519 \\
    375     & 9     & 0.141 & 0.141 & 0.154 & 0.154 \\
    625     & 11    & 0.050 & 0.051 & 0.054 & 0.055 \\
    1125    & 15    & 0.037 & 0.038 & 0.041 & 0.041 \\
    3000    & 21    & 0.087 & 0.167 & 0.148 & 0.136 \\
    10000   & 20    & 0.236 & 0.126 & 0.112 & 0.095 \\
    \hline
    \end{tabular}
\end{center}
\label{tab:phase_screens}
\end{table}

\subsubsection{Artificial and natural sources constellation}
\begin{figure}
\begin{center}
\includegraphics[width=6cm]{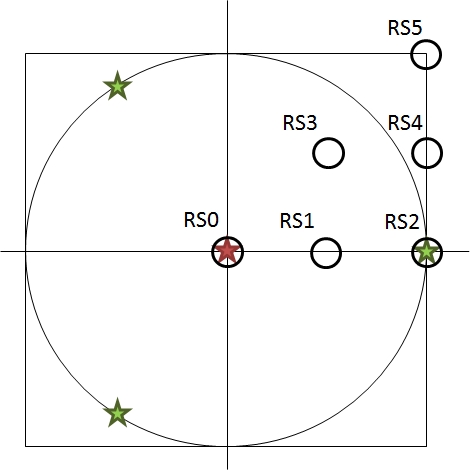}
\end{center}
\vspace{-0.5cm}
\caption{\footnotesize{Arrangement of the reference sources considered in the ARGOS simulations. The objects position is expressed in polar coordinates where the distance from the center of the FoV is given in $arcsec$ and the rotation angle with respect the horizontal axis is given in degrees. The telescope is pointed at Zenith, represented by the intersection of the 2 axes. The external square marks the LUCI FoV of $4\times4arcmin$. The 3 green stars show the position of the LGS: $120arcsec$ off-axis with a $120^{\circ}$ geometry. The red star in the center of the FoV is the NGS used to sense the atmospheric tip-tilt. Finally circles show the positions of the 6 reference sources used to evaluate the performance across the FoV.}}
\label{fig:CAOS_RS_diagram}
\end{figure}

Figure \ref{fig:CAOS_RS_diagram} shows the position of the reference stars in the coordinate system used in simulations. In the diagram the telescope is pointing at Zenith that is also the center of the FoV. The angular distance of the LGS (green stars in figure) is $2arcmin$ in radius. The LGS are modeled as an extended 2D source with a Gaussian profile placed at $12km$ distance from the telescope. The FWHM of the LGS have been evaluated convolving the $0.2arcsec$ FWHM spot generated by the launch telescope by the turbulence profile, to take into account the upward propagation of the laser beacons.
\label{sssec:gs}
\\The propagation of the wavefronts into atmosphere is made under the a geometrical assumption: phases are added linearly, magnifying the portion of phase screen illuminated by the laser beacons. The brightness of the LGS is $m_V\sim7$ that allows a photon flux on the SH subapertures of $1800\gamma \;ms^{-1}$ considering optical transmission coefficient of $0.36$ as expected from the ARGOS design.
\\The NGS used to sense the atmospheric tip-tilt is positioned at the center of the FoV, drawn as a red star in figure \ref{fig:CAOS_RS_diagram}. It is modeled as a point like source at infinity, having a magnitude $m_V=16$, sufficient to ensure an average photon flux of $\sim200\gamma\;ms^{-1}$ on the TT sensor subapertures.
\\Six reference objects cover half of the first quadrant of the FoV and suffice to measure the system performance. The 6 RS are modeled as point like sources at infinity. They are imaged on the focal plane without adding noise contribution or sky background.

\subsubsection{Deformable mirrors}
\label{sssec:dm}
The correction performed by the ASM is simulated by subtracting the wavefront reconstructed on the TT and HO WFS measurements to the wavefront coming from the atmosphere propagation. Since the simulated GLAO loop is running at $1kHz$, one iteration of delay is applied to the HO measurements to represent the time spent by the ARGOS and ASM BCUs to compute the HO slopes from the PnCCD frames and then to reconstruct and apply the correction to the ASM. No clipping is applied to the amplitude of the corrected wavefronts.

\subsubsection{Shack-Hartman WFS}
\label{sssec:SHwfs}
Shack-Hartman sensors measure the phase geometrically propagated from the LGS down to the telescope pupil. The pupil is gridded in $15\times15$ subapertures. Along the pupil perimeter subapertures having less than $0.5$ times the mean illumination value are discarded. The PSF is evaluated for each valid sub-pupil. The focal plane images are then convolved by the 2D profile of the LGS spot and scaled to take into account for the total photon flux on each subaperture. The images are then sampled to the effective number of pixels on the WFS camera. The sky background, the electronics dark current and noise sources (photon and read-out) are added. Finally WFS slopes are evaluated measuring the centroid of the focal plane images without applying any thresholding. Table \ref{tab:SH_WFS_values} resumes the typical parameters used to simulate the ARGOS SH sensors and detector that will be described in detail in chapter \ref{cap:wfs}.
\begin{table}
\caption{\footnotesize{Principal parameters used to represent the 3 SH WFS modules.}}
\vspace{-0.25cm}
\begin{center}
    \begin{tabular}{|l|c|}
    \hline
    \textbf{Parameter} & \textbf{Value} \\
    \hline
    \hline
    Subaperture $\sharp$        &  15   \\
    Sa FoV $[arcsec]$           &  4.6  \\
    Pixel $\sharp$ for sa       &  8    \\
    Sa minimum illumination     &  0.5  \\
    Pixel size $[arcsec]$       &  0.58 \\
    RON $[e^-\;rms]$   &  3.0  \\
    Dark current $[e^-\;s^{-1}]$&  1.0  \\
    \hline
    \end{tabular}
\end{center}
\label{tab:SH_WFS_values}
\end{table}

\subsubsection{Tip-Tilt WFS}
\label{sssec:TTwfs}
\begin{figure}
\begin{center}
\includegraphics[width=8cm]{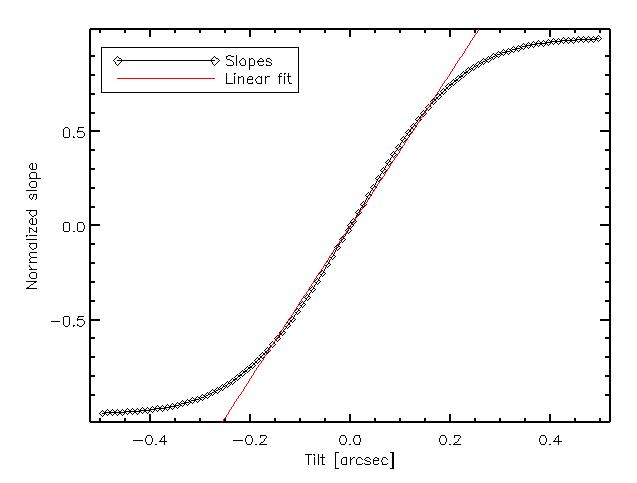}
\end{center}
\vspace{-0.5cm}
\caption{\footnotesize{Calibration curve of the quad-cell sensor. An $0.8arcsec$ FWHM Gaussian source has been moved across the $2arcsec$ FoV of the sensor at $0.01arcsec$ steps. The linear fit is evaluated in the $\pm0.25arcsec$ interval.}}
\label{fig:QC_calib}
\end{figure}

The overall atmospheric tip-tilt error is measured by a quad-cell sensor using an NGS. The sensor has a FoV of $2arcsec$ and it is calibrated using an extended source of Gaussian profile with FWHM comparable with the close-loop PSF FWHM ($\leq0.8arcsec$). Figure \ref{fig:QC_calib} shows the sensor response to a tilt equivalent to $\pm0.5arcsec$ on-sky. The linearity range is between $\pm0.25arcsec$ and it has been calibrated with a linear fit. The noise parameters for the TT sensor reflect the specifications of the APDs having a dark current value of $500e^-\;s^{-1}$.

\subsubsection{Reconstruction and AO loop}
\label{sssec:rec}
The interaction matrix for the 3 SH WFS is recorded applying a wavefront form to the ideal DM and measuring the corresponding signal vectors. These 3 vectors are concatenated and they form the columns of the HO Interaction Matrix (IM). A limited set of Karhunen-Lo\`{e}ve modes, starting from focus up to the 153rd term, are used to record the IM. The modal basis we used is the same employed to study, characterize and optimize the performance of the FLAO system during laboratory acceptance tests \cite{2010_QuirosPacheco_FLAO_char} and on-sky \cite{2010_Esposito_FLAO_SPIE}.
\\The GLAO reconstructor is computed inverting the HO IM using the Singular Value Decomposition (SVD) \cite{1965_Golub_SVD}. This approach corresponds to associate a one-third weight to the measurements of each WFS performing an average the wavefront aberrations integrated in the directions of the 3 LGS.
\\The basic time interval for the numerical simulations is defined by the evolution time step of turbulence ($\Delta t=1ms$ in this case). Any other temporal parameter has to be a multiple of this time interval: for example the integration time of the WFS or the delay applied to the DM correction. The control algorithms used for both the TT and HO loops are pure integrators with a $0.5$ gain factor.

\subsection{Results of ARGOS end-to-end simulations}
\label{ssec:sim_res}
\begin{figure}
\begin{center}
\includegraphics[width=6.7cm]{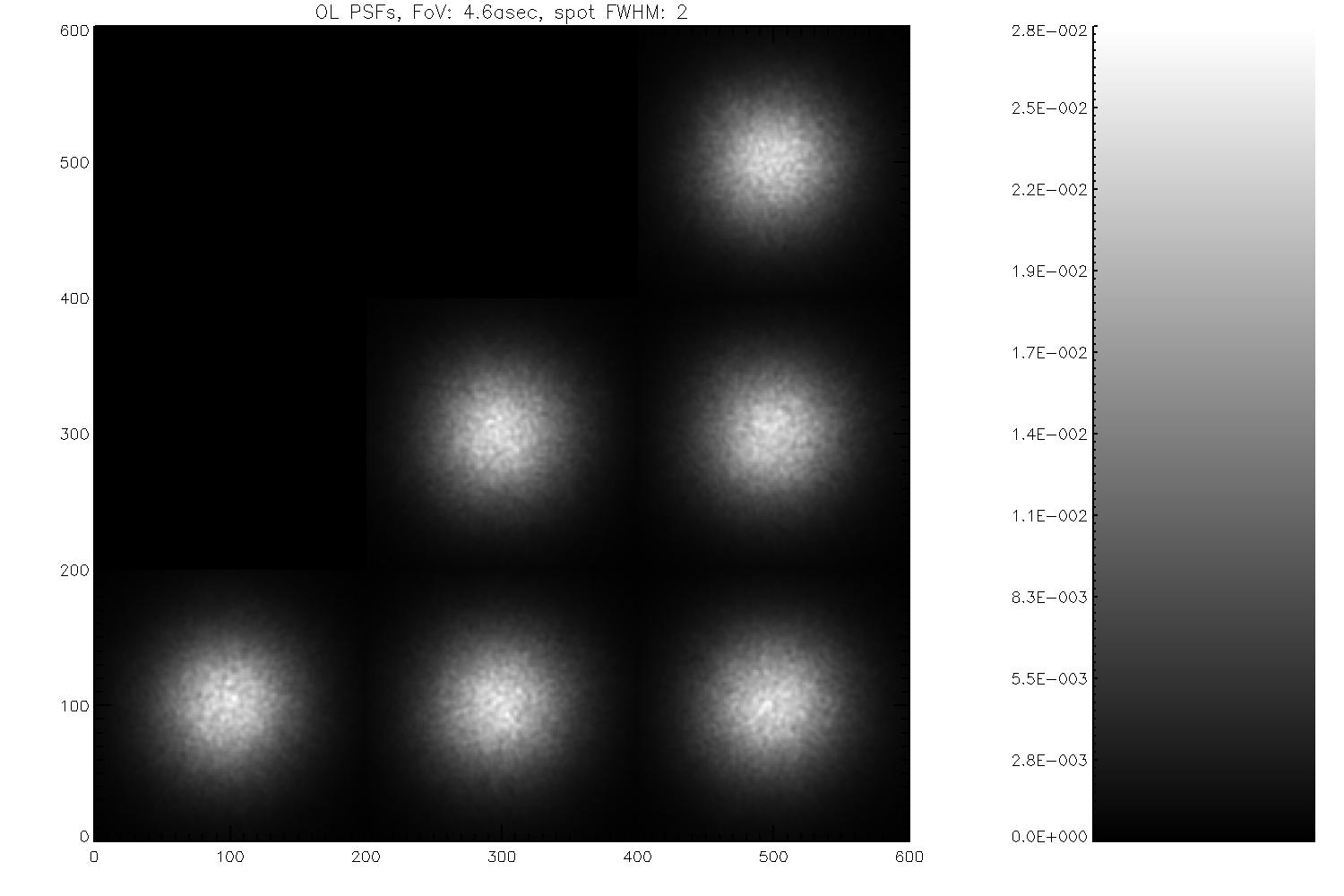}
\includegraphics[width=6.7cm]{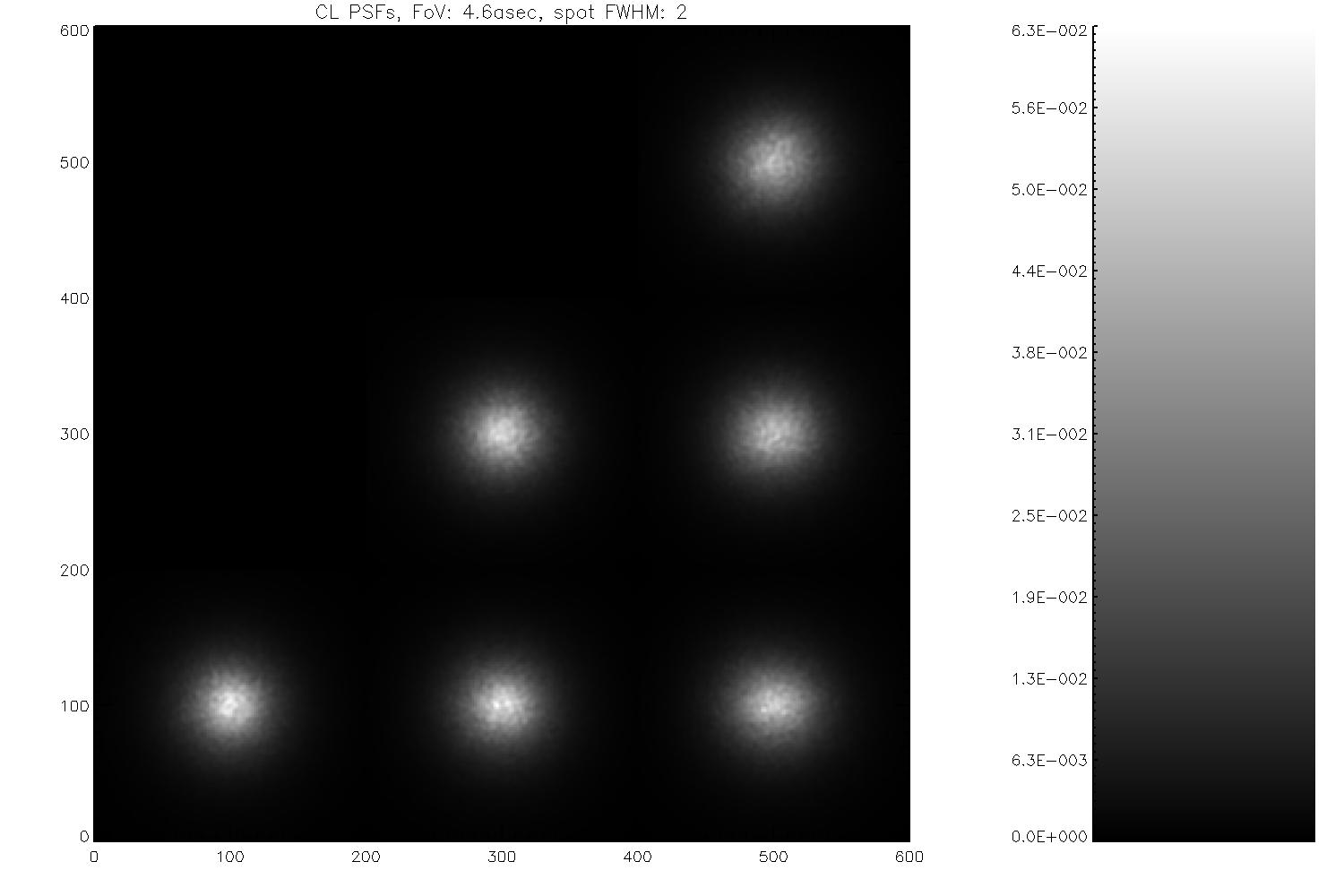}
\end{center}
\vspace{-0.5cm}
\caption{\footnotesize{2D images of the PFSs measured in H band in the direction of the reference sources shown in figure \ref{fig:LGS_asterism}. The PFSs have been sampled with $200\times200$ points corresponding to $2\times2arcsec$ on-sky. The images have been obtained from data without (left) and with (right) subtraction of the GLAO correction in the \emph{bad} atmosphere condition.}}
\label{fig:olVScl_psf}
\end{figure}

\begin{figure}
\begin{center}
\includegraphics[width=13cm]{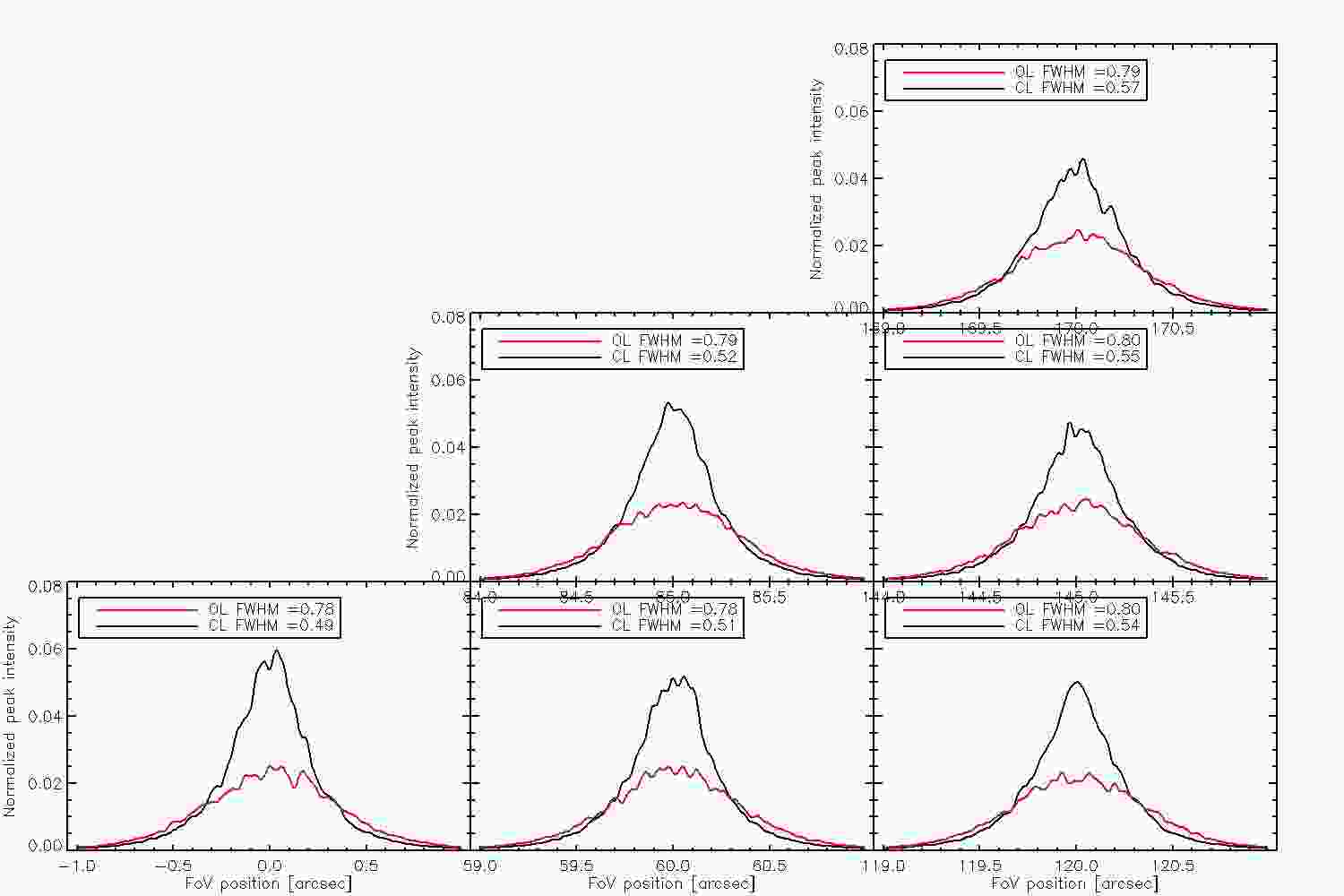}
\end{center}
\vspace{-0.5cm}
\caption{\footnotesize{Cuts across the maxima of the PSFs of figure \ref{fig:olVScl_psf}. The red line shows the open-loop data while the black one refer to the close-loop. The intensity of the PFSs have been normalized to give an integrated value of 1 over the $2\times2arcsec$ area.}}
\label{fig:FWHM_map}
\end{figure}

\begin{figure}
\begin{center}
\includegraphics[width=10cm]{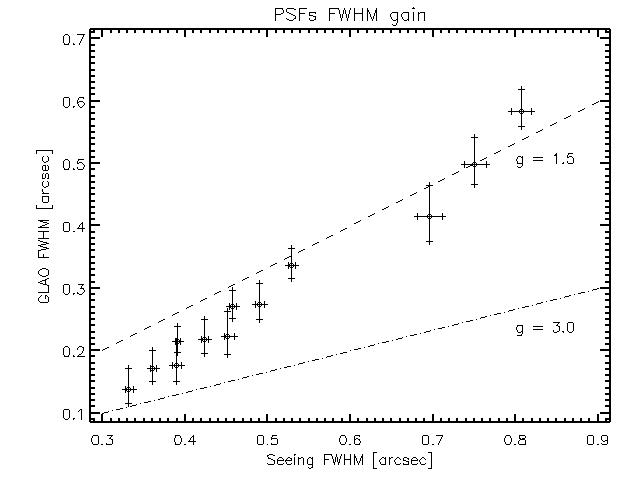}\\
\includegraphics[width=10cm]{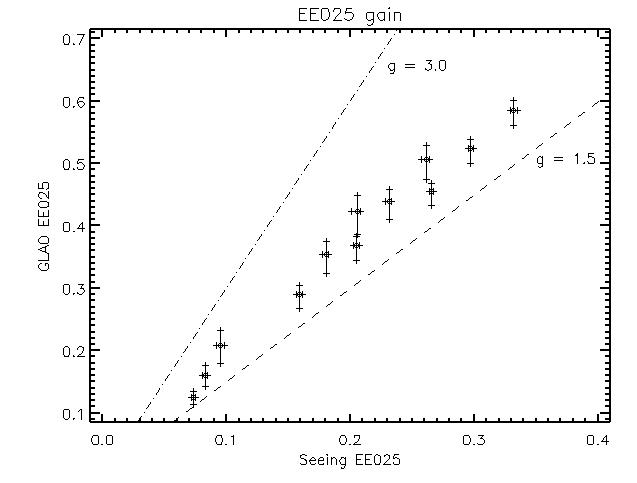}
\end{center}
\vspace{-0.5cm}
\caption{\footnotesize{Summary of the results obtained with ARGOS end-to-end simulations. A gain of a factor $1.5-3$ is visible on both PSF FWHM (above) and EE0.25 (below) in different seeing conditions.}}
\label{fig:FWHM_EE_gain}
\end{figure}

The main contribution to the residual wavefront error in a GLAO system is due to the turbulence of the uncorrected high atmospheric layers \cite{2004_Nicolle_GLAO_SPIE}. This means that a large amount of integration time ($>30s$) is required to obtain well averaged PFSs on which estimate the GLAO loop performance. Considering that the base time step for the simulations is $1ms$ we need $>30k$ iterations and hence phase screens with a very large number of points. Since this procedure is time consuming we choose to perform many measurements using different sets of independently generated phase screens. Each one of these realizations is made of several iterations that are sufficient to ensure the convergence of the loop and to produce stable PFSs. The residual wavefronts produced by these independent loops, considered only after the convergence, are merged together to produce better averaged PSFs. For this study we produced 50 sets of independent phase screens so we could average 50 $0.1s$ simulation runs to evaluate the GLAO performance.
\\As we already discussed the optical quality obtainable in images taken through the atmosphere is measured by the seeing value, that depends both on wavelength and Fried parameter $r_0$. In the simulations were used the same atmospheric profiles employed for the previous studies of ARGOS that are equivalent to $r_0$ values of $0.18$, $0.16$, $0.13$ and $0.09m$, measured at $0.5\mu m$. We refer to these profiles as $25\%,50\%,75\%$ and \emph{bad}. So four different sets of simulations have been run, properly scaling the 50 sets of phase screens we generated to check the performance of the system under different power of the atmospheric turbulence.
\\In an analytical approach it is possible to evaluate the performance of an AO system based on LGS evaluating evaluating the contribution to the AO error budget of the following parameters:
\begin{enumerate}
  \item The spatial wavefront error form un-sampled turbulence above the guide stars and the cone effect (see section \ref{ssec:LGS_lim}).
  \item The finite spatial correction due to the DM actuators pitch and sampling of the WFS.
  \item The temporal wavefront error from delays in the AO loop.
\end{enumerate}
An indicative estimate of the contribution of these parameters to the residual wavefront error can be done using approximated formulas. For example the error due to the cone effect $\sigma_{cone}$ can be expressed as:
\begin{equation}\label{eq:sigma_cone}
    \sigma_{cone}^2 = \Big( \frac{D}{d_0} \Big)^{5/3}.
\end{equation}
Considering a $12km$ height Rayleigh LGS and a good astronomical site a typical value for $d_0$ is between $0.5$ and $1m$. In the approximation of an AO system based on a single Rayleigh LGS and looking at the its performance in direction the LGS we get a residual contribution of $\sigma_{cone}=600-1100nm$ depending on the strength of the considered turbulence profile.
\\The second error value in the list above, the wavefront fitting error, represents the accuracy which the WFS can sample the wavefront on the pupil plane and the DM can fit the reconstructed wavefront form. So this type of error is mainly related to the number of subapertures used to sample the telescope pupil and the distance between 2 adjacent DM actuators. The contribution of the fitting error to the measured wavefront can be written as \cite{1994_Sandler_8m_telescope_AO}:
\begin{equation}\label{eq:sigma_fit}
    \sigma_{fit}^2 = c_{fit}\Big( \frac{d}{r_0} \Big)^{5/3},
\end{equation}
where $d$ is the diameter of a subaperture projected on the telescope primary mirror and $c_{fit}$ a coefficient dependent from the type of DM. Considering a perfect matching between the WFS subapertures and the actuators of a large deformable mirror, such as the ASM, $c_{fit}\simeq0.3$ \cite{1991_Ellerbroek_cfit} giving a residual error in the range of $\sigma_{fit}=100-200nm$.
\\The time delay between measuring and correcting for the wavefront aberrations will introduce a bandwidth error $\sigma_{bw}$. The contribution of this error is significant when the servo bandwidth of the AO loop ($f_{c}$) reaches the \emph{Greenwood frequency}\footnote{Greenwood frequency is the frequency required for optimal correction with an adaptive optics system. It depends on the transverse wind speed as $v_{wind}{5/3}$ and the atmosphere turbulence strength.} ($f_{G}$) \cite{1977_Greenwood_AO_bandwidth}. The contribution of the bandwidth error can be evaluated in an approximated form as:
\begin{equation}\label{eq:sigma_bw}
    \sigma_{bw}^2 = \Big( \frac{f_g}{f_c} \Big)^{5/3},
\end{equation}
where $f_G\simeq 0.426 \overline{v}_{wind} / r_0$ and $\overline{v}_{wind}$ represents the average of the wind speed distribution weighted over the full turbulence profile. Considering the typical ARGOS atmosphere profile $\overline{v}_{wind}\simeq18m\;s^{-1}$, so the bandwidth error results $\sigma_{bw}=40-80nm$.
\\Formulas \ref{eq:sigma_cone}, \ref{eq:sigma_fit} and \ref{eq:sigma_bw} approximate the the error budget of an AO system considering a single LGS. In case of a GLAO system they overestimate the residual wavefront error because the use of many artificial reference stars enlarges the volume of atmosphere sampled above the telescope increasing the performance of the AO correction. This is evident comparing these results with the values reported in table \ref{tab:glao_residuals}. These data have been obtained with a Monte Carlo approach during the ARGOS \emph{Preliminary Design Review} (PDR) \cite{2008_Rabien_ARGOS_PDR_SysDesign}. The main contributor to the residual wavefront error, the cone effect, is reduced by a factor $\sim1.5$ considering multiple LGS.
\\Values reported in table \ref{tab:glao_residuals} can be useful because they represent an upper limit to the results obtained from end-to-end simulations. Last 2 rows in table \ref{tab:glao_residuals} show the rms of the atmospheric and corrected wavefronts measured in the end-to-end simulations. These values have been evaluated as the median of the wavefront rms of the 50 independent runs that we performed for each atmosphere profile. The spread represent the standard deviation over the 50 runs. It is visible that in case of AO system based on Rayleigh beacons the main contribution to the error budget is due to the higher layers of the atmospheric turbulence not sampled by the LGS.
\begin{table}
\caption{\footnotesize{List of the main contributors to the system error budget taken from ARGOS PDR. The rms values have been evaluated considering the 4 different atmosphere profiles, corresponding to $r_0$ values of $0.18, 0.16, 0.13$ and $0.09m$ at $0.5\mu m$ (see table \ref{tab:phase_screens}). Last two rows show the rms of the atmospheric and corrected wavefronts measured from the end-to-end simulations data. All values in table are expressed in $nm$.}}
\vspace{-0.25cm}
\begin{center}
  \begin{tabular}{|l|c|c|c|c|}
    \hline
    \textbf{Contribution} & $\textbf{Bad}$ & $\textbf{75\%}$ & $\textbf{50\%}$ & \textbf{25\%} \\
    \hline
    \hline
    Cone effect         & $800$ & $520$ & $440$ & $360$ \\
    Fitting error       & $200$ & $170$ & $130$ & $100$ \\
    Bandwidth error     & $170$  & $124$ & $106$ & $90$ \\
    \hline
    Total               & $840$ & $560$ & $470$ & $390$ \\
    \hline
    \hline
    Atm. WF rms         & $2200\pm170$  & $1600\pm150$  & $1400\pm130$  & $1300\pm120$ \\
    Corr. WF rms        & $890\pm250$  & $680\pm130$   & $620\pm110$   & $560\pm100$ \\
    \hline
    \end{tabular}
\end{center}
\label{tab:glao_residuals}
\end{table}
\\The results of the ARGOS end-to-end simulations have been evaluated in terms of \cite{2005_Nicolle_GLAO_perf}:
\begin{itemize}
  \item reduction of the PFS FWHM,
  \item uniformity of the PFS shape across the FoV,
  \item increase of encircled energy (EE) within a sensible area.
\end{itemize}
The first 2 parameters are important to check the benefits that a GLAO correction can bring to imaging applications. These are evaluated comparing the PFSs FWHM with and without applying the GLAO correction to the wavefront propagated through the atmosphere. The PFS uniformity is given in terms of standard deviation of the PFS FWHM over the FoV (see figure \ref{fig:CAOS_RS_diagram} for information about the directions in which the FoV have been sampled).
\\The third parameter, the gain in terms of EE, measures the increase in light concentration and it is a quality criterion for spectroscopic applications. EE is evaluated by first determining the total energy of the PSF over the full image plane, then determining the centroid of the PSF and measuring how much intensity fits inside a square of $0.25arcsec$ width, corresponding to the LUCI slit width \cite{2000_Mandel_LUCIFER_SPIE}. This parameter is specified as EE0.25.
\\The images in figure \ref{fig:olVScl_psf} show the 2D PSFs measured at $1.65\mu m$ in the direction of the reference sources of figure \ref{fig:CAOS_RS_diagram}. Each PSF sampled with 200 points, equivalent to a square of $2arcsec$ width on sky. The PSFs are obtained merging, after convergence, the 50 independent simulations run in the \emph{bad} atmosphere profile condition. The images on the left of figure \ref{fig:olVScl_psf} refer to the uncorrected atmosphere while on the right the GLAO correction was applied.
\\On these data we evaluated the difference between the open-loop and closed-loop PFS FWHM in case of the \emph{bad} atmosphere profile, results are summarized in figure \ref{fig:FWHM_map}. The intensity of the PFSs have been normalized to give an integrated value of 1 over the $2\times2arcsec$ area. It is visible that the FWHM is reduced by a factor $1.6$ in this case, giving a closed-loop PSF of $0.5arcsec$ width. The presence of small peaks in the PSFs shape is due to the not perfect averaging of the higher layers turbulence. This effect can slightly reduce the measure of closed-loop PFS FHWM and it can be compensated increasing the number of independent simulation runs, obviously increasing the simulation time requirement.
\\To summarize the results of end-to-end simulations we measured the expected gain in terms of PSF FWHM and EE0.25 between the open-loop and close-loop situations. The open-loop values, plotted in abscissa in figure \ref{fig:FWHM_EE_gain}, represent the seeing limited case of the 4 atmosphere profiles. The close-loop data represent the enhanced seeing condition, when GLAO correction is applied. For each profile we evaluated the PSF in the J, H and K bands, without adding noise parameters. The points in figure \ref{fig:FWHM_EE_gain} represent the average FWHM or EE025 evaluated on the 5 reference sources directions (see figure \ref{fig:CAOS_RS_diagram}). The error bars associated to the points represent the standard deviation of the FWHM or EE025 evaluated in these 5 directions. The gain of GLAO correction is highlighted by the 2 dashed lines and it is visible that it consist of a factor $1.5-3$ under the different seeing conditions and bands. These results, obtained with the most complete end-to-end simulations run with the CAOS code, agree and definitively confirm the previous ones that have been performed for the ARGOS PDR and FDR.

%

%
%
%
%
%

\chapter{The wavefront sensor dichroic}
\label{cap:dich}

We already mentioned in chapter \ref{cap:argos} that AROGS uses a dichroic window to separate the laser light used by the LGS WFS from the scientific light transmitted toward LUCI. Wavelengths lower than $600nm$ are reflected by the dichroic while those over $600nm$ are transmitted.
\\The aim of the optical design of the ARGOS dichroic is to minimize the aberrations it introduces in the instrument focal plane, the laser light loss in reflection and the additional thermal radiation it injects in the infrared instrument. For all these reasons the dichroic is a critical component and the work we have done in designing the optics is described in detail in this chapter.
\\In section \ref{sec:window} we first consider a general case and we analyze the effects of inserting a window in a convergent beam. This analysis is necessary to find the proper way to compensate these effects adding features in the window design. The study we have done to find these features is described in section \ref{sec:win_comp}. Section \ref{sec:dic_fdr} describes the final design of the ARGOS dichroic and it compares the telescope optical quality when the ARGOS dichroic is deployed or it is removed from the optical path. Section \ref{sec:dic_tol} describes the specifications and tolerances we produced for the manufacturing of the optic and the results of the tests we performed on the 2 units that have been produced. In section \ref{sec:dic_coating} same work is done for coating. Section \ref{sec:dic_mech} describes the specifications and the mechanical design of the support structure for the dichroic.

\section{Effects of a window in a convergent beam}
\label{sec:window}
\begin{figure}
\begin{center}
\includegraphics[width=6.5cm]{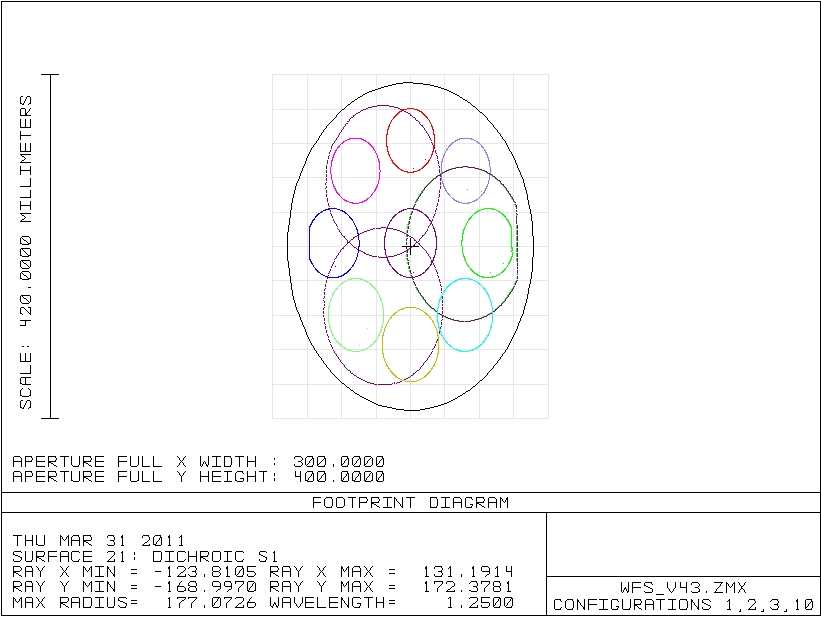}
\includegraphics[width=6.5cm]{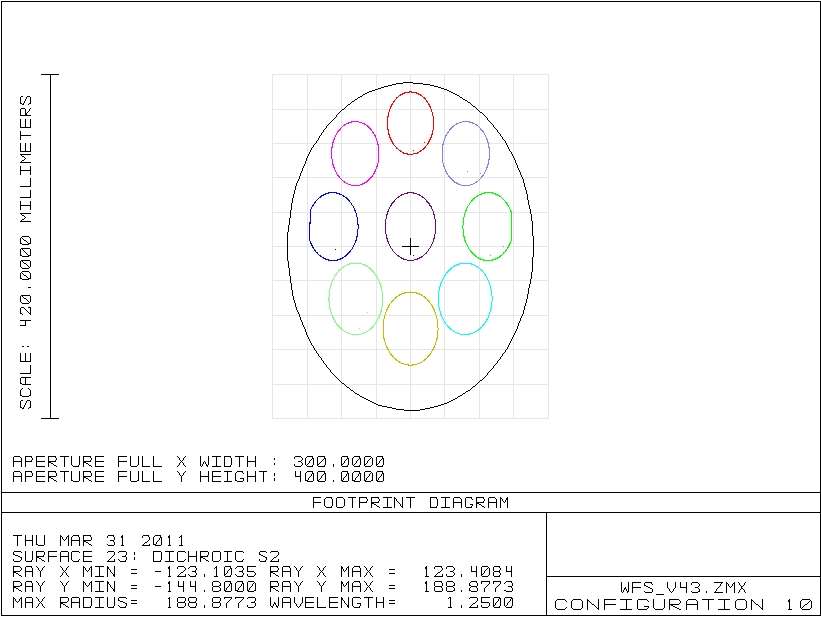}
\end{center}
\vspace{-0.5cm}
\caption{\footnotesize{Left: footprints of the NGS (in colors) and LGS (black) beams on the first surface of the dichroic. The NGS are inscribed in a circle of $2\sqrt{2}arcmin$ radius, that corresponds to the $2arcmin$ FoV of LUCI rotated by $360^{\circ}$. The top of the lens is on the right of figure. Right: NGS footprints on the second surface of the lens. The footprints are shifted by $10.8mm$ with respect the first surface.}}
\label{fig:dic_ftprnt}
\end{figure}

\begin{figure}
\begin{center}
    \includegraphics[width=7cm]{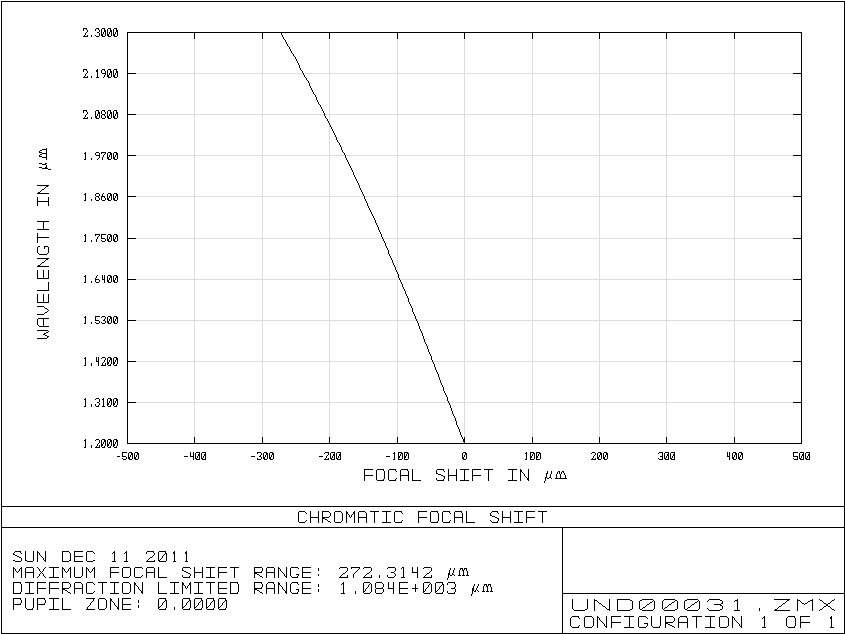}
\end{center}
\vspace{-0.5cm}
\caption{\footnotesize{Plot of the focal plane position in the direction of the optical axis for an $f_{15}$ beam and different wavelengths. Values are given as differential position with respect the shorted wavelength. A chromatic focus shift of $270\mu m$ is introduced in a polychromatic ($1.2\div2.3\mu m$) $f_{15}$ beam by a $40mm$ thick BK7 window.}}
\label{fig:long_chrom_shift}
\end{figure}

\begin{figure}
\begin{center}
\includegraphics[width=6.5cm]{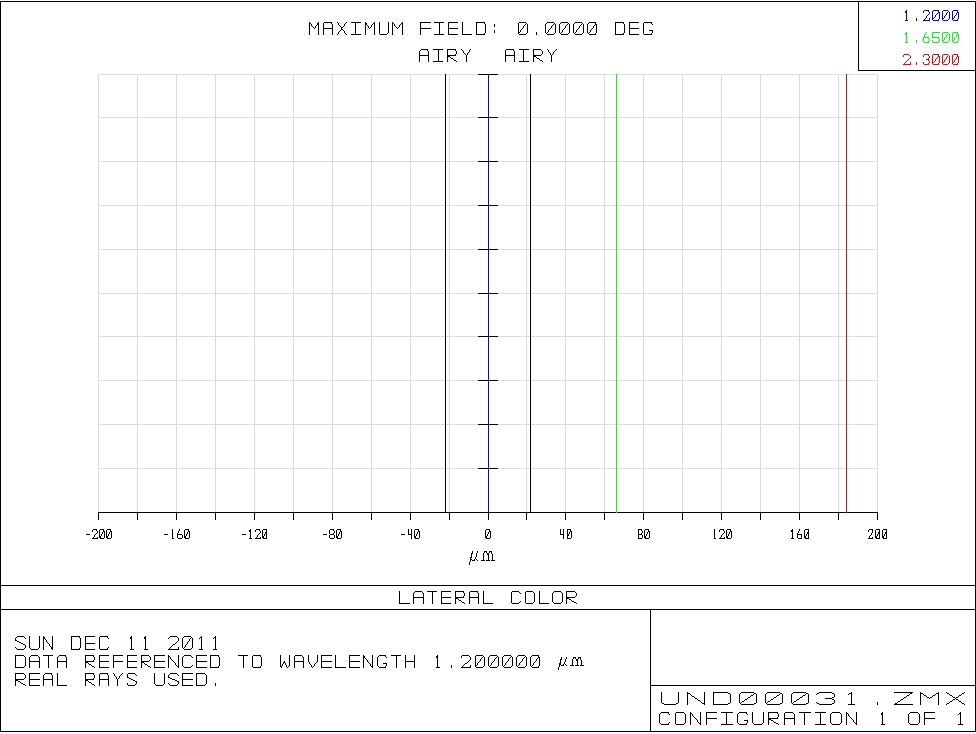}
\end{center}
\vspace{-0.5cm}
\caption{\footnotesize{Plot of the lateral color effect arising in a $f_{15}$ beam passing through a $40mm$ thick BK7 window inclined by $40.5^{\circ}$ angle. Data are plotted as difference with respect the focus position at $1.2\mu m$. Positive values are in the direction toward the focus position when any dichroic is present.}}
\label{fig:lat_chrom_shift}
\end{figure}

\begin{figure}
\begin{center}
\includegraphics[width=8cm]{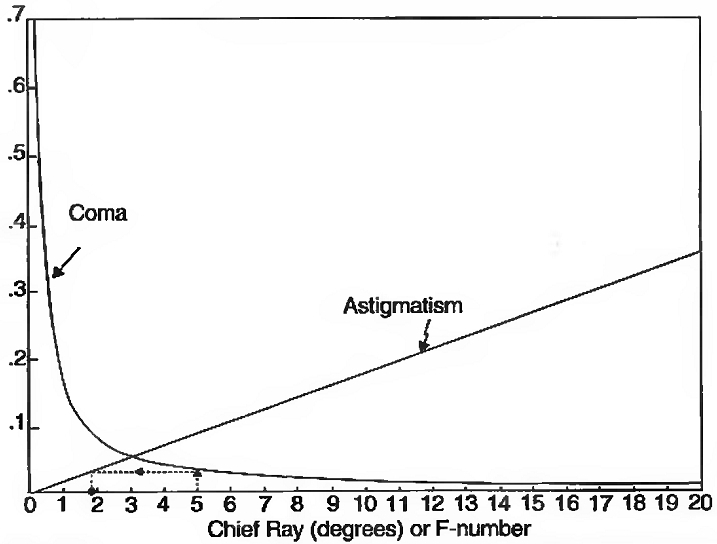}
\end{center}
\vspace{-0.5cm}
\caption{\footnotesize{Plot of the difference factor between coma and astigmatism, taken from \cite{2002_Geary_Lens_Design_Zemax}. For a given beam f-number the incidence angle at which coma equals astigmatism can be evaluated as follows: identify the f-number value on the x-axis, follow the vertical line to the astigmatism curve and then the horizontal line to the coma curve. At this point the vertical intercept with the x-axis gives the angle of incidence at which the 2 aberrations are equal. The opposite scheme can be used to identify for a given incidence angle the f-number at which the 2 aberrations are equal.}}
\label{fig:coma_vs_asti}
\end{figure}

The natural objects light is focussed by the telescope in a $f_{15}$ beam at a $5300mm$ distance from M3. This light has to be separated off the artificial sources focussed in a $f_{16.6}$ beam at a $6720mm$ distance from M3. This task is accomplished placing a tilted dichroic window before the $f_{15}$ focal plane, just in front of the LUCI focal station ($\sim 970mm$ before the telescope $f_{15}$ focus).
\\The minimum window working angle for which it is possible to direct the laser light laterally and to not vignette with the WFS optics the telescope primary mirror is $40.5^{\circ}$. The dimensions of the window are set by the projection of the LUCI FoV on the plane in which it lays. A $300\times400mm$ elliptical window with a clear aperture of $290\times390mm$ is sufficient to contain the entire instrument FoV. Figure \ref{fig:dic_ftprnt} shows the footprints of natural and laser stars on the two dichroic surfaces. The natural stars have a major axis of $85mm$, one of them is placed on-axis while the other 8 are placed $2\sqrt{2}arcmin$ off-axis, corresponding to the corner of the $4\times4arcmin$ LUCI FoV. The artificial stars are simulated with point-like sources placed at $12km$ distance from the telescope on a triangular pattern inscribed in a circle of $2arcmin$ radius. Natural stars footprint are shifted on the rear surface by $\sim10mm$, but a $10mm$ margin is still available between the edges of the FoV and the optical area.
\\The window dimensions places a constrain on the minimum thickness. The dichroic bending due to variations in the gravity vector direction will introduce a tilt in the reflected LGS beams. The tilt has to be low enough to keep the LGS inside the FoV of the WFS. This constrains the maximum sag induced by the dichroic bending to be less than $0.15mm$, with a goal of $0.03mm$. Since rigidity scales as $t^3$ and weight scales as $t$ the resulting bending goes with $t^2$. A finite element analysis performed on a thick plate of glass, varying the gravity vector of $1g$ in the direction perpendicular to the plate surface, shown that a $40mm$ thickness is enough to keep the induced sag within the goal specification.
\\Introducing a parallel plate window in a converging beam is a delicate task, since it will cause different aberrations to arise \cite{2001_Laikin_lens_design}. Some of these aberrations are independent from the window working angle, they are: defocus and spherical aberration.
\\The defocus is introduced because the beam focal plane is shifted along the optical axis. This longitudinal shift $\Delta Z$ can be quantified using the simplified formula:
\begin{equation}\label{eq:long_shift}
    \Delta Z = \frac{(n-1)\; t}{n},
\end{equation}
where $t$ is the plate thickness and $n$ the refractive index of the material from which is produced. Considering the values we set for the ARGOS dichroic window $\Delta Z \simeq 13mm$ for $n=1.5$.
\\The peak aberration coefficient for defocus, equaling the PtV of wavefront error in the paraxial focal plane, can be evaluated as:
\begin{equation}\label{eq:defocus}
    W_{def} = - \frac{\Delta Z}{8\;f_{\sharp}^2},
\end{equation}
where $f_{\sharp}$ is the system f-number, negative for a converging beam. Considering the longitudinal focus shift evaluated above $W_{def}=-3.8\mu m$ wavefront, in case of a $f_{15}$ beam.
\\The longitudinal chromatic focus shift due to the introduction of a window of thickness $t$ between two wavelengths $\lambda_r > \lambda_b$ is given by:
\begin{equation}\label{eq:long_chrom}
    z_b - z_r = \frac{(n_b - n_r)\;t}{n_d^2},
\end{equation}
where the terms $n_b$, $n_r$ and $n_d$ represent the material refraction index respectively at the shorter wavelength, at the longer one and at $587.6nm$ the longitudinal chromatic shift expected in a $f_{15}$ beam passing through a $40mm$ thick BK7 window is showed in figure \ref{fig:long_chrom_shift}.
\\Higher order aberration than defocus are introduced in the converging beam transmitted by the window because rays having larger angle of incidence (as off-axis fields) on the window are displaced more than rays that have a smaller angle of incidence. Spherical aberration however is independent of the field angle, and we can evaluate it using the formula for the Seidel aberration \cite{1992_Wyant_aberr_optical_metro}:
\begin{equation}\label{eq:spherical}
    W_{sphe} = \frac{(n^2-1)\;t}{128n^3\;f_{\sharp}^4},
\end{equation}
that will be still valid in the case the window is tilted. Since spherical aberration strongly depends in the system f-number, considering a slow $f_{15}$ converging beam the Seidel coefficient for the spherical aberration introduced by the $40mm$ window results to be very small: $W_{sphe} = 8nm$ (wavefront).
\\Since the need to direct the laser light aside the LUCI focal station the ARGOS dichroic will be inclined by $40.5^{\circ}$ with respect the optical axis. In case of a tilted window other aberrations are introduced in the converging beam \cite{1975_Prasad_aberr_incl_window}: lateral displacement, coma and astigmatism.
\\The lateral displacement is strongly dependent on the working angle $\theta$ of the parallel plate. A simple formula to quantify this shift is:
\begin{equation}\label{eq:lat_shift}
    \Delta Y = \frac{(n-1) \;t \;\theta}{n} = \Delta Z \; \theta.
\end{equation}
Considering the parameters we used so far $\Delta Y \simeq 9mm$.
\\It is clear that the lateral shift is also a chromatic effect. In polychromatic light bluer wavelengths are displaced more in the lateral direction while redder ones are displaced less, taking as reference the no window case. The lateral chromatic shift experienced between two wavelengths $\lambda_r > \lambda_b$ is proportional to the longitudinal shift $z_b-z_r$ and the window working angle $\theta$, resulting:
\begin{equation}\label{eq:lat_chrom_shift}
    y_b - y_r = \frac{(n_b-n_r)\; t\; \theta}{n_d^2} = (z_b - z_r)\; \theta.
\end{equation}
Figure \ref{fig:lat_chrom_shift} shows the chromatic lateral shift measured for an on-axis field at different wavelengths. A difference in the chief ray position of $180\mu m$ is measured between $1.2$ and $2.3\mu m$ light.
\\The contribution of higher order aberrations (astigmatism and coma) is strongly dependent both on the window working angle and the beam f-number. Given a certain system f-number coma is the dominant aberrations at smaller angles on incidence on the optic surface, while astigmatism dominates at larger angles. For the same reason given a certain incidence angle, coma dominates for faster beams and astigmatism dominates for slower beams. This relation is shown in the plot in figure \ref{fig:coma_vs_asti}, taken from \cite{2002_Geary_Lens_Design_Zemax}. The wavefront aberration introduced by a tilted window can be quantified in terms of third-order coma and astigmatism using the expression for the Seidel aberrations \cite{1992_Wyant_aberr_optical_metro}:
\begin{equation}\label{eq:coma}
    W_{coma} = - \frac{(n^2-1)\; t \; \theta}{16 n^3\; f_{\sharp}^3},
\end{equation}
\begin{equation}\label{eq:asti}
    W_{asti} = - \frac{(n^2-1)\; t \; \theta^2}{8 n^3\; f_{\sharp}^2}.
\end{equation}
Considering the parameters we used so far we get $W_{coma}=0.2\mu m$ and $W_{asti}=-4\mu m$.

\section{Aberration compensation with window shape}
\label{sec:win_comp}
In this section we discuss the features we introduced in the window shape to reduce as much as possible the aberrations caused by the window. In addition we applied a correction on the on the pupil plane, by applying a static shape to the ASM, to null the residual aberrations that are common to the entire LUCI FoV.
\\In this analysis we consider an on-axis $f_{15}$ beam made of three monochromatic wavelengths in correspondence of the central wavelength of the J, H and K working bands of LUCI, correspondent to $1.2$, $1.65$ and $2.3\mu m$ respectively.

\subsection{Effects of a wedge between surfaces}
The chromatic component of the lateral displacement can be compensated by wedging the 2 surfaces of the window. A tilted wedged beam splitter is a special case of a thick wedge prism where the chromatic deviation angle is small compared with the tilt angle of the beam splitter. In this way we can take advantage of the prism deviation angle to compensate for the lateral color. In the thick prism approximation the wedge $\alpha$ required to compensate for the lateral color can be evaluated as:
\begin{equation}\label{eq:wedge}
    \alpha = \frac{y_b - y_r}{(n_b-n_r)\; z},
\end{equation}
where $z$ is the distance between the rear surface of the window-prism and the plane where the lateral color $y_b-y_r$ is evaluated. Considering the full range of wavelengths plotted in figure \ref{fig:lat_chrom_shift} and a distance of $970mm$ between the window and the $f_{15}$ plane, the wedge angle results $\alpha\simeq0.42^{\circ}$.
\\The wedge also helps in lowering the lateral shift of the beam but has the side effect to tilt the directions of the transmitted beams. For example considering a $0.42^{\circ}$ wedge the lateral shift is reduced down to $5mm$ but the rays have been tilted by $\sim6mrad$.
\\Astigmatism and coma introduced by the window can be reduced also using a wedge plate. We can evaluate the wedge angle required to null the third-order coma and astigmatism using the formulas \cite{1985_Howard_asti_coma_wedge_prism}:
\begin{equation}\label{eq:wedge_coma}
    \alpha_{coma} = - \frac{t \; \theta}{2n^2\;y\;f_{\sharp}},
\end{equation}
\begin{equation}\label{eq:wedge_asti}
    \alpha_{asti} = - \frac{t \; \theta}{4n^2\;y\;f_{\sharp}},
\end{equation}
where $y$ is the radius of the beam on the first window surface. Considering that the footprint of the $f_{15}$ beam on the window surface has a major axis equal to the NGS footprint in figure \ref{fig:dic_ftprnt} ($85mm$), the wedge angles for zero coma and astigmatism result: $\alpha_{coma}=0.6^{\circ}$ and $\alpha_{asti}=0.3^{\circ}$ respectively.
\\We have shown here that a wedge between the 2 window surfaces can solve both the chromatics effects and the low order aberrations introduced by the window. Unfortunately the wedge angles required to compensate all these effects are different. We found the best solution to be designing the wedge with a $0.42^{\circ}$ angle to solve for the chromatic effects and to add another feature to the window shape to reduce the residual low order aberrations (see section \ref{ssec:cyl}).
\\To quantify the residual low order aberrations introduced by the wedged window we can imagine it as made of 2 components: a thick plate and a thin prism. Under this approximation the equations \ref{eq:coma} and \ref{eq:asti} can be rewritten to take into account also the contribution of the thin prism:
\begin{equation}\label{eq:coma_thick_prism}
    W_{coma}' = - \frac{(n^2-1)\; t \; \theta}{16 n^3\; f_{\sharp}^3} + \frac{(n+1)\;y\;\delta}{4n\;f_{\sharp}^2},
\end{equation}
\begin{equation}\label{eq:asti_thick_prism}
    W_{asti}' = - \frac{(n^2-1)\; t \; \theta^2}{8 n^3\; f_{\sharp}^2} + \frac{(n+1)\;y\;\delta\;\theta}{2n\;f_{\sharp}},
\end{equation}
where $\delta$ is the prism deviation angle that, in case of a thin prism in air, can be evaluated as: $\delta\simeq\alpha(n-1)$. It is visible from equations \ref{eq:coma_thick_prism} and \ref{eq:asti_thick_prism} that the contribution of the thin prism goes in the opposite direction of the thick plate, effectively reducing the amount of coma and astigmatism introduced. So the low order aberrations for the wedged window result to be: $W_{coma}'=0.08\mu m$ and $W_{asti}'=1.5\mu m$.

\subsection{Effects of a cylindrical surface}
\label{ssec:cyl}
\begin{figure}
\begin{center}
    \includegraphics[width=6.5cm]{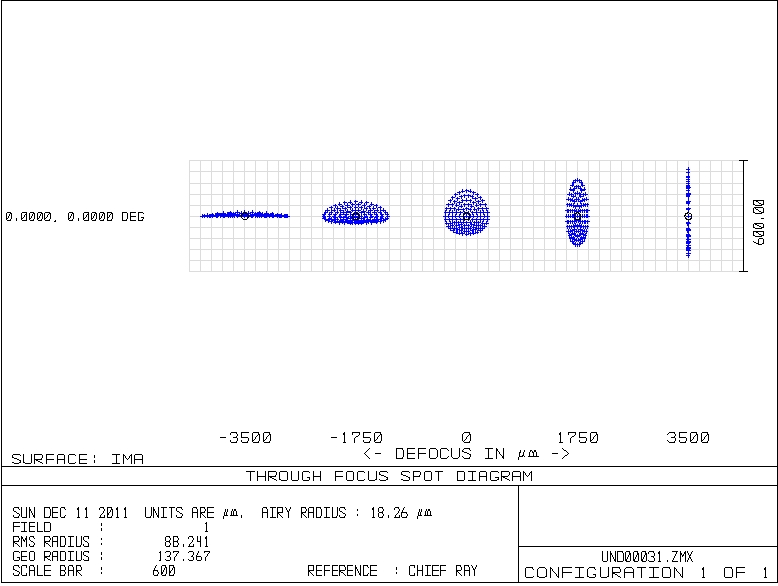}
    \includegraphics[width=6.5cm]{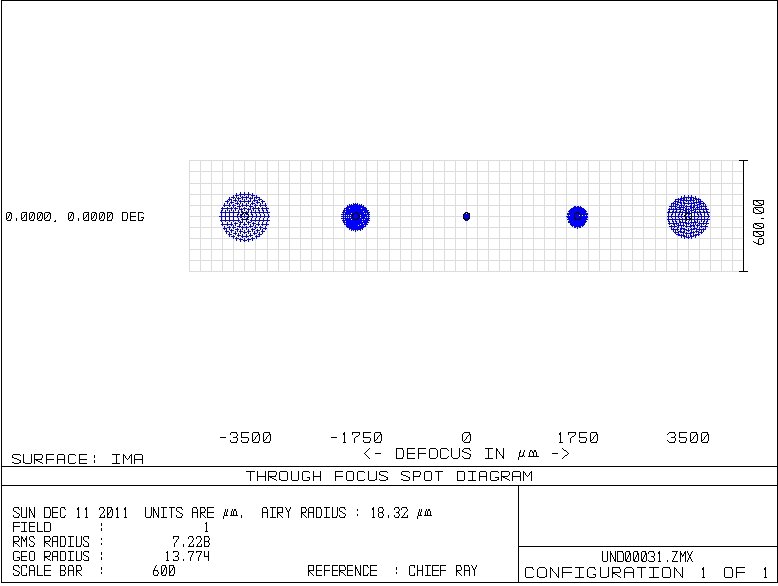}
\end{center}
\vspace{-0.5cm}
\caption{\footnotesize{In line spot diagrams showing the position of the sagittal and tangential focal planes when a $f_{15}$ beam is transmitted through a tilted window having a $0.42^{\circ}$ wedge (left). The $7mm$ distance between the 2 planes can be compensated designing the rear surface of the window as a concave cylinder with a radius of curvature of $150m$.}}
\label{fig:wedge_thr_focus}
\end{figure}

Since astigmatism causes the sagittal and tangential rays to have different focal plane positions along the optical axis, this aberration can be compensated designing the window as a lens. This lens however must have optical power only in one plane leaving untouched the rays in the other one, this means it has to be designed as a cylindrical lens.
\\Considering the wedge angle of $0.42^{\circ}$, necessary to compensate the lateral chromatism, the residual astigmatism for the on-axis ray on the $f_{15}$ plane resulted $2\mu m$ from equation \ref{eq:asti_thick_prism}. Figure \ref{fig:wedge_thr_focus} shows the distance from the best focus position of the tangential and sagittal focus planes for the beam transmitted by the wedged window. The $7mm$ distance between the 2 focal planes can be compensated with a concave cylinder of $150m$ radius of curvature.
\\Also the cylinder helps in reducing the field dependent component of the astigmatism. This effect is due to the different angle of incidence of the off-axis fields on the dichroic surface. Because the system pupil is placed at a finite distance the off-axis fields are tilted with respect the on-axis chief-ray direction. This tilt amount to $0.5^{\circ}$ in case of a $2\sqrt{2}$ field, corresponding to the corner of the $4\times4arcmin$ LUCI FoV. So the most external fields on the tangential plane hits the dichroic with an angle variable in the $40-41^{\circ}$ range. This causes a variation of the astigmatism introduced on these fields of $\pm5\%$ considering equation \ref{eq:asti_thick_prism}. The $150m$ radius cylinder is used also to get rid of this effect.

\subsection{Static correction with the ASM}
\label{ssec:ASM_comp}
Most of the dichroic induced aberrations on the instrument focal plane can be compensated adding proper features to the window design (wedge and cylinder), but not all of them can be nulled at the same time. LBT has the adaptive optic corrector in the telescope optical path, so applying a static correction to the mirror it is possible to compensate for the residual aberrations of the window design.
\\The residual low order aberrations we expect are mainly defocus and coma, plus a little contribution of spherical and residual astigmatism. In addition the beam lateral displacement could not be fully compensated by the window wedge. Table \ref{tab:ASM_comp} resumes the static correction, expressed in terms of the hexapod degrees of freedom and surface Zernike coefficients, that have to be applied to the ASM to null the on-axis beam aberrations and re-optimize the optical quality on the focal plane. These corrections have been evaluated in Zemax, considering the $40mm$ thick BK7 window wedged by $0.42^{\circ}$ having the rear surface as a concave cylinder with $150m$ radius of curvature. The hexapod adjustments are necessary to recenter the on-axis beam on the instrument rotator axis, and to compensate for the defocus term. The Zernike coefficients express the deformation needed in the thin shell to null the residual low order aberrations (astigmatism, coma and spherical).
\begin{table}
\caption{\footnotesize{Summary of the static correction applied to the ASM to re-optimize the optical quality on the $f_{15}$ plane when it passes through a wedged and cylindrical window (RoC and angle are shown in table). Position adjustments are expressed as hexapod degrees of freedom, while variations in the ASM shape are expressed in terms of Zernike surface coefficients.}}
\vspace{-0.25cm}
\begin{center}
    \begin{tabular}{|l|c|}
    \hline
    \textbf{Parameter} & \textbf{Value} \\
    \hline
    \hline
    Cylinder RoC            & $150m$ \\
    Wedge angle             & $0.42^{\circ}$ \\
    \hline
    Hexapod piston          & $-90\mu m$  \\
    Hexapod decenter        & $[35;-470]\mu m$   \\
    Hexapod tilt            & $[-0.025;0.001]^{\circ}$  \\
    ASM astigmatism         & $Z5\&6: 11nm$ \\
    ASM coma                & $Z7\&8: -20nm$ \\
    ASM spherical           & $Z11: 18nm$ \\
    \hline
    \end{tabular}
\end{center}
\label{tab:ASM_comp}
\end{table}

\section{Final design of the ARGOS dichroic}
\label{sec:dic_fdr}
\begin{figure}
\begin{center}
\includegraphics[width=6.5cm]{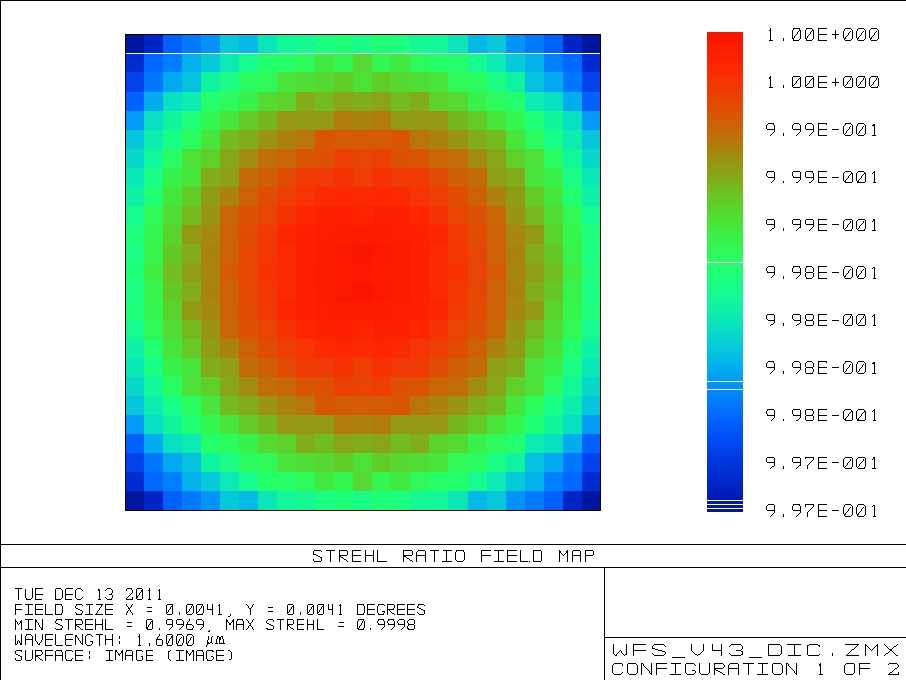}
\includegraphics[width=6.5cm]{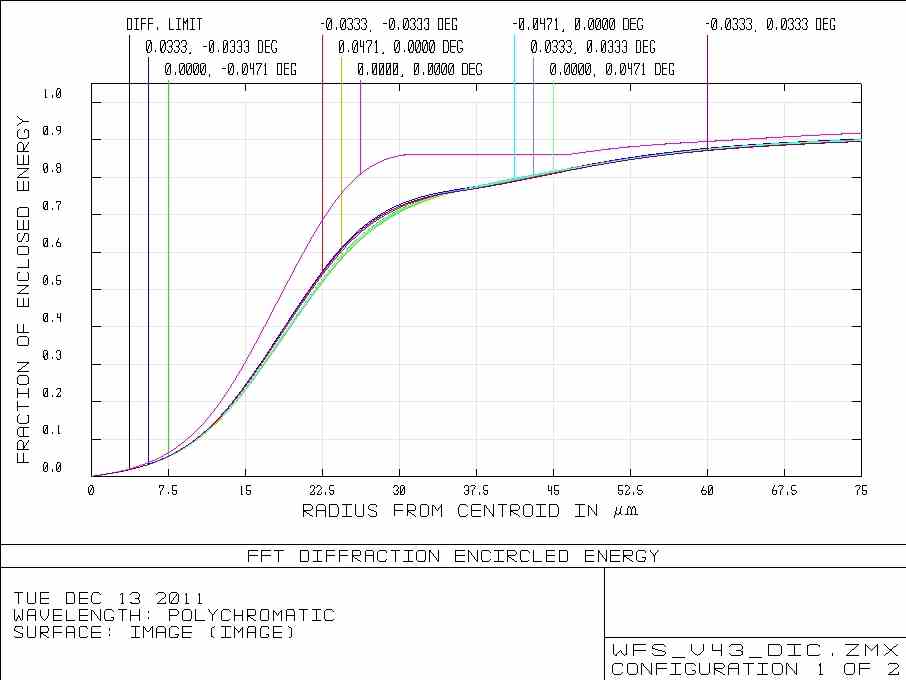} \\
\includegraphics[width=6.5cm]{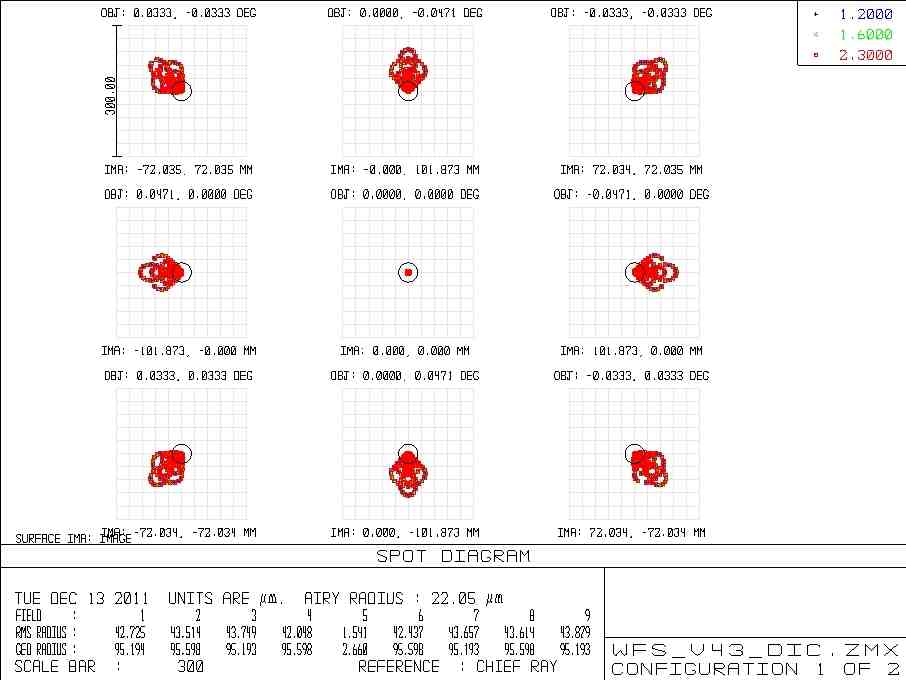}
\includegraphics[width=6.5cm]{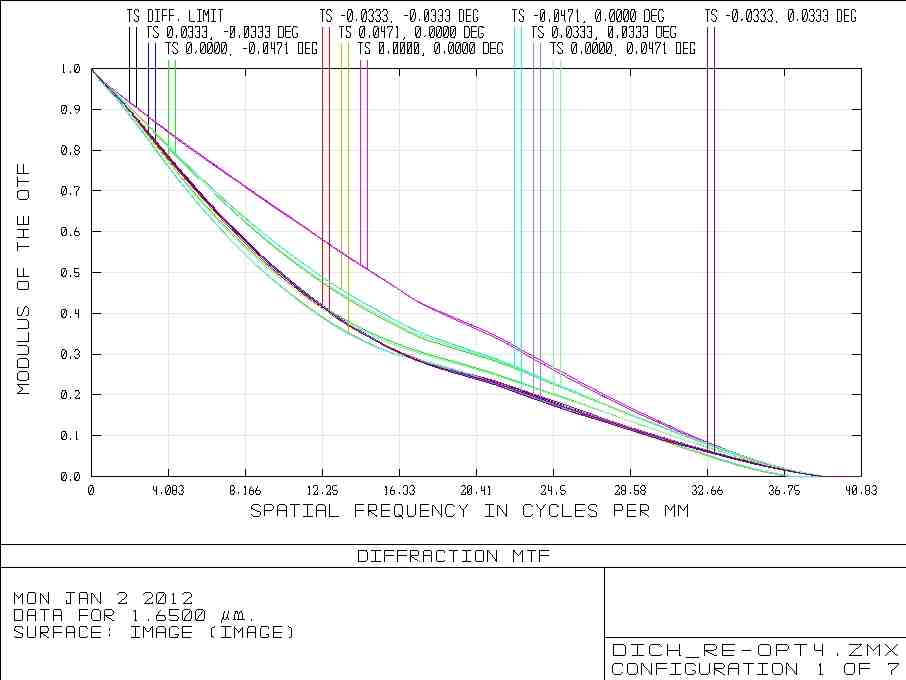}
\end{center}
\vspace{-0.5cm}
\caption{\footnotesize{Optical quality of the LBT telescope evaluated on the $f_{15}$ plane. Strehl ratio, shown in the upper left corner, is evaluated on a field of $30\times30arcsec$. Encircled energy, spot radius and system MTF are evaluated instead at the center of the FoV and on 8 directions distributed on a circle of $2\sqrt2arcmin$ radius.}}
\label{fig:LBT_ima_quality}
\end{figure}

\begin{figure}
\begin{center}
\includegraphics[width=6.5cm]{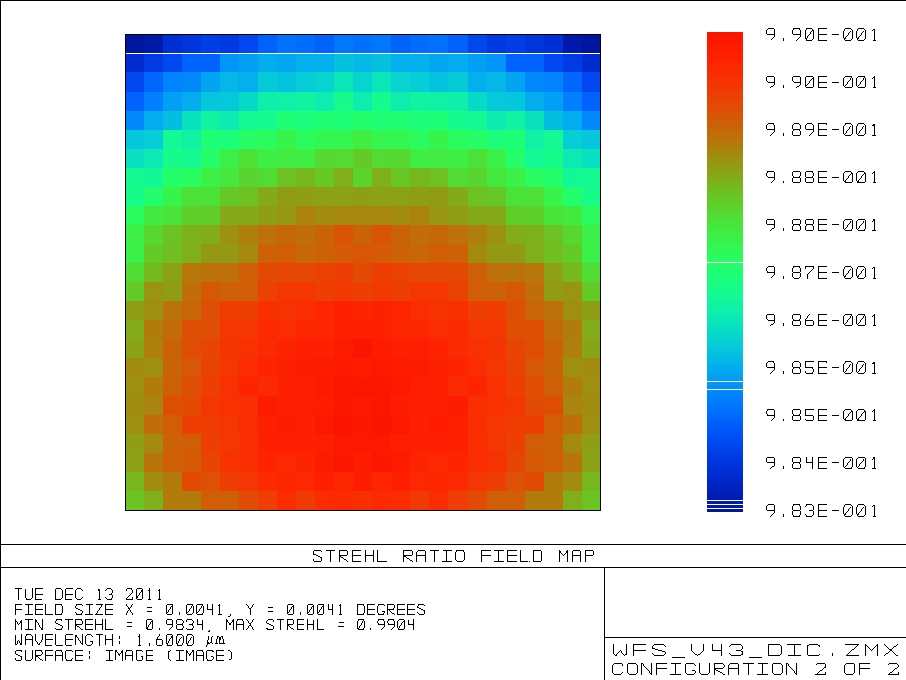}
\includegraphics[width=6.5cm]{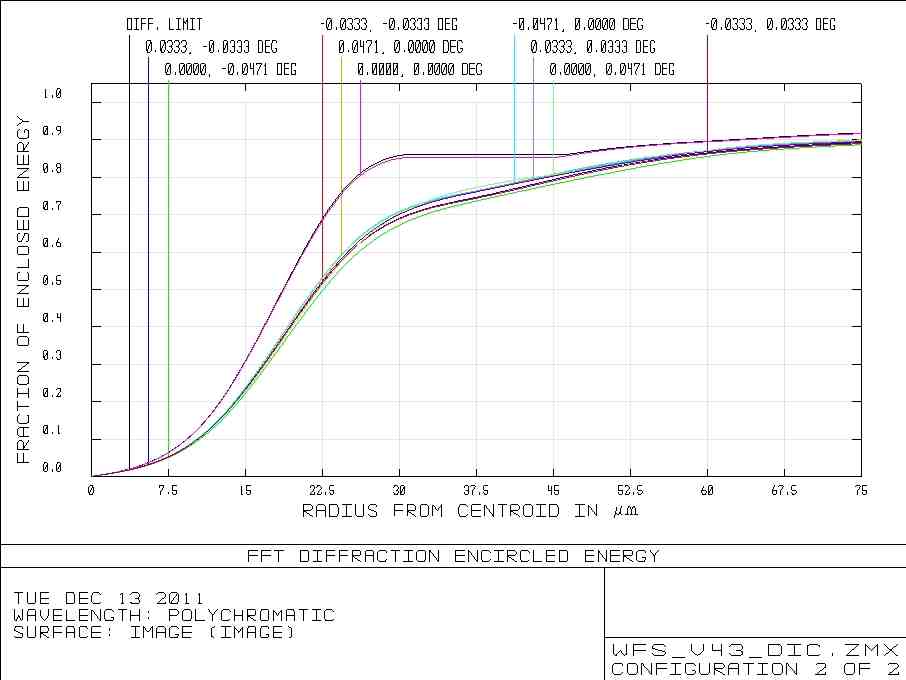} \\
\includegraphics[width=6.5cm]{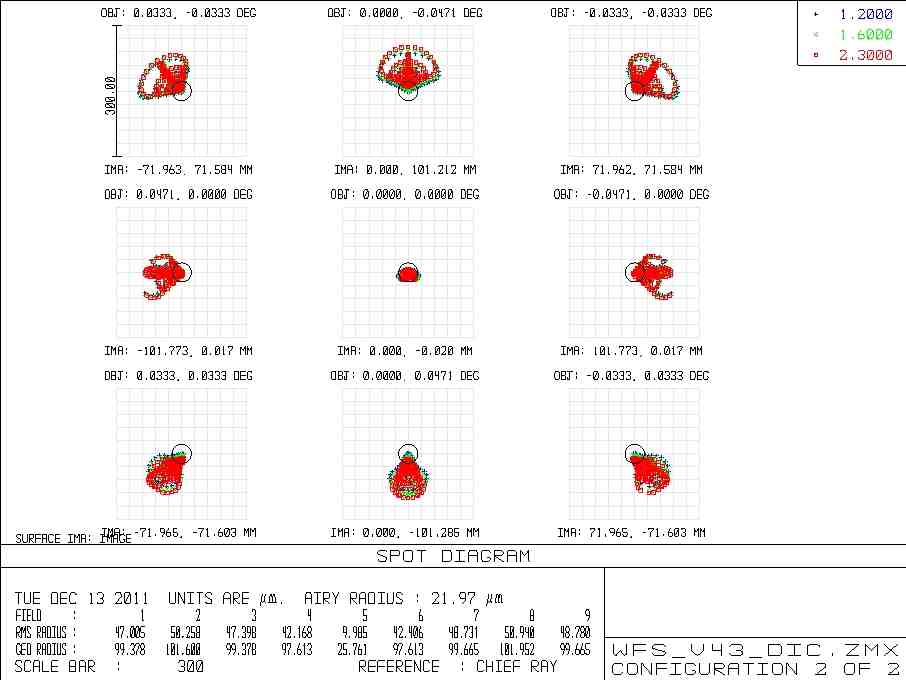}
\includegraphics[width=6.5cm]{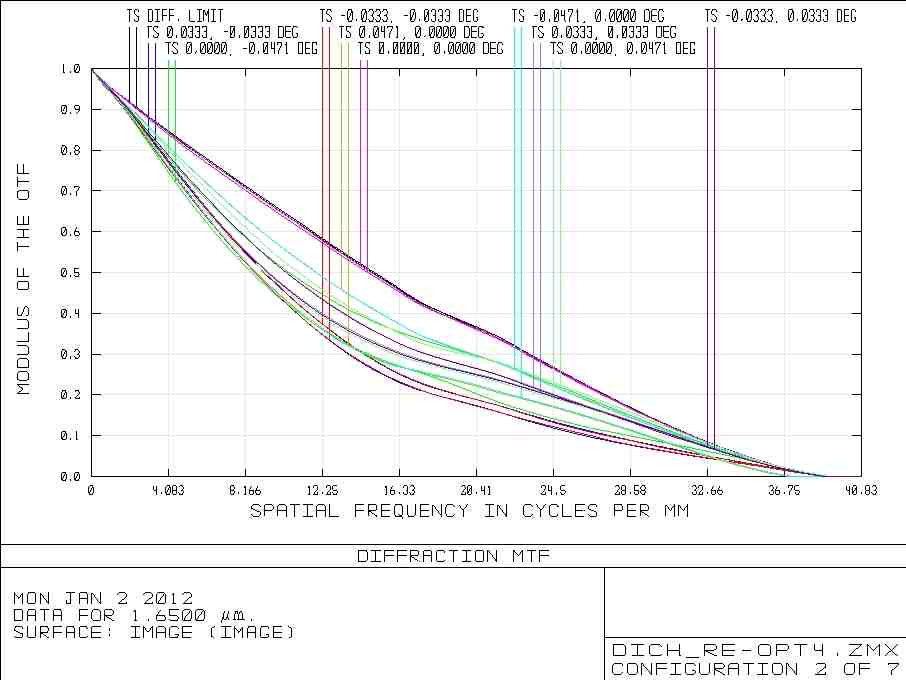}
\end{center}
\vspace{-0.5cm}
\caption{\footnotesize{Optical quality of the LBT evaluated on the LUCI focal plane when the ARGOS dichroic is inserted in the optical path. The parameters checked are the same of figure \ref{fig:LBT_ima_quality}.}}
\label{fig:dic_ima_quality}
\end{figure}

In section \ref{sec:window} and \ref{sec:win_comp} we analyzed the aberrations that we expect a dichroic window will introduce on the LUCI focal plane and we shown how we could compensate for them. However we determined the compensator values considering each aberration individually. This work can be improved performing the dichroic design optimization with an optical design software, as Zemax, that allow to optimize the dichroic shape considering all the effects. In this section we describe this optimization process and the results we obtained for the final design of the dichroic.
\\Figure \ref{fig:LBT_ima_quality} shows the image quality obtainable on the LBT $f_{15}$ plane, when no dichroic is inserted in the optical path. We will evaluate the results of the optimization process taking as reference these values. The quality parameters we used in the optimization process are:
\begin{itemize}
  \item SR values higher than 99\% over the entire $30\times30arcsec$ FoV of LUCI diffraction limited camera.
  \item EE on a radius $75\mu m$ (equivalent to a diameter of $0.25arcsec$ on-sky) higher than 90\% over a FoV diameter of $4arcmin$, that corresponds to the area available for spectroscopy with the enhanced-seeing camera optics of LUCI.
  \item PSF FWHM less than $150\mu m$ over a FoV of $4\times4arcmin$ that allow to do imaging with the $125mas$ resolution of the enhanced-seeing camera of LUCI.
\end{itemize}
As substrate for the ARGOS dichroic we selected INFRASIL 302\footnote{INFRASIL 302 is an optical quartz glass produced by Heraeus. It is manufactured by fusion of natural quartz crystals is an electrically heated furnace. This production process guarantee that the index of homogeneity is $\Delta n \leq 6\times10^{-6}$ over the total substrate. The optical homogeneity is the main criterion to ensure low transmitted wavefront distortions.}. This type of glass has negligible emissivity and minimal absorption at infrared bands, a feature that minimizes thermal background radiation injected from the telescope environment into the instrument reducing the impact of the new optic on the instrument signal and background noise.
\\The optimization has been done inserting a $40mm$ window made of INFRASIL in the LBT optical project, $970mm$ before the telescope focus. The window has been tilted by $40.5^{\circ}$ with respect the optical axis and we let Zemax vary the wedge angle between window surfaces, the radius of curvature of the second cylindrical surface and the ASM degrees of freedom and shape. The optimization has been done using a Merit Function that constrained the optical quality (expressed in terms wavefront rms) on a FoV of $4arcmin$ diameter and an interval of wavelengths between $1.2$ and $2.3\mu m$.
\\From Zemax optimization we get a radius of curvature for the cylinder of $237m$ and a wedge angle between the surfaces of $0.557^{\circ}$. The ASM has been shifted by $70\mu m$ in the direction of M3, it has been decentered by $145\mu m$ and tilted by $0.005^{\circ}$. The static shape applied to the thin shell, expressed as Zernike surface coefficients, resulted to be: $335nm$ of astigmatism, $263nm$ of coma and $10nm$ of spherical aberration.
\\Figure \ref{fig:dic_ima_quality} shows the optical quality on the $f_{15}$ plane of LBT when the optimized ARGOS dichroic is in use. We can see that:
\begin{itemize}
  \item SR is almost unchanged, the pattern in figure \ref{fig:dic_ima_quality} evidences a PtV variation of 1\% with respect to the no dichroic case.
  \item The EE within $75\mu m$ radius, equivalent to a slit width of $0.25arcsec$ on-sky, is bigger than 90\%.
  \item Spot patterns are slightly changed but the PSF size is comparable with the no dichroic case.
\end{itemize}

\section{Tolerances for the dichroic production}
\label{sec:dic_tol}
We evaluate the tolerances allowed for the dichroic manufacturing from two arguments. First we check the maximum allowed variation of the design parameters that keep the MF used for optimization within a given limit and second we evaluate the maximum surface errors, caused by the finite precision of the optical polishing procedure, that produce a certain effect on the Rayleigh WFS or instrument focal planes.
\\The first analysis is aimed to investigate the tolerance on the physical properties of the dichroic and it has been carried out with Zemax. We allowed a maximum increase of $5\%$ in the rms of the wavefront transmitted through the dichroic and we have checked the maximum variation in the design parameters that kept the MF within this limit. Table \ref{tab:dic_param} resumes the results of the analysis: they constrain mainly the tolerances on the radius of curvature of the cylinder, the wedge angle, the dichroic central thickness and the glass optical properties.
\begin{table}
\caption{\footnotesize{Summary of the manufacturing parameters and tolerances of the dichroic.}}
\vspace{-0.25cm}
\begin{center}
    \begin{tabular}{|l|c|}
    \hline
    \textbf{Parameter} & \textbf{Value} \\
    \hline
    \hline
    Material                & INFRASIL 302  \\
    Refraction index        & $1.458\pm0.001$   \\
    Abbe parameter          & $67.7\pm0.5$  \\
    Shape 					& Elliptical: $(300\times400\pm1)mm$ \\
    Clear aperture (CA)		& Elliptical: $(290\times390\pm1)mm$ \\
    Central thickness		& $(40.0\pm0.8)mm$ \\
    Wedge angle				& $(0.560\pm0.017)^{\circ}$ \\
    S1 shape				& Flat \\
    S2 shape				& Concave cylindrical along minor axis \\
    S2 radius of curvature 	& $(230\pm30)m$ \\
    \hline
    \end{tabular}
\end{center}
\label{tab:dic_param}
\end{table}
\\The tolerances on the dichroic surface quality instead have been retrieved on the following arguments:
\begin{enumerate}
  \item To ensure the flatness of dichroic S1 the local tilt of the surface, measured on the area of a single WFS subaperture (equivalent to $13mm$), must be low enough to not displace the correspondent spot on the SH focal plane by more than $1/20$ of the WFS dynamic range. Since the WFS FoV is equivalent to $4.7arcsec$ we defined acceptable a maximum spot displacement of $0.25arcsec$. Considering the telescope plate scale of $0.66mm \; arcsec^{-1}$ on the $f_{16.6}$ plane this means that the acceptable sag of the dichroic S1 is $0.67\mu m$ over a patch of $13mm$, which corresponds to a maximum derivative of $26nm\;mm^{-1}$.
  \item The optical quality of the dichroic S1 in reflection must be better than $\lambda/10$ over the $13mm$ subaperture footprint. This is needed to ensure that the WFS is able to produce diffraction limited spots on the SH focal plane. So the specification for the S1 optical quality is a maximum surface error (SFE) of $50nm$ rms measured on patches of $13mm$ diameter and removing the local tilt.
  \item The optical quality of the 2 dichroic surfaces in single pass transmission must allow a $SR > 90\%$ over the NGS footprints to not waste the telescope optical quality and the results obtained with dichroic shape optimization. This requirement, expressed in terms of transmitted wavefront error, is equivalent to measure a maximum rms of $63nm$ on patches of $85mm$ diameter (set by the major axis of the NGS footprints) distributed on the whole optical area of the dichroic. Considering that the dichroic has a working angle of $40.5^{\circ}$, the maximum allowed wavefront error (WFE) is reduced to $48nm$ rms.
  \item To ensure that the transmitted beams are not displaced by surface local tilts, the sum of the low order aberrations over an NGS footprint must result in a tilt low enough to maintain star image within LUCI's slit width. Considering also that the instrument field rotates during the observation while the dichroic is fixed this requirement sets the maximum tilt that can be introduced in the transmitted wavefront. Since the LUCI slit width is equivalent to $0.125arcsec$ on-sky, the maximum tilt introduced on a $85mm$ diameter beam has to be less than $0.025arcsec$ to avoid the effects described above. This is requirement means that the low order aberration, evaluated summing the first 26 Zernike modes fitted on the transmitted wavefront, must contribute to a tilt smaller than $9nm\;mm^{-1}$.
\end{enumerate}
The specifications listed above and the parameters of table \ref{tab:dic_param} have been submitted to several companies. After a call for tender the production of the ARGOS dichroic has been assigned to the Soci\'{e}t\'{e} Europ\'{e}enne de Syst\`{e}mes Optiques (SESO). The units produced have been tested at company premises in April 2011 as we describe in section \ref{ssec:dic_tests}.

\subsection{Results of the optical test}
\label{ssec:dic_tests}
\begin{figure}
    \begin{center}
    \includegraphics[width=6.5cm]{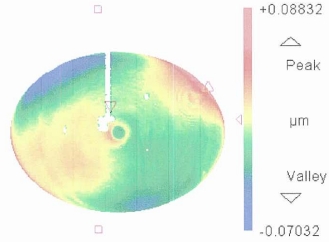}
    \end{center}
    \vspace{-0.25cm}
    \caption{\footnotesize{Interferogram of the dichroic first surface measured in units of $\mu m$. 21 different sub-patches have been selected on this surface to test the optical quality in reflection over diameters of $13mm$ (courtesy of L. Guestin, SESO).}}
    \label{fig:dich_refl}
\end{figure}

\begin{figure}
    \begin{center}
    \includegraphics[width=6.5cm]{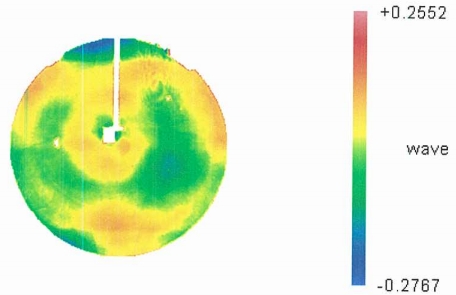}
    \includegraphics[width=6.5cm]{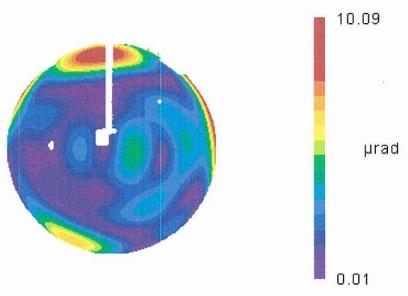}
    \end{center}
    \vspace{-0.25cm}
    \caption{\footnotesize{Analysis of the optical quality of the dichroic in transmission. Left: wavefront error over a patch of $85mm$ diameter. Right: wavefront tilt retrieved fitting the first 26 Zernike polynomials on the interferogram of the dichroic when cavity is subtracted (courtesy of L. Guestin, SESO).}}
    \label{fig:dich_transm}
\end{figure}

The test we performed to accept the manufactured dichroic consisted of:
\begin{enumerate}
  \item a 3D measurement with a scanner to check for the elliptical contour of the lens, its thickness and the radius of curvature of the cylinder.
  \item Interferometry to measure the optical quality parameters listed in section \ref{sec:dic_tol} and resumed in table \ref{tab:dic_specs}.
  \item Visual inspection with a microscope to check for surface imperfections of the two optical surfaces.
\end{enumerate}
The results of the 3D metrology measurements done on the dichroic are within the tolerance values resumed in table \ref{tab:dic_param}.
\\The visual inspection shown the presence of hairlines on the center of the cylindrical surfaces of the units produced but in any case these defects are within the specification we provided of $5/6\times0.4$\footnote{According to ISO 10110-7 standard the tolerances for surface imperfection are defined using the code $5/N\times A$, where N represents the maximum number of defects allowed within an area of $A^2 mm^2$ of the optic clear aperture\cite{2006_Handbook_OptSys_Gross}.}.
\begin{figure}
    \begin{center}
    \includegraphics[width=13cm]{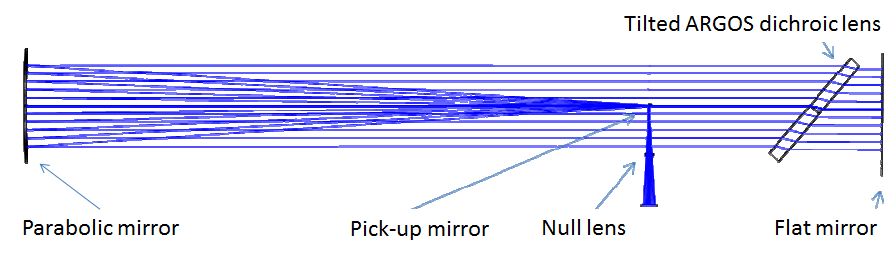}\\
       \includegraphics[width=4.6cm]{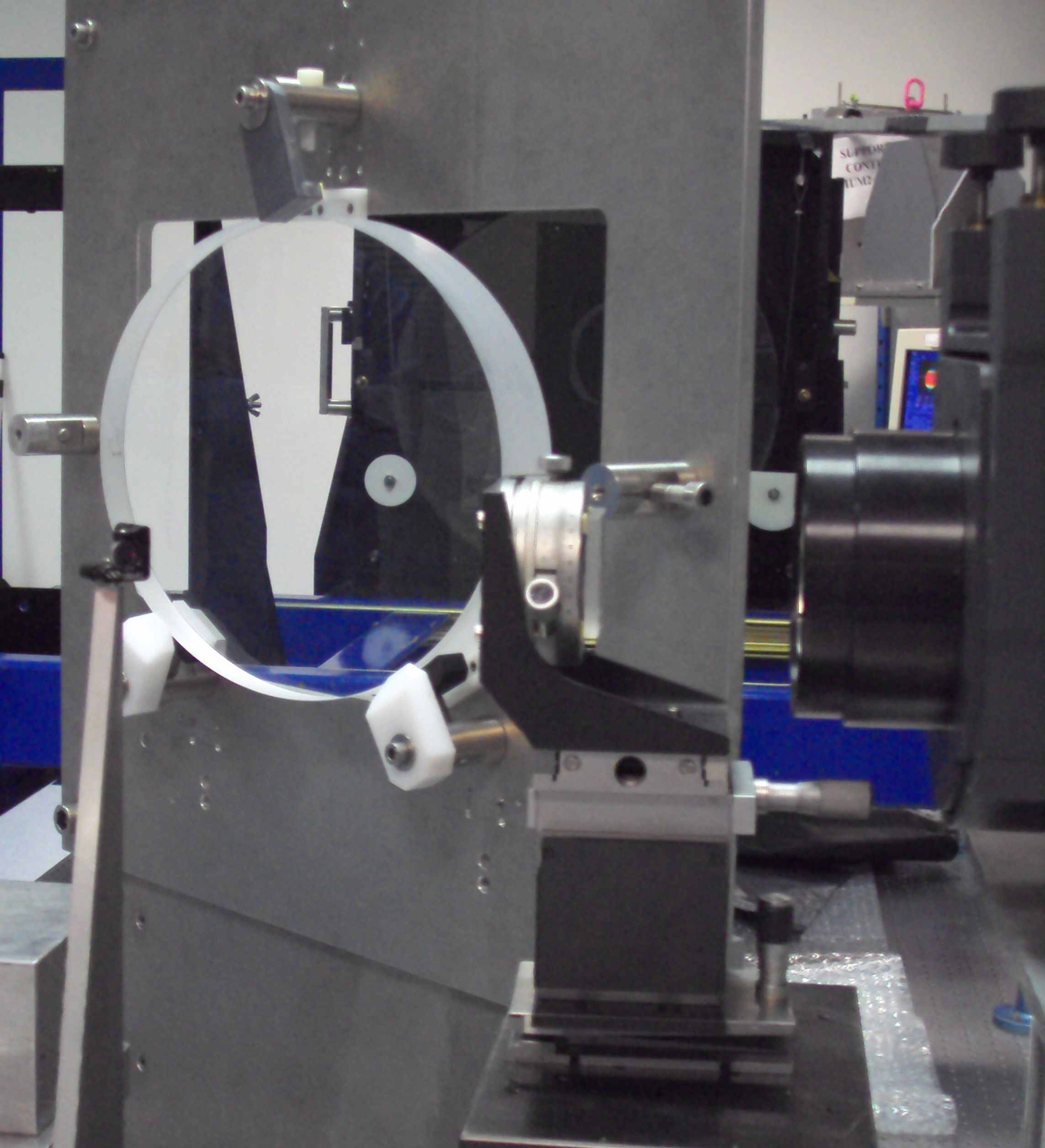}
    \includegraphics[width=8.3cm]{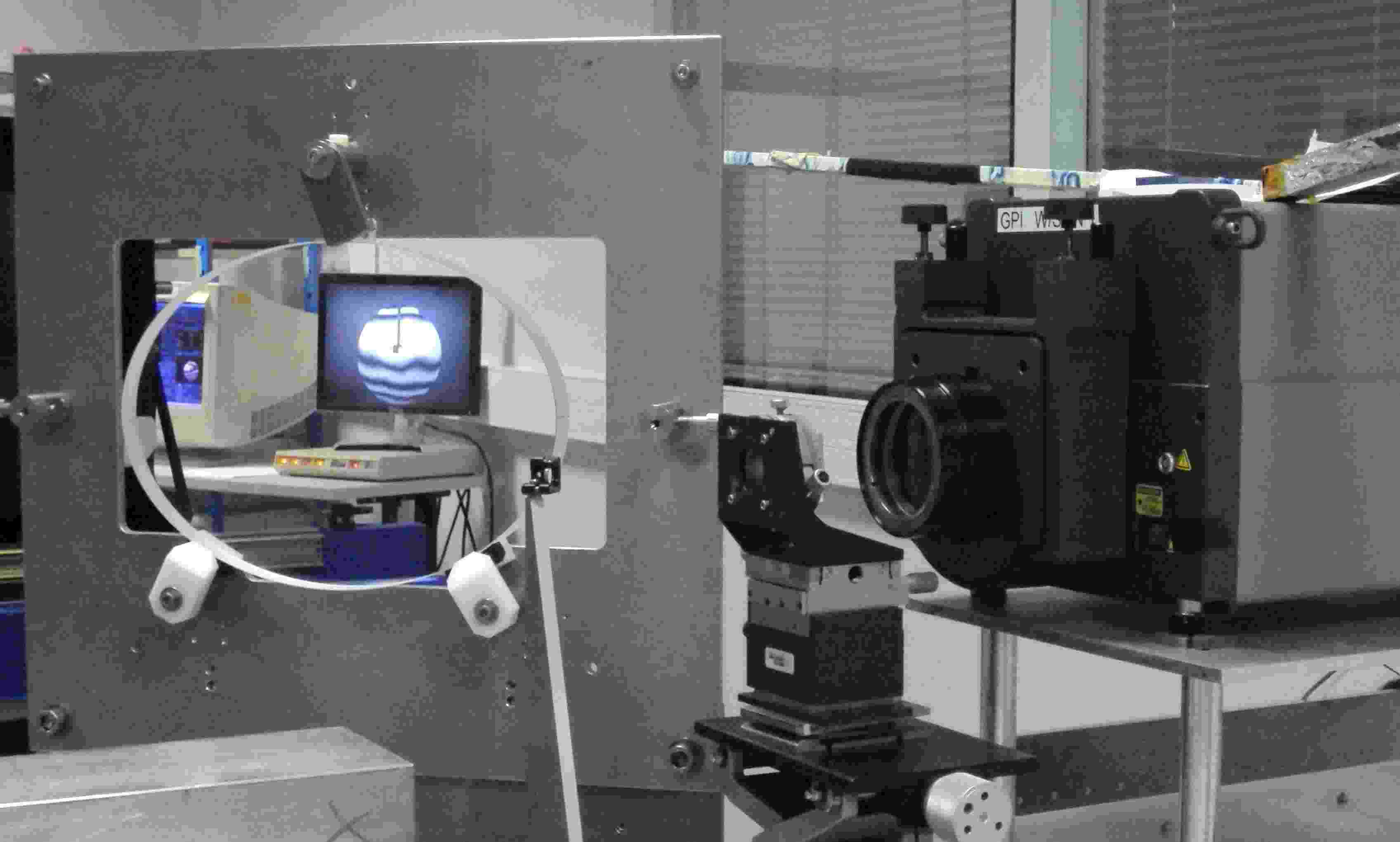}
    \end{center}
    \vspace{-0.25cm}
    \caption{\footnotesize{Above: Zemax project of the optical layout used to test the WFE in transmission of the dichroic lenses. The beam emitted by the interferometer passes through a null lens to compensate for the cylindrical shape of surface 2. The beam is then collimated by a parabolic mirror and it is transmitted through the dichroic lens. A flat mirror, tilted by $0.4^{\circ}$ to compensate for the lens wedge, reflects back the beam toward the parabola and the interferometer. Bottom: picture of the interferometric test setup produced at SESO premises (courtesy of L. Guestin, SESO).}}
    \label{fig:dich_seso}
\end{figure}
\\Figure \ref{fig:dich_seso} shows the optical setup realized at SESO premises to test interferometrically the optical quality of the dichroic units. The setup uses a full aperture beam produced by a Zygo GPI interferometer and collimated by a parabolic mirror. The optical quality of dichroic S1 in reflection is tested at normal incidence, while the dichroic is tilted by $40.5^{\circ}$ to test the optical quality in transmission. In the second case a null lens is introduced in the optical path to compensate for the cylinder on the dichroic S2 and the test are done in double pass by placing a $1m$ diameter $\lambda/20$ flat mirror tilted by $0.4^{\circ}$, to compensate for the wedge, after the dichroic. In this test the aberrations due to the cavity are subtracted from the measurements done on the dichroic by the use of fiducials on the flat mirror surface. Sub-patches of the dichroic optical area have been sampled adding a software mask to the images obtained from the interferometer.
\\The image in figure \ref{fig:dich_refl} shows the interferometric measurement of the surface tilt on dichroic S1. By selecting 21 sub-patches of $13mm$ diameter an average surface tilt of $(4.0\pm1.0)\mu rad$ have been measured. From this measure it has also been evaluated the S1 surface error that resulted to be $\sim150nm$ PtV over the full optical area (subtracting tilt and power) and $(5.0\pm2.0)nm$ on average over the 21 sub-patches.
\\The wavefront error in transmission has been measured on 7 different patches of $85mm$ diameter. Figure \ref{fig:dich_transm} on left shows one of these measurements. An average wavefront rms of $(20\pm5)nm$ have been measured. The image on right shows the wavefront tilt introduced by the low order surface aberrations of the dichroic. The image represents the sum of the first 26 Zernike polynomials fitted on the interferogram of the dichroic in transmission (after cavity subtraction). A PtV tilt of $10\mu rad$ is visible over the full optical area, while it reduces to $(5\pm2)\mu rad$ over sub-patches of $85mm$ diameter.
\\We resumed the results obtained in the optical tests in table \ref{tab:dic_specs}. The measures show that the tested optics are within the specifications described in section \ref{sec:dic_tol}.
\begin{table}
\caption{\footnotesize{Summary of the polishing specifications for the dichroic lenses provided to SESO and results of the interferometric measurements done on the units produced.}}
\vspace{-0.25cm}
\begin{center}
    \begin{tabular}{|m{4.5cm}|c|c|}
    \hline
    \textbf{Parameter} & \textbf{Specification} & \textbf{Measure} \\
    \hline
    \hline
    Average surface tilt of any sub-area \mbox{$\oslash=13mm$} inside the CA   & $<26nm\;mm^{-1}$ PtV    & $(4.0\pm1.0)nm\;mm^{-1}$ PtV \\
    \hline
    SFE of any sub-area \mbox{$\oslash=13mm$} inside the CA         			& $<50nm$ rms       & $(5.0\pm2.0)nm$ rms \\
    \hline
    WFE in transmission on any sub-area \mbox{$\oslash=85mm$} inside the CA	& $<48 nm$ rms      & $(20.0\pm5.0)nm$ rms \\
    \hline
    Sum of 26 fitted Zernike polynomials over CA				        & $<9nm\;mm^{-1}$         & $(5.0\pm2.0)nm\;mm^{-1}$ \\
    \hline
    \end{tabular}
\end{center}
\label{tab:dic_specs}
\end{table}

\section{Coating specifications}
\label{sec:dic_coating}
\begin{figure}
    \begin{center}
    \includegraphics[width=6.5cm]{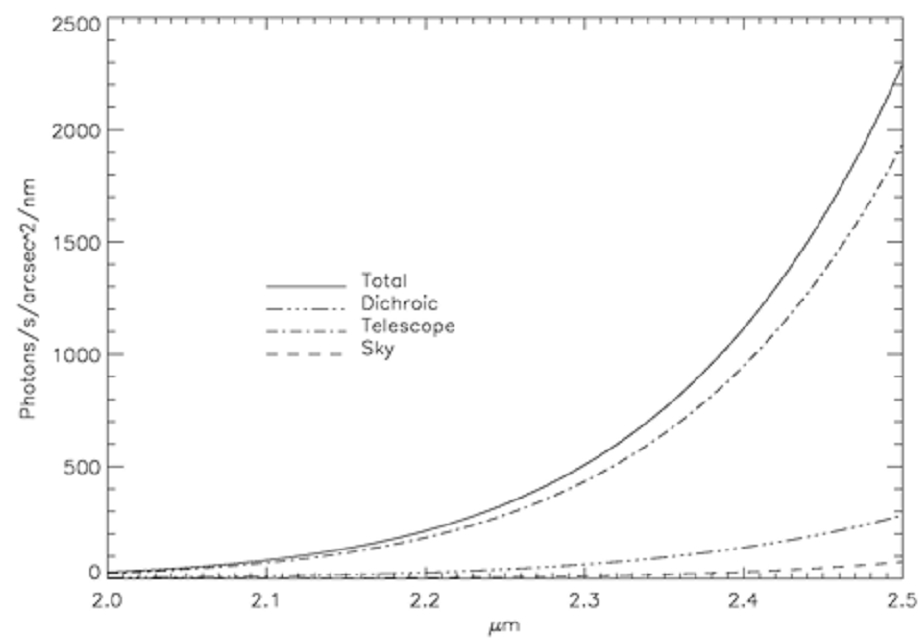}
    \end{center}
    \vspace{-0.5cm}
    \caption{\footnotesize{Plot of the different contribution to the infrared background measured on the instrument focal plane. The sources have been parameterized as black body sources at different temperatures. In this approximation the dichroic reflects an additional $2\%$ of the telescope ambient radiation toward the instrument.}}
    \label{fig:dic_backgrnd_sources}
\end{figure}

The dichroic lens coating has been designed to transmit the wavelengths larger than $0.6\mu m$ and to reflect the shorter ones, being optimized to have peaks of reflectivity at $532$ and $589nm$ in correspondence of the Rayleigh and Sodium lasers wavelengths. \\Since the ARGOS dichroic is placed in proximity of the infrared instrument it will impact the signal and background noise in different ways:
\begin{itemize}
  \item the dichroic substrate will absorb and reflect both the science objects and sky background radiation by the same amount at any wavelength,
  \item it will reflect thermal radiation from the telescope environment into the instrument increasing the background,
  \item dust and contaminants settled on the dichroic surfaces will both scatter science light and produce thermal radiation.
\end{itemize}
The choice of INFRASIL 302 as dichroic substrate will reduce the effects of absorbtion. Dust and contaminants are controlled with periodic optic cleaning. Reflection losses and thermal background injection instead are controlled with a proper coating design.
\\Since the emissivity at K band wavelengths of the telescope enclosure ambient is higher than that of the sky, the coating has to minimize the amount thermal radiation the dichroic will reflect to the instrument. Figure \ref{fig:dic_backgrnd_sources} compares the contribution of different background sources measured on the instrument focal plane. The plots have been produced considering black body sources at different temperatures: $220K$ for the sky and $273K$ for the telescope environment and dichroic. For the dichroic, in addition to the black body emission, we considered a reflection of $2\%$ of the telescope ambient emission toward the instrument. This is equivalent to consider a coating reflectivity in the K band $<2\%$. Figure \ref{fig:dic_backgrnd_sources} shows that providing the dichroic with an anti-reflection coating with $T>98\%$ in K band is sufficient to give a negligible contribution of the optic to the infrared background measured by the instrument.
\\On these assumptions we retrieved the specifications for the coating of the ARGOS dichroic. Both optic surfaces have to be coated with the same substrate to avoid bending due to differential stresses. The coating process has been assigned to Layertec GmbH. The optic is coated using ion beam sputtering, a technique that avoid the heating of the substrate to deposit the coating layer ensuring at the same time an high thermal and climatic stability of the coating deposited.

\subsection{Results of the coating test}
\label{ssec:coating_tests}
\begin{figure}
    \begin{center}
    \includegraphics[width=6.5cm]{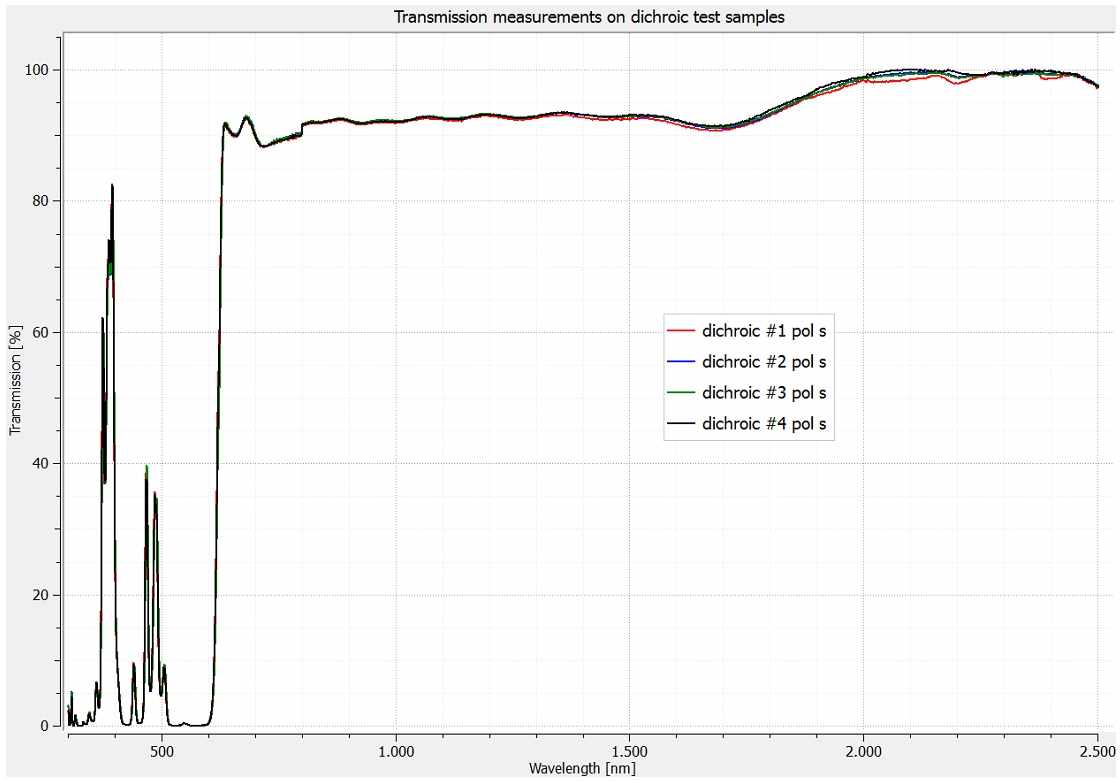}
    \includegraphics[width=6.7cm]{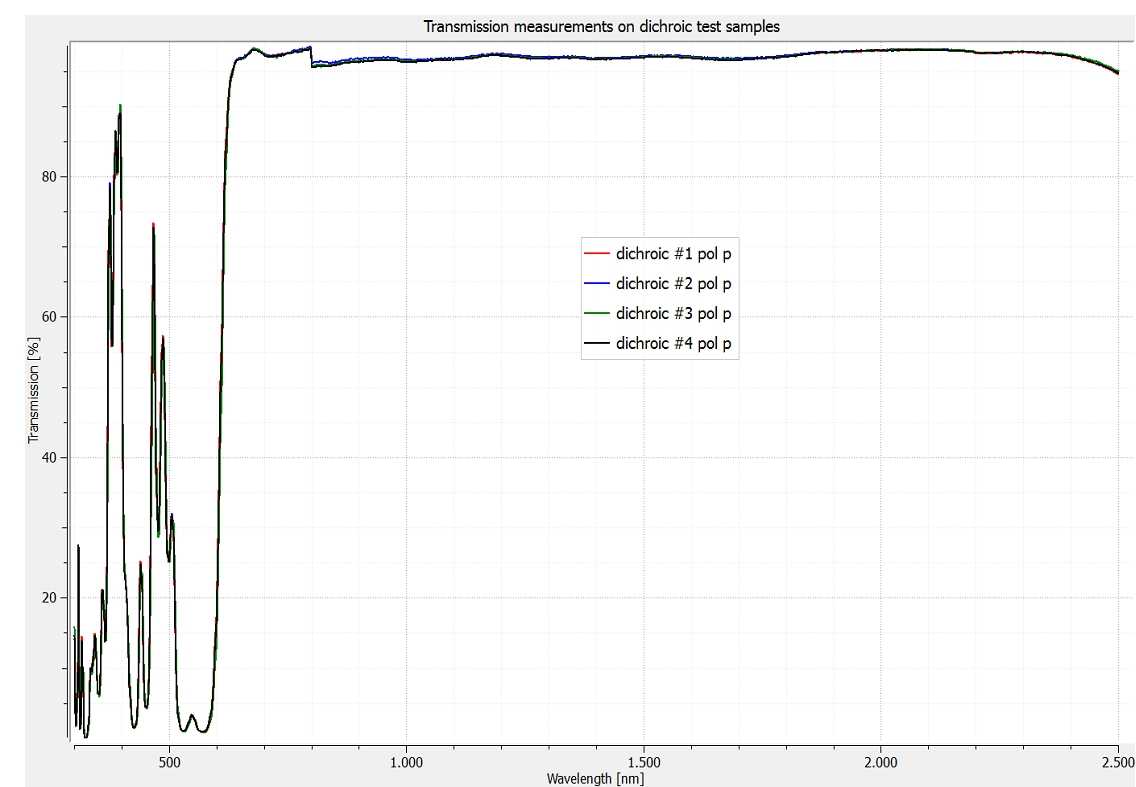}
    \end{center}
    \vspace{-0.5cm}
    \caption{\footnotesize{Transmission curves in the $300-2500nm$ wavelength range of the 4 samples of dichroic coating provided by Layertec (courtesy of L. Bahl, MPE).}}
    \label{fig:dic_coating}
\end{figure}

\begin{figure}
    \begin{center}
    \includegraphics[height=5cm]{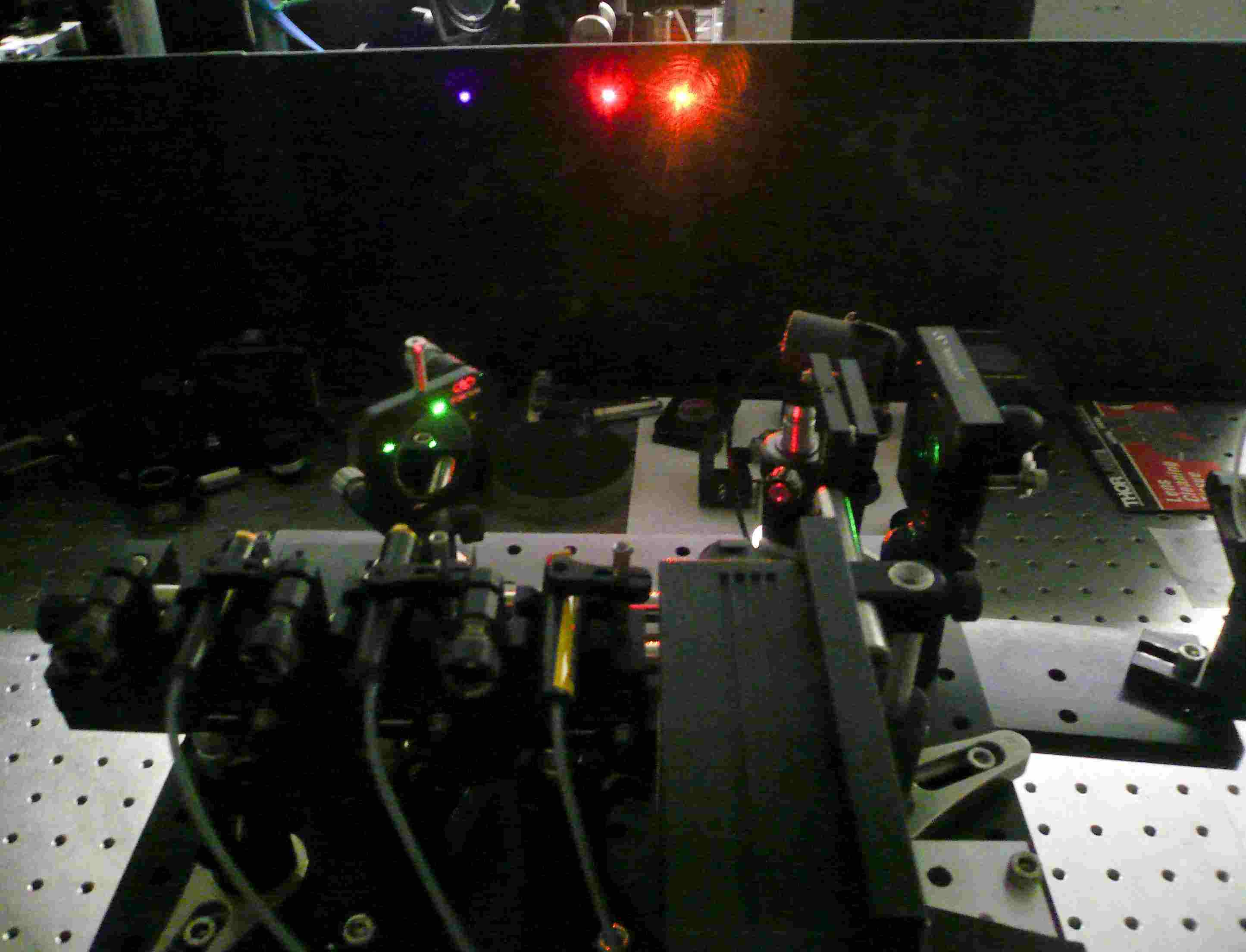}
    \includegraphics[height=5cm]{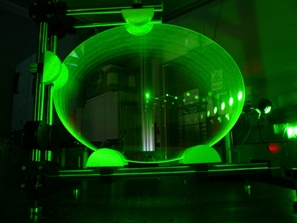}
    \end{center}
    \vspace{-0.5cm}
    \caption{\footnotesize{Left: particular of the laser sources used to measure the transmissivity and reflectivity of the dichroic coating. The green spot is not visible on the screen because it is shined in the direction of the dichroic by the fold mirror. The successive elements are a depolarizer and a linear polarizer used to select s or p polarization. On the extreme right it is partially visible the window used to reflect a portion of the light toward the reference photodiode. Right: picture taken when the dichroic was illuminated with the $532nm$ laser source. Images courtesy of S. Rabien, MPE.}}
    \label{fig:coating_test}
\end{figure}

To verify the performance of the coating deposited on the dichroic different tests have been performed at Max-Plank-Institut f\"{u}r Extraterrestrische Physik (MPE) premises.
\\The overall reflectance and transmissivity of the coating has been measured using a spectrophotometer. Since the dimension of the dichroic prevented to put the optic in this instrument, Layertec provided four samples of the dichroic coating deposited on small commercial INFRASIL 302 windows of $50mm$ diameter. Using the spectrophotometer it has been possible to measure the transmission of the windows at an angle of incidence of $40.5^{\circ}$ on a wide range of wavelengths. A polarizer was inserted in the optical path before the window to test the transmission of the s or p-polarized light (the $532nm$ laser light reflected by dichroic is p-polarized). Figure \ref{fig:dic_coating} shows the results of the test: the cutoff wavelength for reflection is $605nm$, while transmission reaches values $>90\%$ at $620nm$. The range of transmitted wavelengths is up to $2.5\mu m$. Note that the measurements $>100\%$ are caused by the $\pm1\%$ accuracy of the instrument.
\\We have then measured the reflectivity and transmissivity at specific visible wavelengths using the optical setup showed in figure \ref{fig:coating_test}. A laser source projects a small collimated beam directed toward a photodiode. The coating reflectivity (R) and transmissivity (T) are measured evaluating the ratio between the photodiode voltage with and without inserting the optic in the beam path. To filter out the laser oscillations we inserted a window just in front of the laser output and we referred the measurements to the voltage measured on this \emph{reference} photodiode. So T and R are evaluated as:
\begin{equation}\label{eq:T/R}
    T, R = \frac{V_{in} }{V_{out}},
\end{equation}
where $V_{in}$ and $V_{out}$ are given by the average of 1000 measurements of the photodiode in transmission (or reflection) over the reference one: $V_{in}, V_{out} = V_{T,R}/V_{ref}$.
\\Inserting also a linear polarizer in front of the laser output we could select s or p-polarized light. The measurements have been repeated at 532, 589 and $632.8nm$ using different laser sources with an accuracy of $\pm0.1\%$ in the measurements. The results we have obtained are shown in table \ref{tab:coating_specs} and they confirm that the coating produced by Layertec is within the specifications we provided.
\begin{table}
\caption{\footnotesize{Summary of the specifications and measures for the dichroic coating.}}
\vspace{-0.25cm}
\begin{center}
    \begin{tabular}{|l|c|c|}
    \hline
    \textbf{Requirement} & \textbf{Specification} & \textbf{Measure} \\
    \hline
    \hline
    Reflected wavelengths                         & $<0.6\mu m$       &  $<0.605\mu m$ \\
    S reflectance at 0.532 and $0.589\mu m$       & $>90\%$           &  $98\%, 98\%$ \\
    Transmitted wavelengths                       & $0.6\div2.5\mu m$ &  $0.62\div2.5\mu m$ \\
    Un-pol transmission at $0.632\mu m$           & $>85\%$           &  $88.5\%$ \\
    Un-pol transmission at $1\mu m$               & $>90\%$           &  $93.5\%$ \\
    Un-pol transmission at $2.45\mu m$            & $>98\%$           &  $98.3\%$ \\
    \hline
    \end{tabular}
\end{center}
\label{tab:coating_specs}
\end{table}

\section{Mechanical design of the dichroic support}
\label{sec:dic_mech}
\begin{figure}
  \centering
  \includegraphics[width=6.5cm]{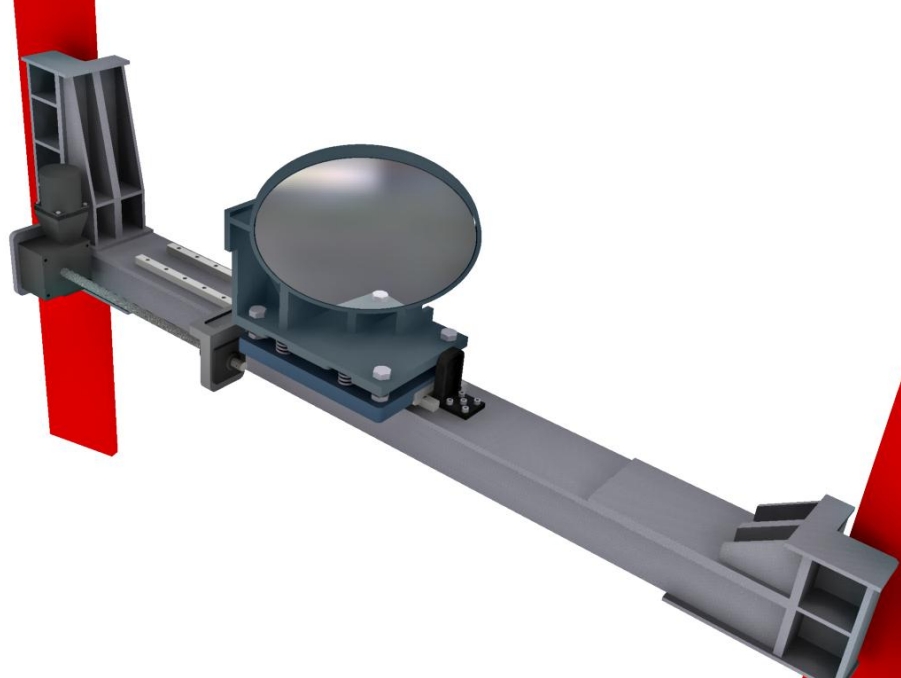}
  \includegraphics[width=6.5cm]{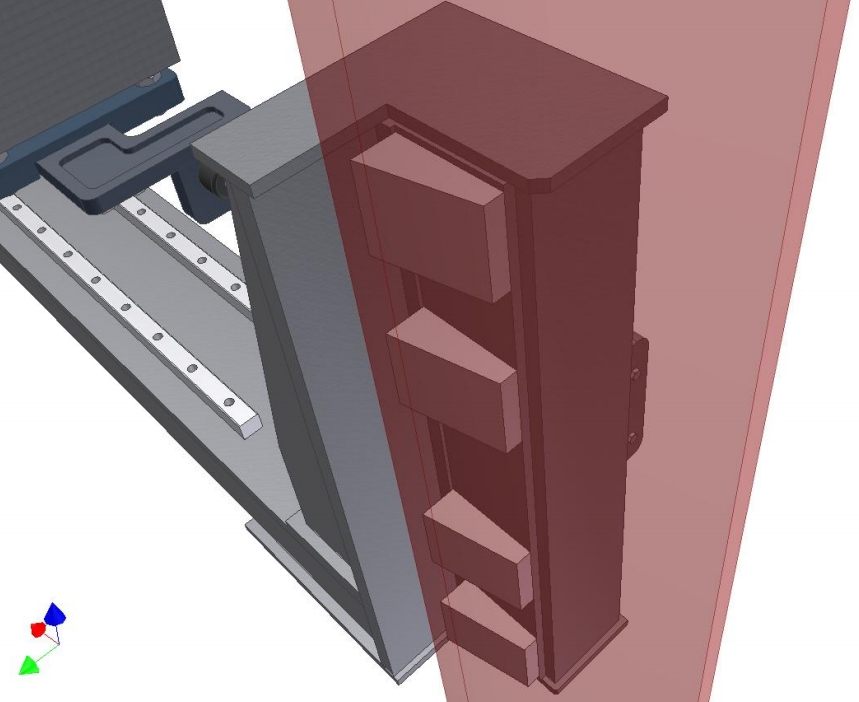}
  \vspace{-0.25cm}
  \caption{\footnotesize{Left: view of the dichroic lens support structure. The two red flanges represent the LUCI rotator structure to which the dichroic support is attached. The metal cover that protects the optic when parked has been removed to show the two railways. Right: close up view of the wedge system used to attach the dichroic support to the rotator structure.}}
  \label{fig:dic_support}
\end{figure}

\begin{figure}
  \centering
  \includegraphics[width=6.5cm]{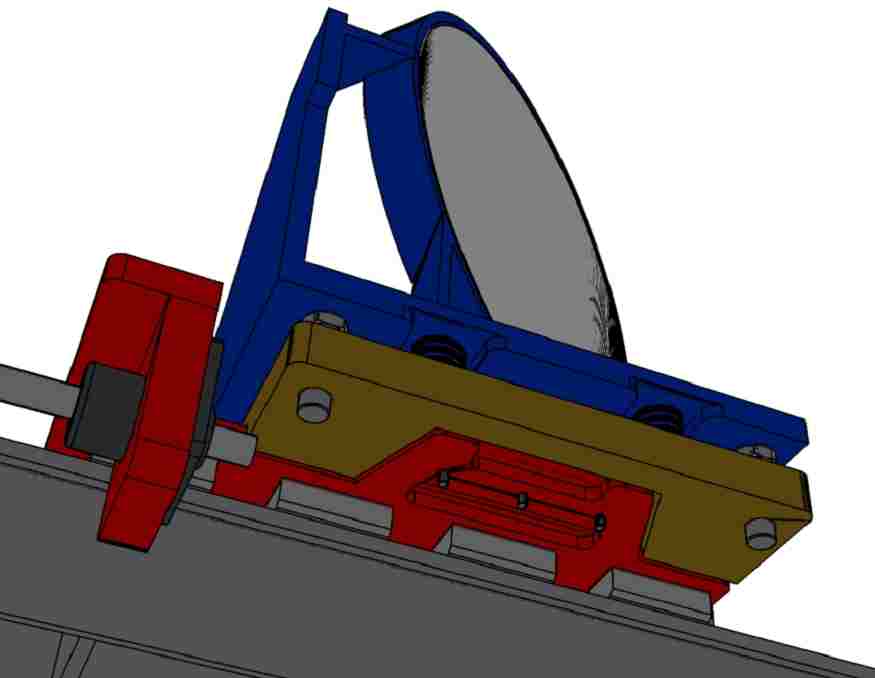}
  \includegraphics[width=6.5cm]{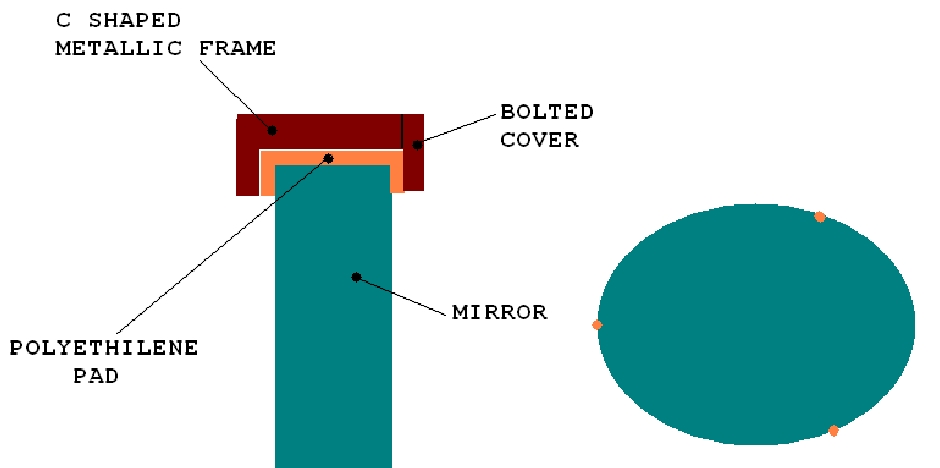}
  \vspace{-0.25cm}
  \caption{\footnotesize{Left: view of the three frames that hold the dichroic. The colors distinguish between the different adjustments. Right: scheme of the disposition of the three polyethylene pads interposed between the glass and the metallic frame.}}
  \label{fig:dic_frame}
\end{figure}

When ARGOS is in use the dichroic lens is moved in front of the LUCI focal station. Instead when ARGOS is not in use, or the LUCI calibration unit has to be deployed, the dichroic has to be removed from the telescope optical path and to be parked aside the instrument focal station by a remotely controlled deploying mechanism.
\\The dichroic support structure and its deploying mechanism have to work at different telescope elevations. Considering that the LUCI focal station forms an angle of $64^{\circ}$ with respect the telescope elevation axis, when the telescope is pointing at Zenith the gravity vector lays in a plane parallel to the focal plane and it is directed toward M1. When the telescope is pointing at horizon, the gravity vector lays in a plane orthogonal to M1 axis and it forms an angle of $15^{\circ}$ with respect the dichroic flat surface.
\\The requirements on the stiffness of the dichroic support structure and the repeatability of the deploying mechanism can be evaluated considering that the Rayleigh WFS has 2 input fields:
\begin{enumerate}
  \item a large one of $60arcsec$ diameter on-sky, that corresponds to the field imaged on the patrol cameras that are used to acquire the LGS position and to recenter them on the nominal position $2arcmin$ off-axis.
  \item a small one of $4.7arcsec$ diameter on-sky. This is the field transmitted in the WFS and represents the dynamic range of the WFS.
\end{enumerate}
The first WFS field sets the requirement on the positioning repeatability: in fact to ensure that the LGS fall inside the field patrolled by the acquisition cameras the dichroic tip-tilt introduced by looses in the deploying mechanism must be within $\pm0.1^{\circ}$. For the same reason $\pm2mm$ are allowed in the position repeatability along the optical axis and the decenter of the optic in the direction on deployment. To have a proper control on the working position of the dichroic a resolution better than $0.1mm$ is necessary in the deployment mechanism.
\\The second WFS field sets the requirement on the stiffness of the support structure. When the telescope is tracking the change in elevation, so in the direction of the gravity vector, can induce bending on the dichroic shape that can move the LGS out of the WFS FoV causing the interruption of the adaptive optics correction loop. So the dichroic mechanical mount has to ensure a total sag induced by gravitational effects less than $0.15mm$, with a goal specification of $0.03mm$.
\\Table \ref{tab:mech_specs} resumes the complete set of requirements for the design of the dichroic mechanical structure.
\begin{table}
\caption{\footnotesize{Summary of the specifications for the design of the dichroic mechanical support.}}
\vspace{-0.25cm}
\begin{center}
    \begin{tabular}{|l|p{8cm}|}
    \hline
    \textbf{Requirement} & \textbf{Specification} \\
    \hline
    \hline
    Deploying repeatability                 & Tip-tilt: $\pm0.1^{\circ}$; position: $\pm2mm$   \\
    Deployment resolution                   & $0.1mm$ \\
    Deployment time                         & $<60s$ \\
    Alignment range                         & Tip-tilt: $1^{\circ}$; position: $2mm$ \\
    Stability                               & Total sag due to elevation change $<0.15mm$ (baseline), $<0.03mm$ (goal) \\
    Mass                                    & $<250kg$ \\
    Safety                                  & Interlocks required to prevent collisions \\
    Dust removal                            & Dry air flow to prevent dust deposition when parked \\
    Temperature                             & Operating: $-20\div30^{\circ}C$, storage: $-30\div50^{\circ}C$ \\
    \hline
    \end{tabular}
\end{center}
\label{tab:mech_specs}
\end{table}
\\The final design of the dichroic support has been produced by A.D.S. International, figure \ref{fig:dic_support} shows the 3D model of the unit. The system is composed by:
\begin{enumerate}
  \item A massive mounting frame that provides a stiff support for the optic and the mechanics.
  \item Two railways to drive the optic from its park position to the operative one and back.
  \item A steel frame sliding on the railways, whose relative position can be changed to correct the alignment of the dichroic.
  \item An electric motor with a reduction gear and a screw to pilot the sliding motion of the otpic on the railways.
  \item Two sets of shims, shaped as wedges, to permit a clear interface of the mounting frame with the beams of the LBT rotator gallery.
  \item A metal cover to protect the optic when it is parked.
\end{enumerate}
The interface of the dichroic support of the rotator gallery is obtained by a series of metallic wedges as shown on right of figure \ref{fig:dic_support}. Such a system allows, by small changes in the profile of the wedges, the finalization of the design despite the uncertainties on the real geometry of the interface, allowing a $20mm$ tolerance in the interface. The wedges moreover are maintained when the dichroic support is detached from the gallery to allow the primary mirror aluminization and they serve as reference surface plate to speed up the reassembling of the system.
\\The dichroic slides along the railways held by three frames (see left of figure \ref{fig:dic_frame}). The first frame (red) is completely fixed to the carriages, the second one (brown) can slide above the first permitting to adjust the position of the dichroic in the same plane of the railways. The third frame (blue), during the alignment, lays over four springs that keep it suspended above the second. A set of bolts, acting against the springs, permit to adjust the vertical position of the dichroic and the tilt around the horizontal axis passing by its center. Once that the alignment is concluded the frames are strongly fixed one to each other by standard bolts. The position of the dichroic can be adjusted by translations up to $3mm$ and by rotations up to $3^{\circ}$.
\\The dichroic is mounted inside an elliptic frame of aluminium where it lays above three small pads built from Polyethylene (see right of figure \ref{fig:dic_frame}). These pads, with a thickness of $2mm$, allow the minimum needed compensation for what concerns the difference of thermal deformations between the glass and the steel, and provide at the same time a sufficient stiff basis for the mirror.

\subsection{Finite element analysis}
\label{ssec:fea}
\begin{figure}
  \centering
  \includegraphics[width=5.7cm]{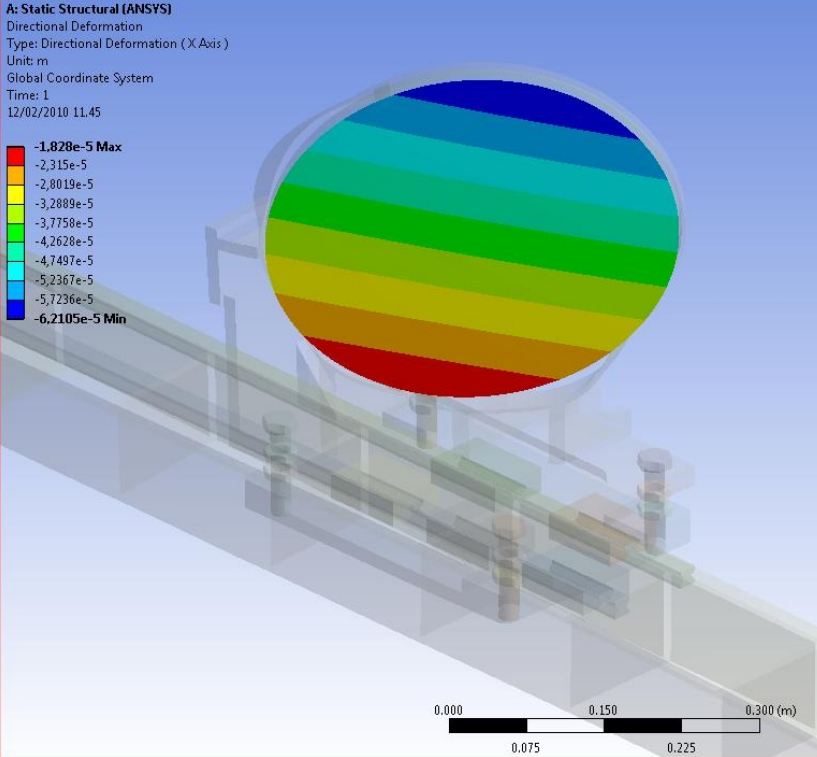}
  \includegraphics[width=7.5cm]{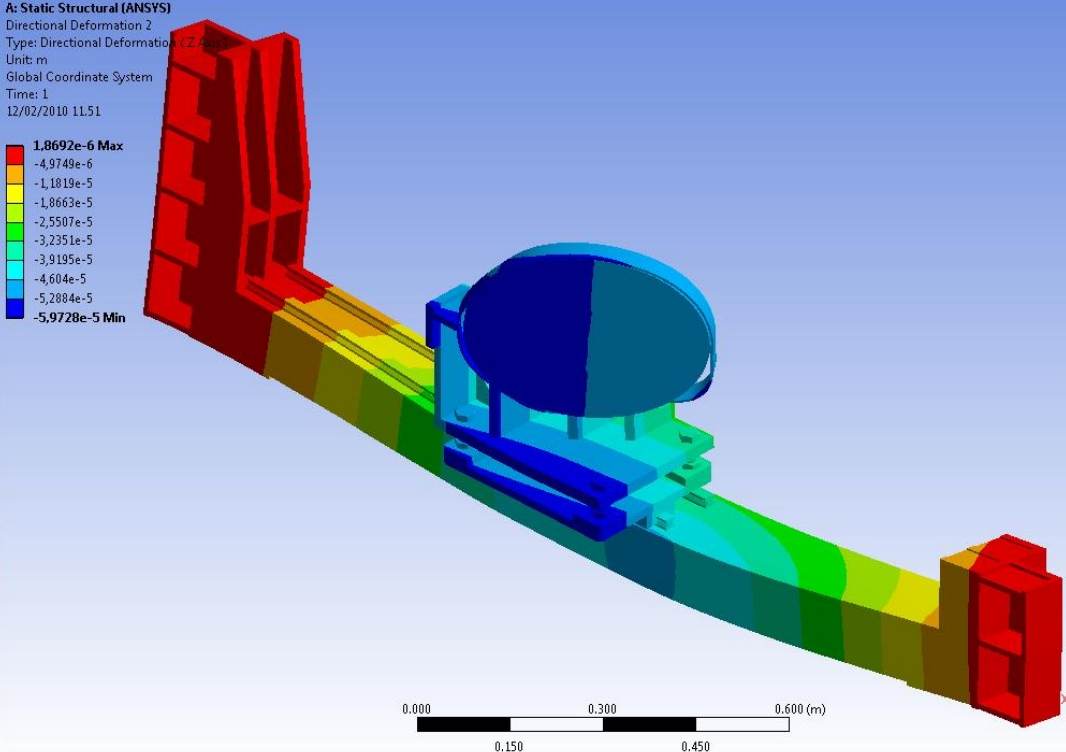}
  \vspace{-0.25cm}
  \caption{\footnotesize{Left: displacement of the dichroic lens along LUCI optical axis when telescope points at horizon. Right: displacements of the support structure when the telescope points at Zenith. In these figures the x axis is directed towards the left, the y axis is vertical and directed toward the top of figures and the z axis is coincident with the optical path and it points inside the page.}}
  \label{fig:dic_fea}
\end{figure}

A finite elements analysis has been performed by A.D.S. to check the mechanical stability of the dichroic support under the effects of gravity. The model has been developed using only shell and beam elements. The results resumed table \ref{tab:fea_results} refer to a coordinate system where the z axis coincides with the optical axis of LUCI, the y axis is parallel to the primary mirror axis and the x axis is directed towards the LGS WFS (see caption of figure \ref{fig:dic_fea}). The center of the coordinate system is located in the center of the dichroic first surface. Note that the values of displacements and rotations reported are the highest detected over the entire dichroic surfaces. The first eigenmode of the structure is at $47Hz$. These results demonstrate that the model of the dichroic mechanical structure is compliant to the design specifications.
\begin{table}
  \centering
  \caption{\footnotesize{FEA analysis results. Tilt and displacement of the dichroic under the effects of gravity are inside the acceptable range for telescope pointing both at Zenith and horizon.}}\label{tab:fea_results}
\begin{tabular}{|m{2cm}|m{2.2cm}|m{2.5cm}|m{2.5cm}|m{2.5cm}|}
  \hline
  \textbf{Telescope pointing} & Tilt around y [$^{\circ}$] & Displacement in x [$\mu m$] & Displacement in y [$\mu m$]& Displacement in z [$\mu m$] \\
  \hline
  \hline
  Zenith  & 0.0053 & -59.7 & 10.0 & 44.1 \\
  Horizon & 0.0084 & -10.0 & 86.0 & 62.1 \\
  \hline
\end{tabular}
\end{table}

%

%
%
%
%

\chapter{The wavefront sensor design}
\label{cap:wfs}

\begin{figure}
  \centering
  \includegraphics[width=12cm]{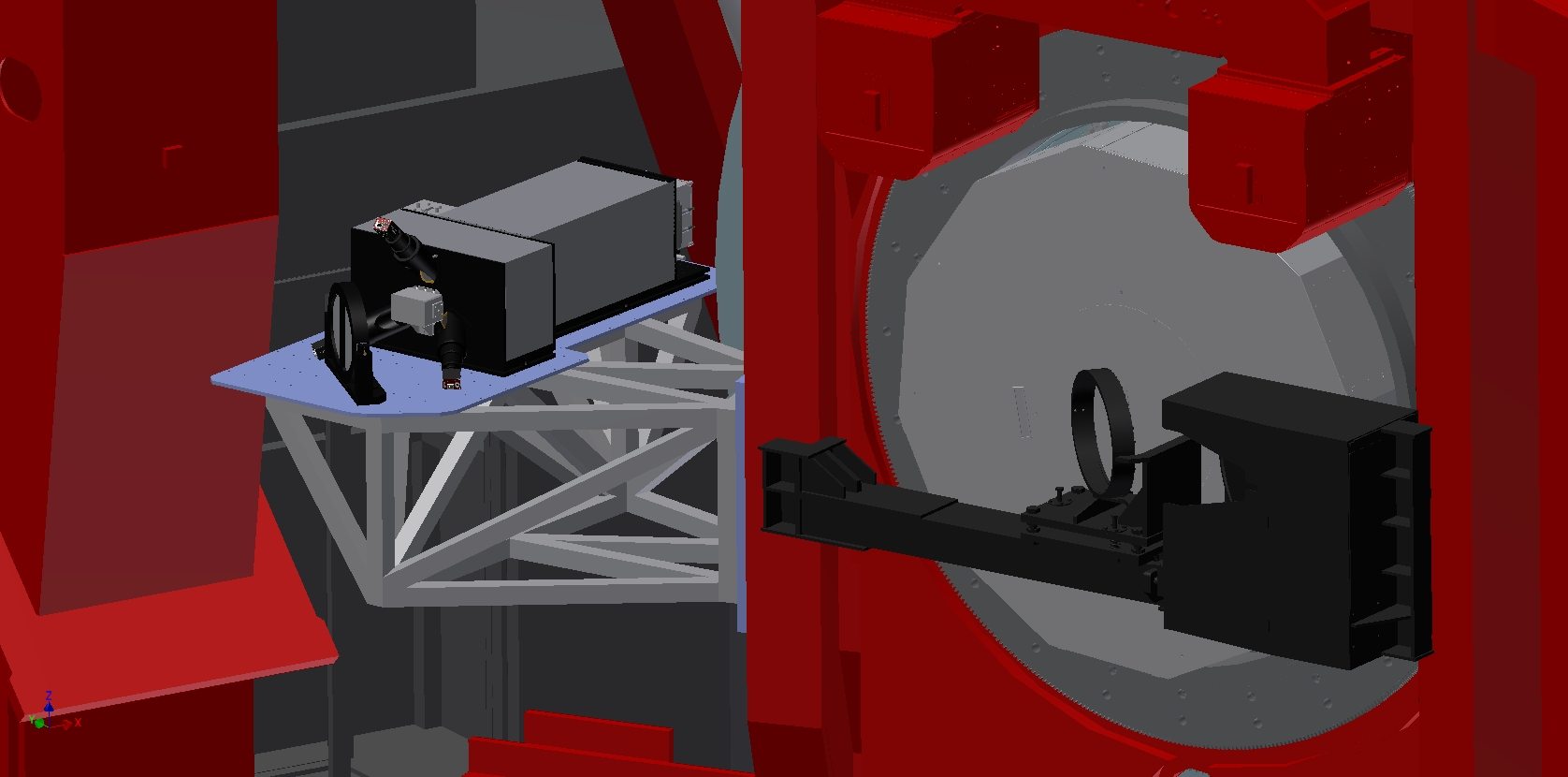}
  \vspace{-0.25cm}
  \caption{\footnotesize{3D model of the telescope showing the LUCI focal station, the dichroic mount attached in front of it and the Rayleigh WFS installed on its support structure bolted aside the instrument.}}
  \label{fig:wfs_cap_overview}
\end{figure}

\begin{figure}
  \centering
  \includegraphics[width=12cm]{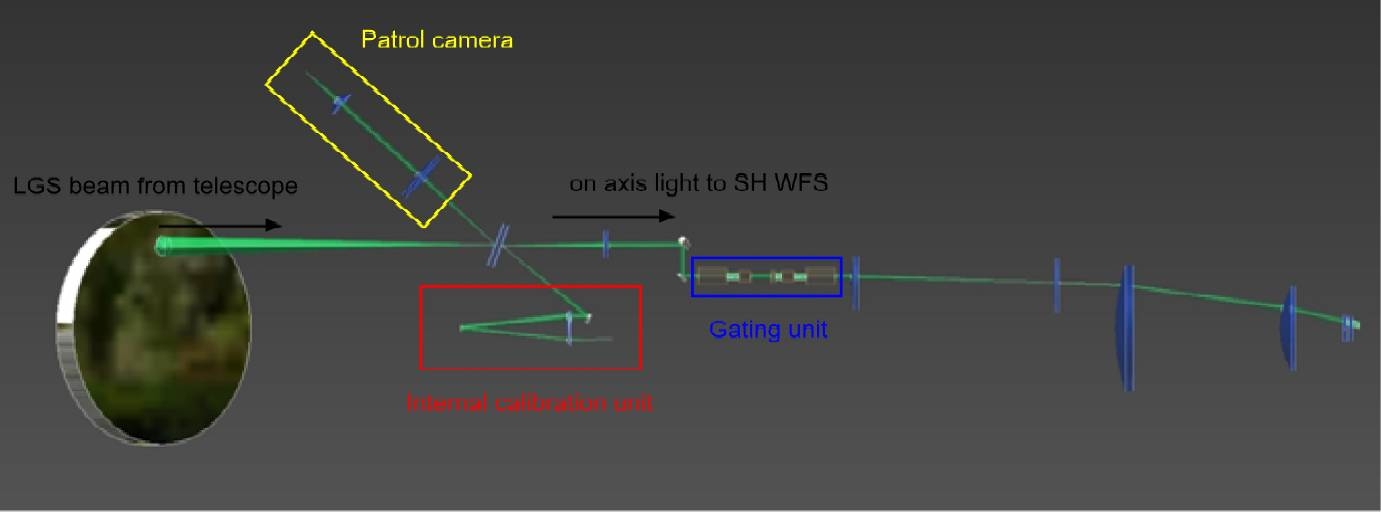}
  \vspace{-0.25cm}
  \caption{\footnotesize{3D model showing the optics that compose a single arm of the WFS. Different subsystems are highlighted and they will be discussed in detail in next sections.}}
  \label{fig:wfs_1arm_view}
\end{figure}

\begin{figure}
  \centering
  \includegraphics[width=11cm]{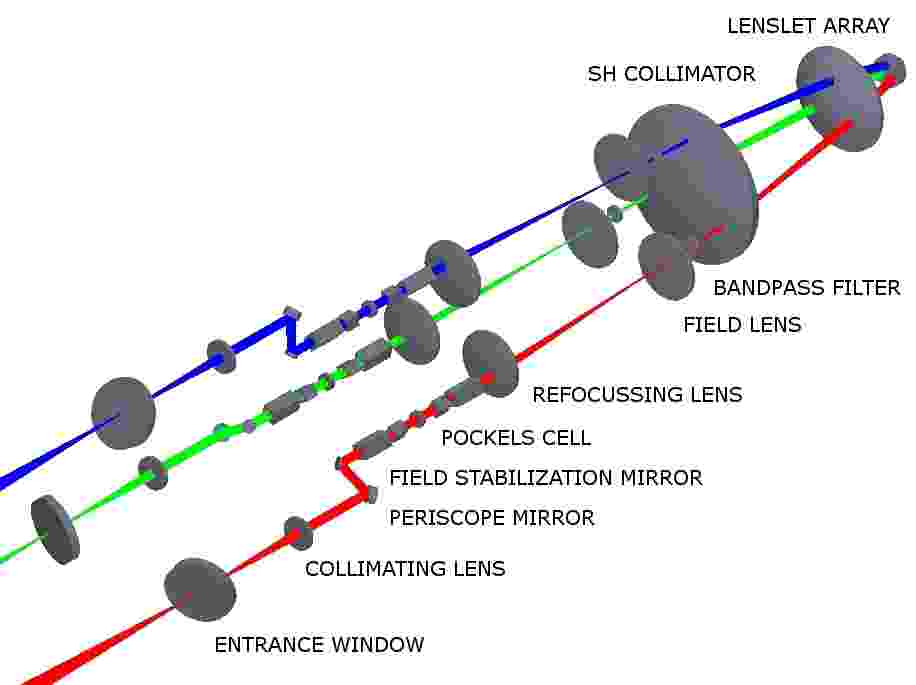}
  \vspace{-0.25cm}
  \caption{\footnotesize{Scheme of the arrangement of the 3 LGS beams inside the WFS. The optical element up to the detector are marked and the light path of the 3 beams is drawn in different colors.}}
  \label{fig:wfs_optics}
\end{figure}

In this chapter we describe the final design of the ARGOS wavefront sensor. The main requirements that we have considered designing the WFS are:
\begin{itemize}
  \item It has to sense independently the wavefront aberrations in the direction of the 3 LGS arranged on a triangular asterism of $2arcmin$ radius with 3 Shack-Hartman (SH) type WFS.
  \item Each WFS has to sample the pupil with $15\times15$ subapertures to ensure that almost 150 modes can be corrected. This number is sufficient to reduce the contribution of the fitting error below the residuals of the uncorrected higher layer turbulence.
  \item The WFS FoV should be larger than $4.5arcsec$ to allow operations also in bad seeing conditions.
  \item The 3 SH patterns have to be arranged on a single $256\times248$ pixel detector with $48\mu m$ pixel-size.
  \item To shrink the spot elongation effects on the detector the WFS has to host gating units to gate the range of altitudes from which the backscattered light can reach the sensor. These gating units have to be feeded with a collimated beam of $6mm$ diameter.
  \item The WFS must be provided with 3 large field cameras ($1arcmin$ FoV diameter) to track the laser spots on sky and to control the laser pointing system in closed-loop.
  \item The WFS must provide field stabilization devices to compensate for the jitter of the laser spots on sky.
  \item The WFS is fed with s-polarized monochromatic light at $532nm$.
  \item To compensate any mechanical flexures that can affect the position of the pupil in the WFS, the system must implement a control loop that estimates the position of the pupils from the CCD frame. This has to be done independently for each one of the 3 beams. The system has then to re-adjust the position of the beams on the lenslet array that creates the SH pattern on the CCD.
  \item An internal calibration unit is needed to check the WFS alignment in laboratory and at the telescope. This device must be also provided with a Deformable Mirror (DM) that will be used during the laboratory characterization of the WFS.
  \item A shutter and a flat field illuminator are needed to calibrate the WFS detector dynamics.
  \item Because of the high number of optical elements that compose the WFS system their transmission or reflection coefficient at $532nm$ must be as high as possible.
  \item To reduce aging of the coatings and dust contamination the WFS enclosure will be sealed and it will be flushed with a flow of dry air to keep the dust out of it. Critical optics, as the WFS entrance windows or the detector window, will be continuously flushed with dry air when ARGOS is not in use.
\end{itemize}
As described in chapter \ref{cap:argos} the Rayleigh WFS sits on a dedicated bench, bolted aside the LUCI focal station. Figure \ref{fig:wfs_cap_overview} shows a 3D model of the WFS placed in correspondence of the $f_{16.6}$ plane where the telescope optics focus objects at $12km$ distance. The large flat mirror in front of the WFS folds the laser light coming from the dichroic toward the WFS itself.
\\For shake of simplicity figure \ref{fig:wfs_1arm_view} shows the layout of only one arm of the WFS optics, optic mounts and holders have been hidden. The laser light folded by the flat mirror (section \ref{sec:fold_mirror}) is focussed in correspondence of the WFS entrance window. This element, described in detail in section \ref{sec:ew}, acts as field stop for the WFS: a $4.7 arcsec$ FoV around the nominal laser focus position is transmitted into the WFS, while an annular region of $60arcsec$ diameter is reflected and imaged on a wide-field camera to control the laser pointing on sky (see section \ref{sec:pc}). The entrance window back surface is also used to reflect toward the WFS few percents of the light emitted by an internal light source used to check the functionality of the WFS and to perform the laboratory tests (section \ref{sec:cal_unit}).
\\In figure \ref{fig:wfs_optics} are shown the optics employed to sense all the 3 laser beams. The first element after the entrance window is a plano-convex lens of $100mm$ focal length that collimate the beam for the gating unit and it images the telescope pupil, placed $15.1m$ before the lens, on the steering mirror surface. The position of this lens is remotely controlled using two axis motorized stages to place its axis on the LGS focus. In such a way the lens steers the off-axis beam and it makes it parallel to the WFS mechanical axis. More details on this element will be given in section \ref{sec:coll_lens}.
\\To reduce the distance between the 3 LGS beams and to reduce the dimension of the final collimator optics a periscope assembly has been designed (see section \ref{sec:pi}). This element is composed by a $12.5mm$ diameter flat mirror and a piezo-driven mirror. The latter device is conjugated to the telescope pupil by the collimating lens and it is used to compensate for the LGS beam jitter on-sky.
\\Because the WFS detector can expose at maximum frequencies of $1kHz$ and in this time interval it will integrate the backscattered light from all altitudes it is necessary to provide the WFS with a device able to properly shutter the laser light. This device is the gating unit that is inserted in the optical trail of the WFS after the piezo-driven mirror and it is described in section \ref{sec:pockels}. Triggering the gating unit with the laser pulses allows the WFS detector to integrate only the light coming from a short range of altitudes ($300m$) centered at $12km$.
\\From the begin of the launch system up to the gating units the 3 laser beams are arranged on a $120^{\circ}$ rotational symmetry. This symmetry has to be broken because they have to fit on the square geometry of the lenslet array and detector. As a consequence of that, the refocusing lenses and field lenses work in a decentered position. In section \ref{sec:ref+field_lenses} it will be described that the main task of this elements is to center both pupil and the spot pattern on the proper position on lenslet array and detector respectively.
\\The last independent optic of the 3 beams is a $532nm$ bandpass filter to avoid the background light to reach the detector. After this element the 3 LGS beams are collimated by a group of 2 lenses, section \ref{sec:SH_coll}, and they are arranged on a single lenslet array and detector, sections \ref{sec:lenslet} and \ref{sec:ccd} respectively.

\section{Fold mirror}
\label{sec:fold_mirror}
\begin{figure}
  \centering
  \includegraphics[width=8cm]{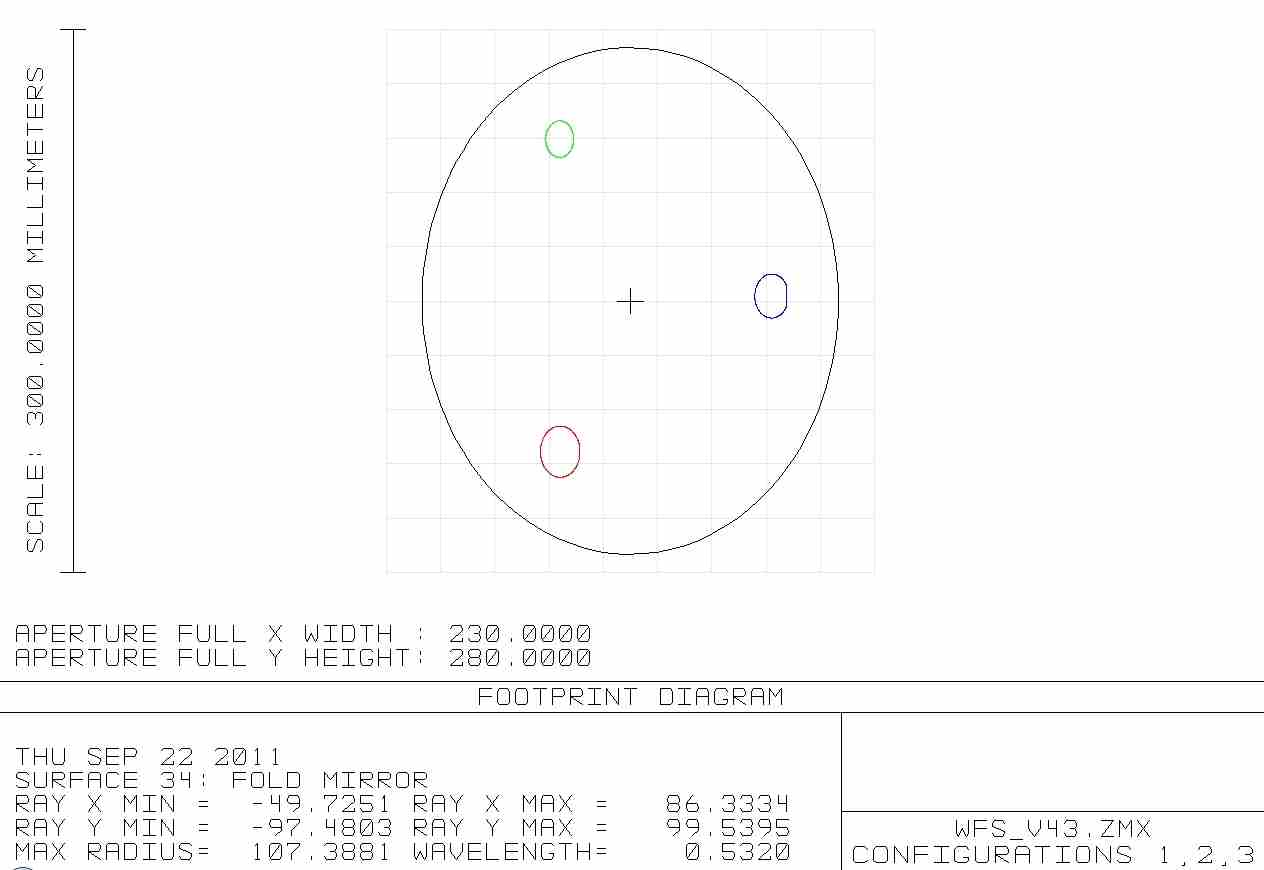}
  \vspace{-0.25cm}
  \caption{\footnotesize{Position of the 3 LGS footprints on the WFS fold mirror. Patches major axes are between $24.5mm$ (green) and $47.8mm$ (red). The top of the mirror is towards right of figure.}}
  \label{fig:fold_layout}
\end{figure}

\begin{figure}
  \centering
  \includegraphics[height=4.2cm]{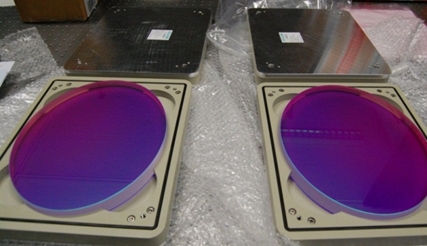}
  \includegraphics[height=4.2cm]{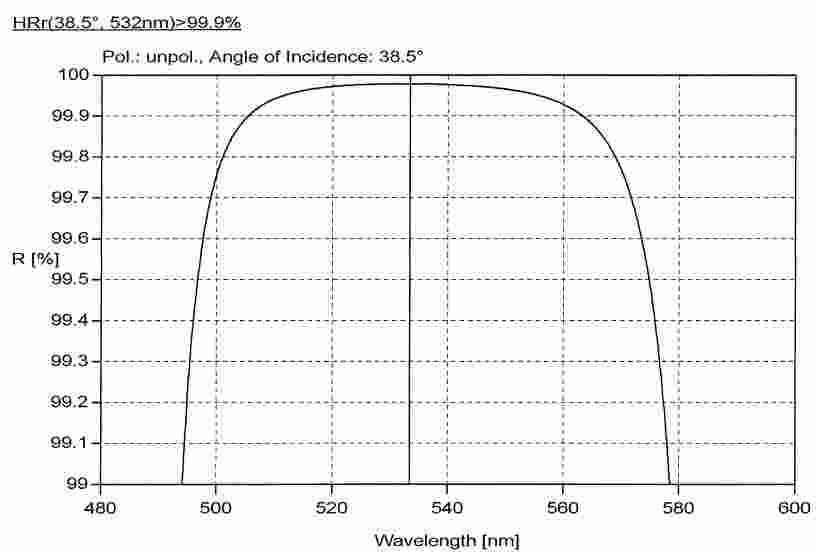}
  \vspace{-0.25cm}
  \caption{\footnotesize{Left: Picture of the 2 units of fold mirrors produced by Layertec. Right: reflectivity curve measured by Layertec on a sample patch of the fold mirror. A $R>99.9\%$ was measured at $532nm$ with an angle of incidence of $38.5^{\circ}$.}}
  \label{fig:fold_coating}
\end{figure}

The first element of the WFS is a flat mirror made of BK7 glass having an elliptical shape with $280\times230mm$ axes and $25mm$ thickness. Two units of this element have been produced by Layertec GmbH that take into account both the polishing of the substrates and subsequently the coating of the units.
\\The choice of having a single reflective surface for the 3 beams instead of separated mirrors is dictated from the fact that it avoids to provide the WFS with an independent focus system for the 3 arms. Figure \ref{fig:fold_layout} shows the arrangement of the 3 laser beams on the fold mirror. The beam footprints have different dimension because the mirror is inclined by $38.5^{\circ}$ with respect the direction of the beams. The bigger footprint has a major axis of $47.8mm$ the smaller one $24.5mm$.
\\The requirements for the mirror polishing were specified in terms of surface quality, that had to be better than $\lambda/4$ over the full optical area reaching $<\lambda/10$ over any circular patch of $50mm$. Two units of these mirrors were produced by Layertec (see figure \ref{fig:fold_coating}). Layertec was also able to produce a dielectric coating reaching $R>99.9\%$ at $532nm$ as it is shown by the reflectivity plot on right of figure \ref{fig:fold_coating}.

\section{Entrance window}
\label{sec:ew}
\begin{figure}
  \centering
  \includegraphics[width=8cm]{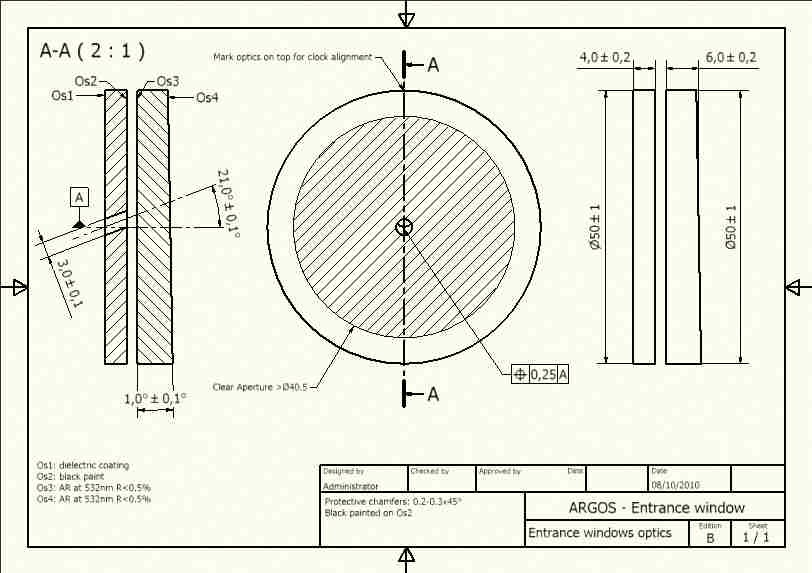}
  \vspace{-0.25cm}
  \caption{\footnotesize{Drawing of the entrance window with specifications and tolerances required in manufacturing.}}
  \label{fig:ew_sketch}
\end{figure}

The 3 entrance windows of the WFS are placed in correspondence of the $f_{16.6}$ plane. They are used to feed the light from and to the different sub-systems of the WFS, acting as:
\begin{itemize}
  \item Field stop for the LGS beams, transmitting to the SH sensor a circular FoV of $4.7 arcsec$ diameter.
  \item Back reflecting surface: their back side is used to direct the light from the internal light source toward the SH sensor.
  \item Front reflecting surface: the first surface of the windows reflects an annular FoV of $60 arcsec$ to an imaging camera.
\end{itemize}
To ease the production a single entrance window is composed by 2 elements: the first one is a $4mm$ thick plane-parallel window drilled in the center of the front surface. The second one is a $6mm$ thick wedged window. The $1^{\circ}$ wedge in the window allows to avoid double reflections of the laser light. Figure \ref{fig:ew_sketch} resumes the physical characteristics of the two elements and the tolerances required for manufacturing. The optical surfaces of the 2 elements have the following specifications:
\begin{itemize}
  \item OS1: it is polished to ensure a $\lambda/4$ surface quality over the CA. It is dielectric coated to ensure a reflectivity of $R>99\%$ at $532nm$.
  \item OS 2: the rear surface of the first element of the EW is commercially polished and black-painted to block the few percent of light transmitted by OS 1 to reach the WFS.
  \item OS 3: it is polished to ensure a $\lambda/10$ surface quality and it is coated to ensure $T>99.5\%$ at $532nm$ over a $30mm$ diameter patch, centered on the optical axis.
  \item OS 4: same polishing and coating of OS 3.
\end{itemize}

\subsection{Entrance window test}
\label{ssec:EW_test}
\begin{figure}
  \centering
  \includegraphics[width=6.5cm]{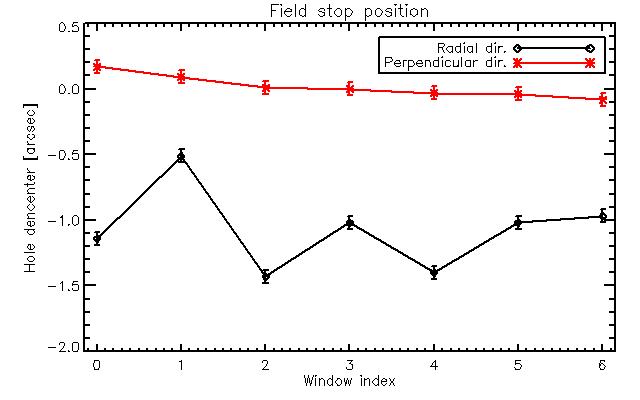}
  \includegraphics[width=6.5cm]{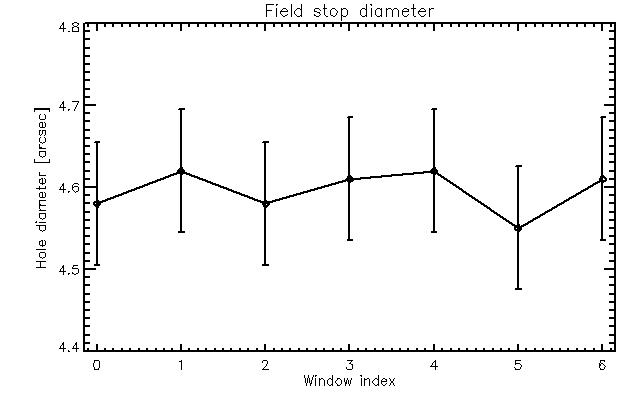}
  \vspace{-0.25cm}
  \caption{\footnotesize{Measurements of the position and diameter of the field stop on the first element of the 7 entrance windows.}}
  \label{fig:ew_hole_pos_dim}
\end{figure}

\begin{figure}
  \centering
  \includegraphics[width=10cm]{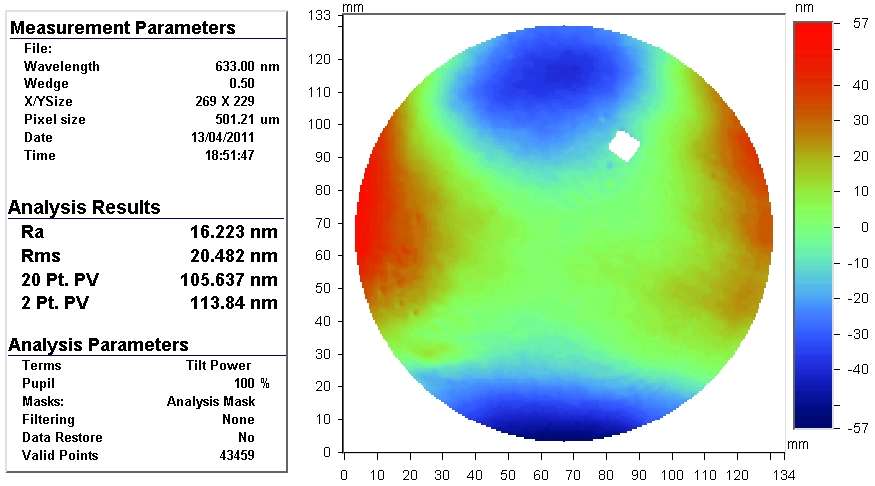}
  \vspace{-0.25cm}
  \caption{\footnotesize{Surface error of the second element of the entrance window measured with the Wyko interferometer. The white square is the fiducial used for the cavity subtraction.}}
  \label{fig:ew_interf}
\end{figure}

\begin{figure}
  \centering
  \includegraphics[width=5cm]{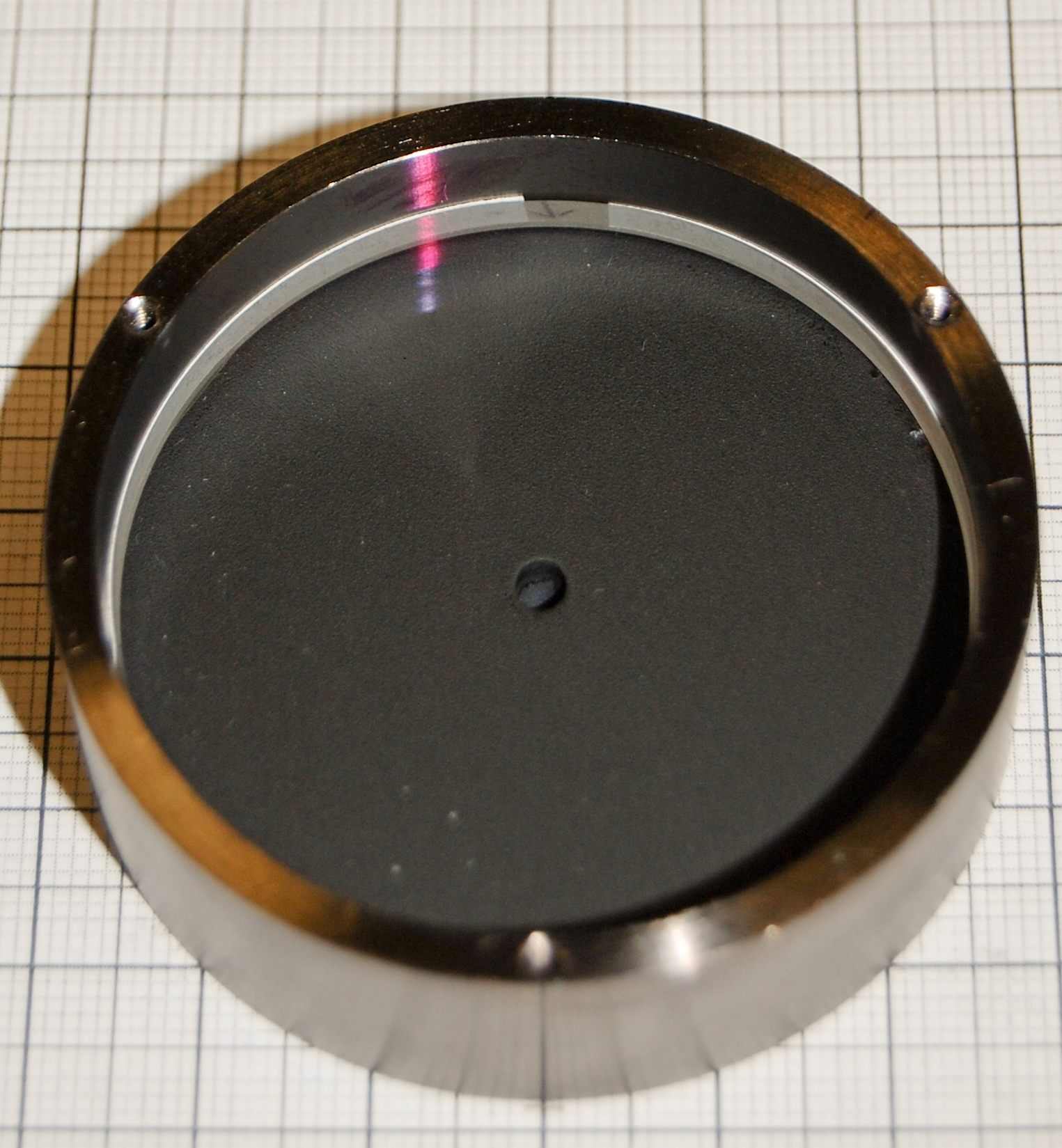}
  \includegraphics[height=5cm,width=7cm]{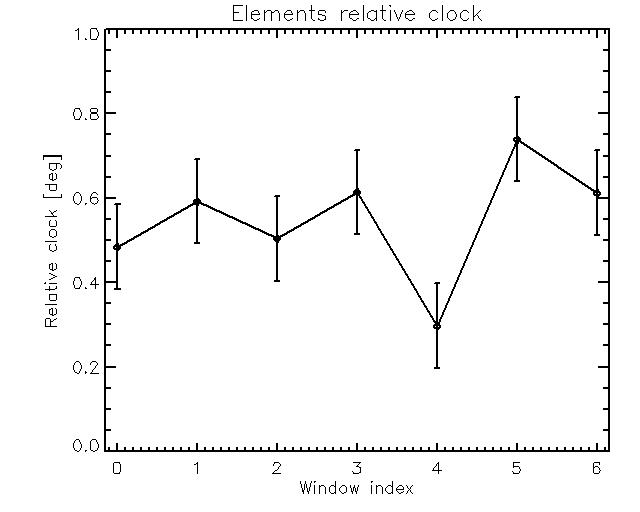}
  \vspace{-0.25cm}
  \caption{\footnotesize{Left: picture of an EW mounted in the barrel taken with the camera at an angle of $21^{\circ}$ with respect the bench. Right: plot of the measured clock error between the 2 elements of the windows.}}
  \label{fig:ew_clock}
\end{figure}

The production of 7 units of entrance window has been assigned to Custom Scientific, USA. The company took care both of manufacturing and coating the units. The most critical features of the entrance window that have been tested in laboratory are:
\begin{itemize}
  \item The diameter and position of the hole on the first element. These parameters constrain the dimension of the WFS FoV and the position on sky of the LGS. As shown in figure \ref{fig:ew_sketch} the hole diameter has to be $(3.0\pm0.1)mm$.
  \item The $(1.0\pm0.1)^{\circ}$ wedge between the two optical surfaces of the second element: this feature has been introduced to avoid the double reflection of the light coming from the calibration unit toward the WFS.
  \item The surface quality of the second element, that has to be better than $\lambda/10$.
  \item The clock error between the two elements once they are held in the metallic barrel. The barrel clock will be adjusted once mounted on the first WFS flange to ensure that the light from the internal source reaches the WFS, but a relative clock between the elements will make WFS FoV elliptical. Allowing maximum clock error of $1.0^{\circ}$ it is sufficient to ensure that the FoV diameter shrinks less than $1\mu m$.
\end{itemize}
The position and diameter of the hole has been measured using a microscope capable of a factor $4$ in magnification. Figure \ref{fig:ew_hole_pos_dim} resume the measurements. Considering the telescope plate scale of $0.66mm\;arcsec^{-1}$ at the Rayleigh focal plane, the field stop position is decentered on average by $(1.00\pm0.05)arcsec$ toward the center of the FoV. According to this value the entrance window holder has been designed to place the center of the hole in the nominal LGS position. The measured field stop dimension is $(4.6\pm0.1)arcsec$. From the ratio of the major and minor axes of the field stop we measured the inclination of the thru-hole that is $(20.5\pm0.5)^{\circ}$. These results agree with the values stated above.
\\The wedge angle between the two optical surfaces have been tested measuring the deflection of a $532nm$ laser beam with a CCD camera mounted on a rail. Recalling Snell's law:
\begin{equation}\label{eq:snell}
    n_{BK7}\; sin\theta_{i} = sin \theta_{o},
\end{equation}
$n_{BK7}=1.52$ is the refraction index of BK7 glass at $532nm$, $\theta_i$ is the incidence angle on the window and $\theta_{o}$ is the output direction of the beam. Since $\theta_o=\theta_{wedge}+\theta_{mes}$ where $\theta_{wedge}$ is the wedge angle and $\theta_{mes}$ is the the deflection angle measured moving the CCD along the rail. So we can evaluate the wedge angle as:
\begin{equation}\label{eq:wedge}
    \theta_{wedge} = \frac{\theta_{mes}}{n_{BK7}-1}.
\end{equation}
From this test we get a wedge angle of $(0.991\pm0.005)^{\circ}$ that is within specification.
\\The distortion of a plane wave transmitted through the second element of the window has been tested using a Wyko 4100 RTI interferometer. We used a standard setup with a reference flat and a return mirror. Because the window is wedged the position of the return flat must be adjusted when the window is inserted in the optical trail. In this case the cavity subtraction has been done using fiducial. Figure \ref{fig:ew_interf} shows the results of the analysis, values are referred to surface. After tilt and power subtraction the window surface rms is better than $\lambda/30$ that is compliant with specification stated above.
\\After we checked that the optical and geometrical properties of the windows are in specifications we aligned them inside a metallic barrel. The wedge direction on the second element of the windows has been marked by the producer with an arrow, pointing to the thinner edge of the window to allow its clock alignment in the barrel.  The residual clock error of the alignment procedure will reduce the FoV of the WFS and it cannot be compensated with any other degree of freedom in the system.
\\To measure the clock error we used a photographic camera inclined of $21^{\circ}$ with respect the optical bench. We aligned the barrel and the transmissive window (considering the mark provided by the producer) along the same direction and then we measured the difference of the 2 axis of the hole to retrieve the clock error. The measure results are summarized in the graph on figure \ref{fig:ew_clock} and they are within specifications.

\section{Patrol camera}
\label{sec:pc}
\begin{figure}
  \centering
  \includegraphics[width=6.5cm]{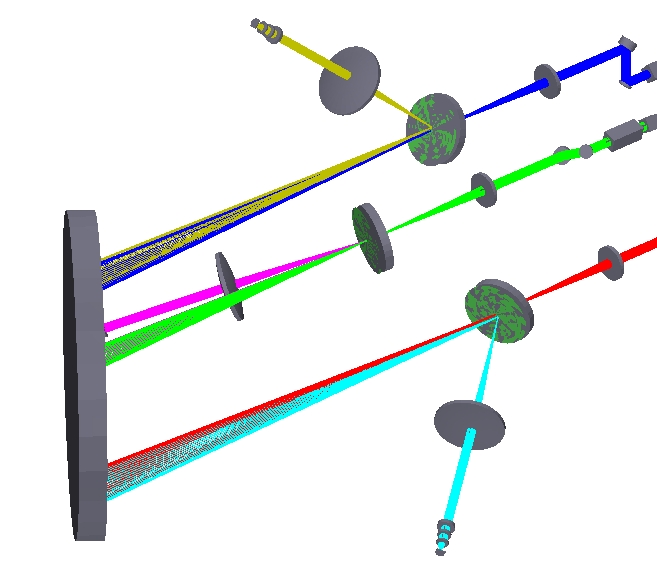}
  \includegraphics[width=6.5cm]{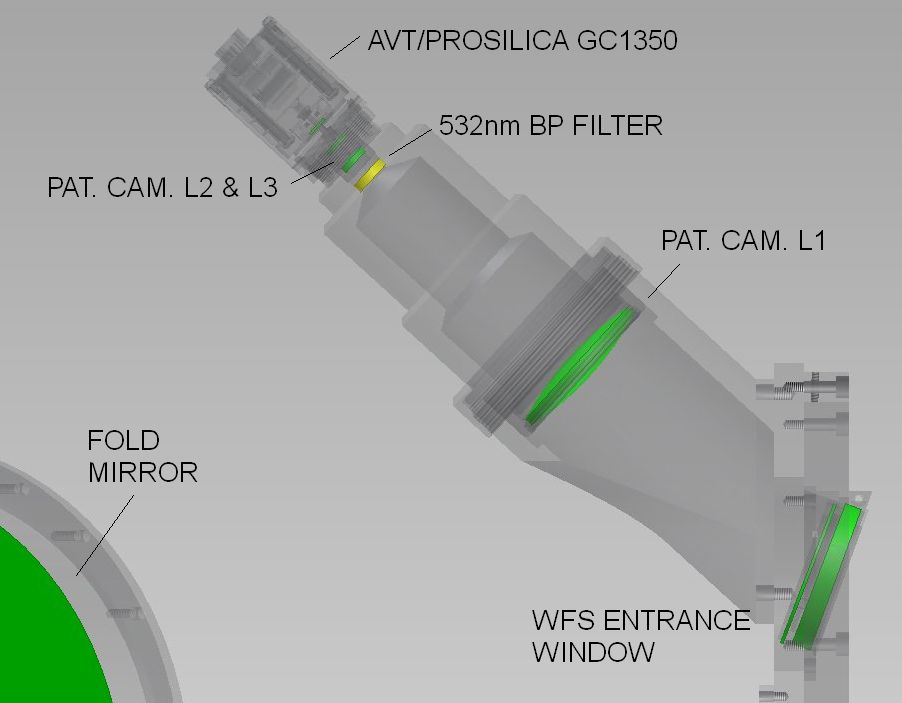}
  \vspace{-0.25cm}
  \caption{\footnotesize{Left: scheme of the arrangement of the 3 patrol cameras of the WFS. Backscattered laser light is arriving from the fold mirror on the left side of the image. The beams drawn here represent the 3 LGS in their nominal position (that enter the WFS) and at a $5arcsec$ distance from it. Right: see-through view of the patrol camera mechanics. Different optical elements that compose the system are marked.}}
  \label{fig:pc_optics}
\end{figure}

\begin{figure}
  \centering
  \includegraphics[width=7cm]{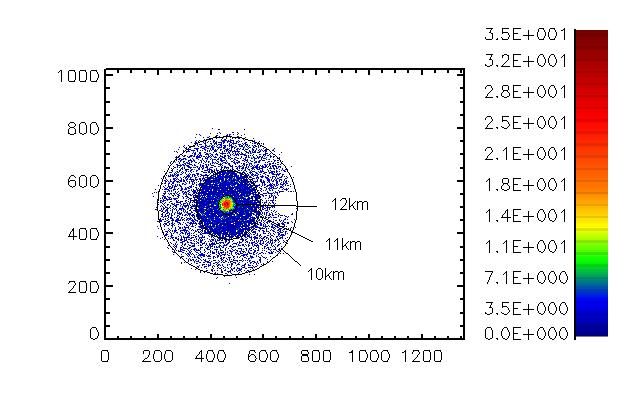}
  \includegraphics[width=6.5cm]{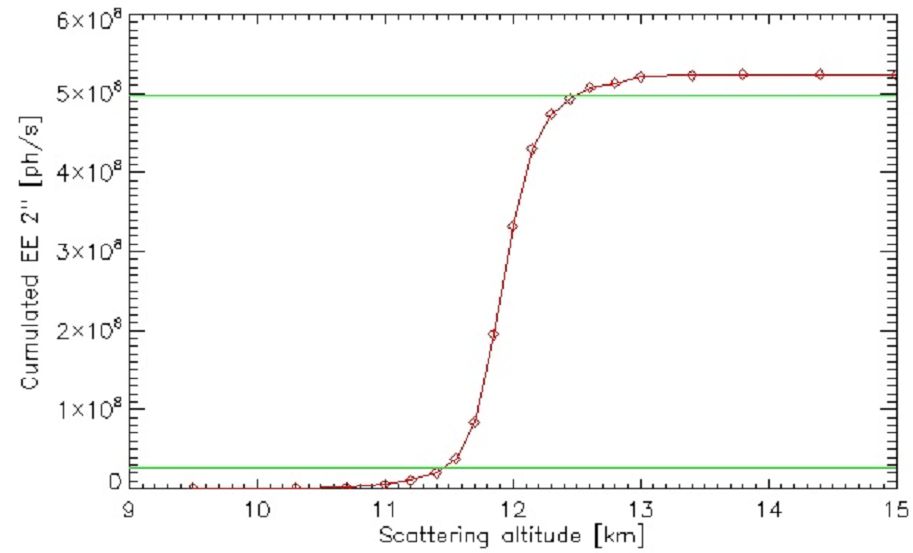}
  \vspace{-0.25cm}
  \caption{\footnotesize{Left: patrol camera images of light backscattered from different altitudes (10, 11 and $12km$ respectively). The laser beacon position is $20arcsec$ far from the nominal position. Intensities are not in scale. The white pattern on the $10km$ light area is due to the hole in the WFS entrance window. Right: flux in a $2\times2 arcsec$ square around the peak. Almost $90\%$ of the intensity of peak comes between 11.5 and $12.5 km$. Intensities are scaled to correspond to the nominal return flux expected for the WFS: $300Mph\;s^{-1}$ integrated from 11.85 to $12.15km$.}}
  \label{fig:pc_bksctr_ima}
\end{figure}

In case one of the LGS is not positioned within a radius of $2.35arcsec$ from the nominal position the backscattered light will fall on the reflective area of the entrance windows and it will be reflected toward the patrol cameras. These devices are used to measure the distance between the LGS position and the nominal one allowing to recenter the LGS acting on the optics inside the ARGOS laser system. Each LGS has a dedicated patrol camera that re-images the LGS focal plane on a CCD through the optical system showed in figure \ref{fig:pc_optics}.
\\The cameras are designed to have a circular FoV of $60arcsec$ and a magnification of 0.1286, that corresponds to have a plate scale of $0.08mm\;arcsec^{-1}$. The detector is AVT/Prosilica GC1350 that has a Sony ICX205 progressive CCD of $1360\times1024$ pixels with a cell size of $4.65\mu m\;px^{-1}$. The scale of the patrol cameras is $58mas\;px^{-1}$.
\\The patrol cameras are placed before the gating unit of the WFS, therefore they integrate the Rayleigh light backscattered from the entire atmosphere. Figure \ref{fig:pc_bksctr_ima} shows the spreading of light coming form different altitudes on the patrol camera image plane. Only light backscattered from a $1000m$ range around $12km$ height is properly focused and mostly contributes to the measure of the laser pointing position.
\\Considering a spot FWHM of $2arcsec$ almost $90\%$ of the LGS flux is within an area of $80\times80$ pixels. The integration time required to not saturate the camera can be estimated re-scaling the expected LGS return flux on the WFS. Considering the contribution of light in the gated range of $11850\div12150m$ the flux over the entire telescope pupil is $300Mph\;s^{-1}$. This yields to a low estimate for the average flux on the patrol cameras of $50kph\;px^{-1}\;s^{-1}$. An exposure time of $0.1s$ should provide a good signal ($\sim1500ADU\;px^{-1}$, considering the measured gain of $4ph\;ADU^{-1}$ of the GC1350 camera) without reaching the saturation of the sensor ($\sim4000ADU$) at full frame.

\subsection{Analysis of the patrol cameras optical quality}
\begin{figure}
  \centering
  \includegraphics[width=6.5cm]{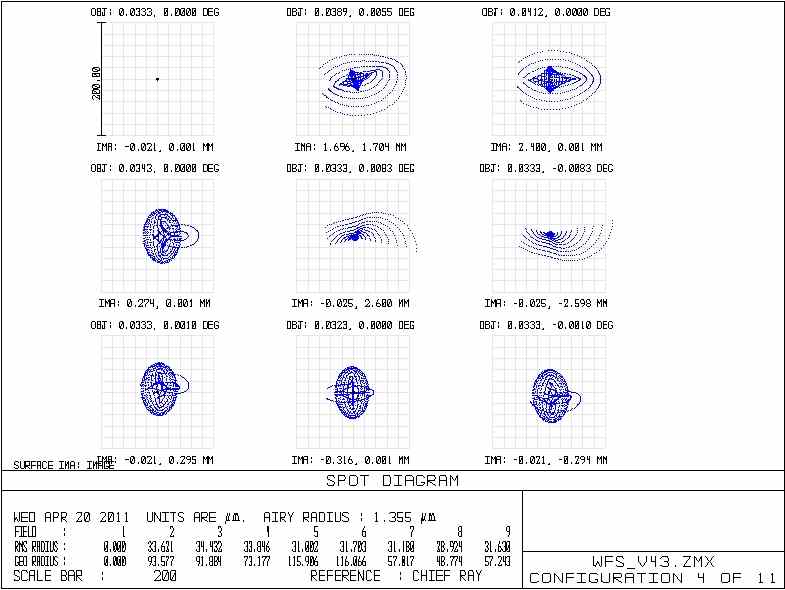}
  \includegraphics[width=6.5cm]{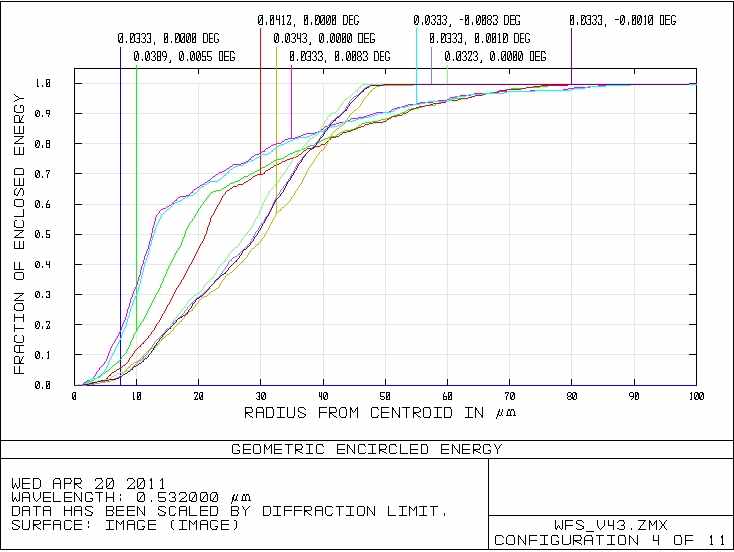}
  \vspace{-0.25cm}
  \caption{\footnotesize{Left: spot pattern on the patrol camera image plane when a point-like source is at different positions over the FoV. Right: encircled energy as a function of the same spot position.}}
  \label{fig:pc_spots}
\end{figure}

\begin{figure}
  \centering
  \includegraphics[width=10cm]{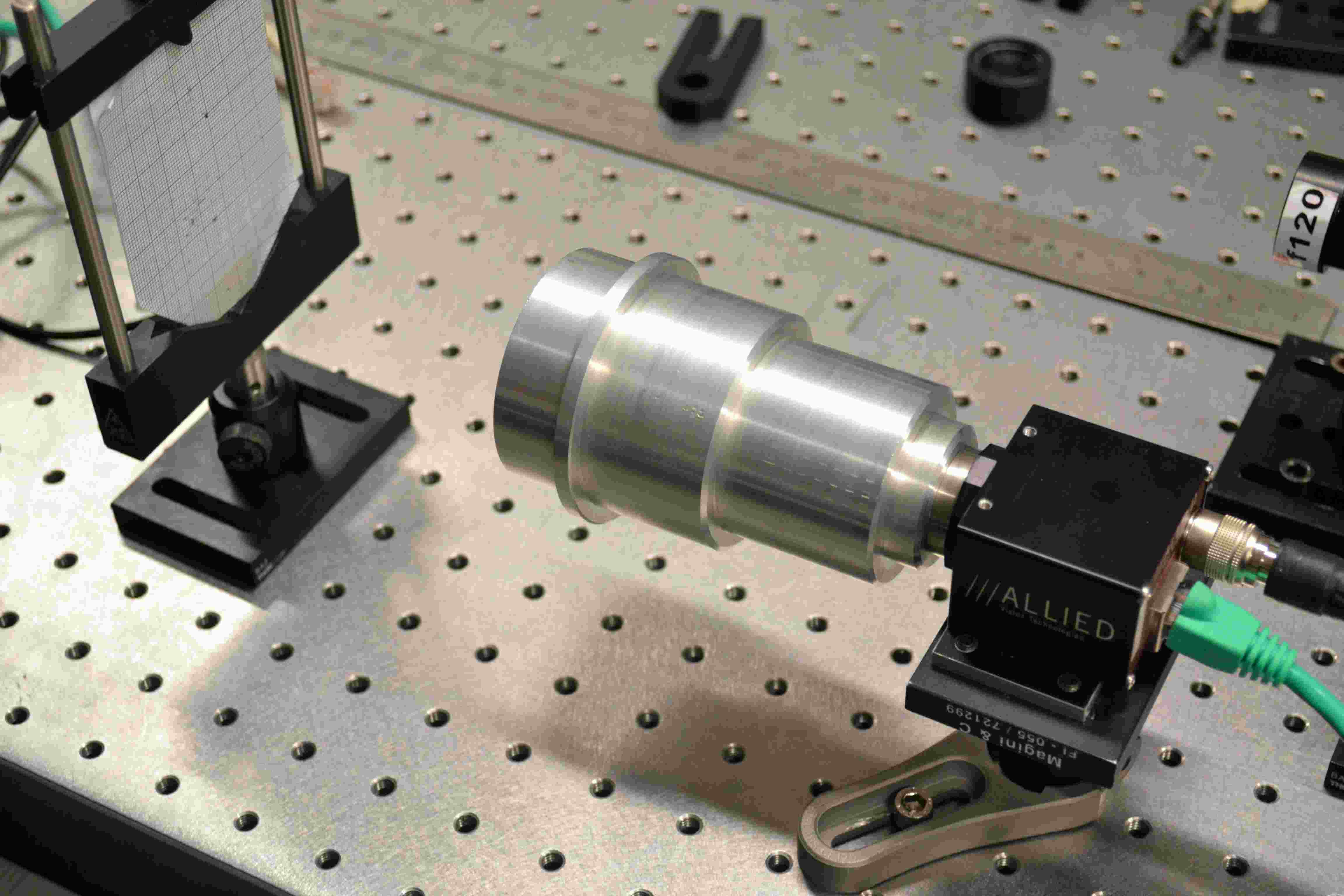}
  \vspace{-0.25cm}
  \caption{\footnotesize{Picture of the first unit of patrol camera assembled and tested in laboratory.}}
  \label{fig:pc_layout}
\end{figure}

\begin{figure}
  \centering
  \includegraphics[width=6.7cm]{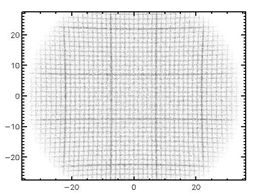}
  \includegraphics[width=6.5cm]{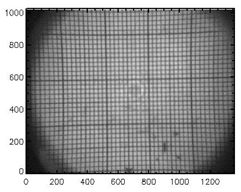}
  \vspace{-0.25cm}
  \caption{\footnotesize{Comparison between the Zemax simulated field distortion of the patrol camera and the one measured in laboratory with the test unit.}}
  \label{fig:pc_field_dist}
\end{figure}

Figure \ref{fig:pc_spots} shows the Zemax simulated spots and encircled energy on the patrol camera image plane for various positions on the FoV when a point-like source is positioned at $12km$ distance from the telescope. In worst seeing conditions (spot FWHM $\sim2arcsec$) the LGS will be imaged through the patrol camera optics over $0.16mm$ on the chip. This corresponds to 3 times the dimension of a point-like source, as shown on left in figure \ref{fig:pc_spots}. This means that the patrol camera detector can be binned up to a factor 3 without loosing optical quality.
\\Figure \ref{fig:pc_layout} shows the first unit of patrol camera assembled in laboratory for tests. All the camera optics are held in fixed position inside an aluminium barrel. The precision of machine tooling was enough to place them within tolerances. The only optical degree of freedom of the camera is the position of the CCD along the optical axis to proper re-image the EW surface on the CCD. This alignment has been done looking at the sharpness of a graph paper sheet on the camera live view software.
\\Figure \ref{fig:pc_field_dist} shows the comparison between the Zemax simulated field distortion of the patrol camera and the one measured in laboratory during the alignment procedure. The measured field of view of the camera is $48\times37mm$, equivalent to $72\times55arcsec$ on sky.

\section{Collimating lens}
\label{sec:coll_lens}
\begin{figure}
  \centering
  \includegraphics[width=6.5cm]{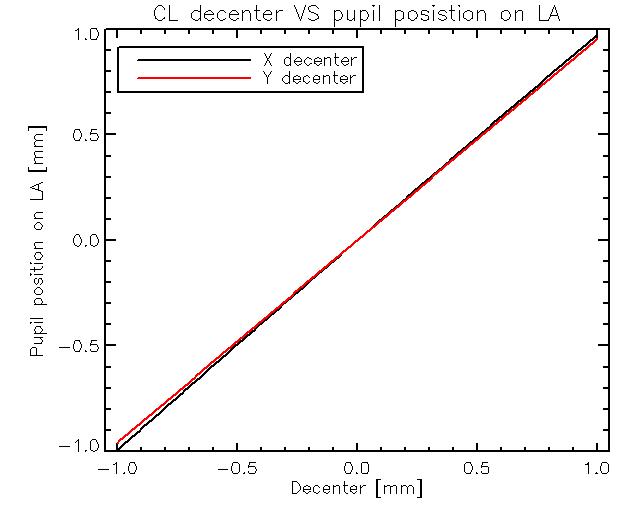}
  \vspace{-0.25cm}
  \caption{\footnotesize{Zemax simulation of collimating lens decenter and relative pupil displacement on the LA surface. The data plotted refers to the blue WFS arm.}}
  \label{fig:LCdecVSpupLA}
\end{figure}

\begin{figure}
  \centering
  \includegraphics[width=6.5cm]{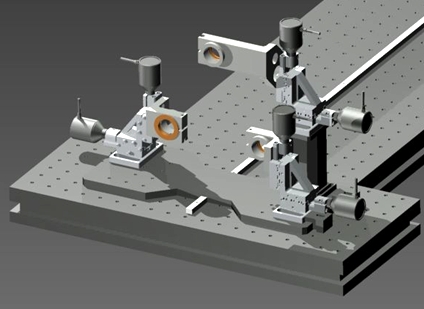}
  \includegraphics[width=6.5cm]{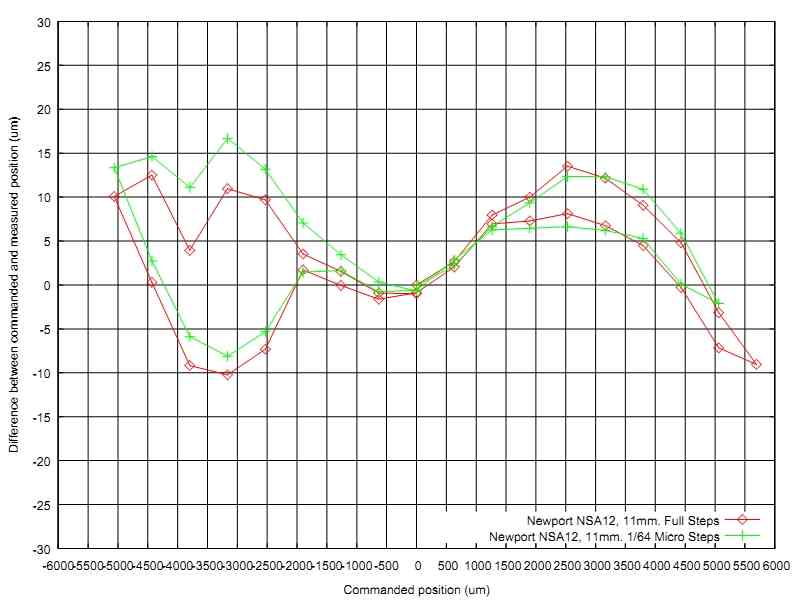}
  \vspace{-0.25cm}
  \caption{\footnotesize{Left: 3D model of the pupil recentering system of the 3 arms. Right: accuracy of the Newport NSA12 motor evaluated over the full $11mm$ stroke (courtesy of J. Ziegleder, MPE).}}
  \label{fig:coll_lens}
\end{figure}

The first WFS element with optical power is a commercial plano-convex lens of $100mm$ focal length. This element makes the $f_{16.6}$ laser light coming from the telescope a $6mm$ diameter collimated beam. The chief-ray tilt due to the off-axis position of the LGS is removed aligning the axis of the lens with the position of the spot on the focal plane. The collimating lens also re-images the telescope pupil, corresponding to the ASM placed at a $\sim15.1m$ distance from the lens, on a mirror mounted on a piezo driven tip-tilt stage.
\\The collimating lens is held on a motorized stage able to displace it by $\pm5.5mm$ in the XY plane with a $6\mu m$ resolution. Considering that the telescope plate scale on the $f_{16.6}$ plane is $\sim0.66mm\;arcsec^{-1}$ this means a resolution of $9mas$ on sky. This feature is mainly necessary to displace the pupil on the lenslet array matching the proper pattern of subapertures.
\\Figure \ref{fig:coll_lens} on the left shows the opto-mechanical assembly of the 3 beams collimating lenses. The XY stages are placed on a common board to ease the alignment of the lenses in the Z direction. The stage used to decenter the lens is a Newport M-461-XZ-M actuated by 2 Newport NSA12 stepper motors. The NSA12 motor has been tested at MPE showing that an accuracy of $\pm 15\mu m$ can be reached over the motor full range, see right of figure \ref{fig:coll_lens}. Figure \ref{fig:LCdecVSpupLA} shows the pupil displacement on the LA surface evaluated in Zemax when a decenter is applied to the collimating lens. Because the ratio is close to 1 the $15\mu m$ accuracy in the lens positioning is sufficient to ensure a resolution of $\pm1/20$ of subaperture in the pupil position.
\\The 6 NSA12 stepper motors inside a WFS unit are driven by a Motor Controller (MoCon) developed by MPIA and used also in LUCI and LINC/NIRVANA instruments. The MoCon design is modular, so it can host different amplifier boards capable to control different devices. The 6 NSA12 stepper motors for example are controlled by a single SMD8 V2 board able to provide up to eight $2.1A$ lines.

\section{Periscope and steering mirrors}
\label{sec:pi}
\begin{figure}
  \centering
  \includegraphics[width=6.5cm]{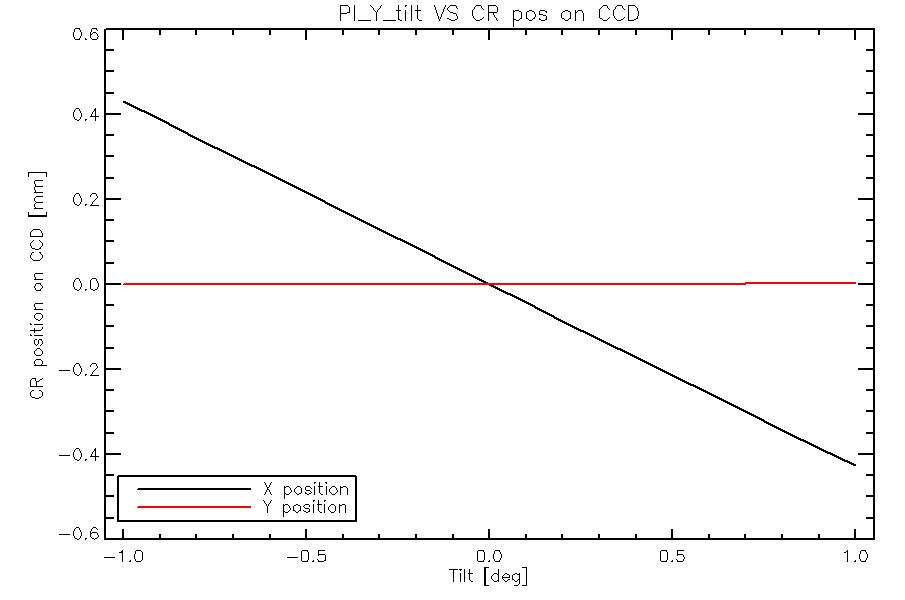}
  \includegraphics[width=6.5cm]{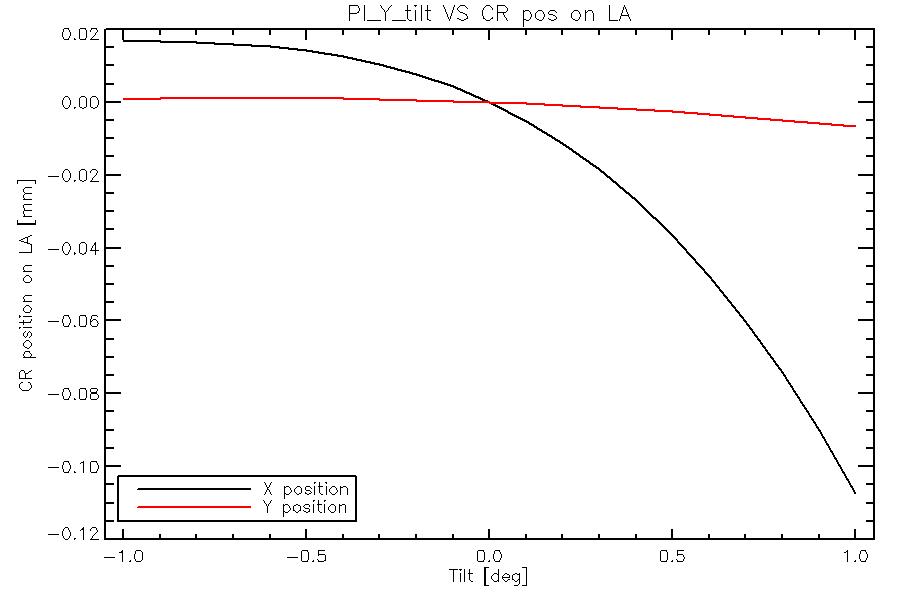}
  \vspace{-0.25cm}
  \caption{\footnotesize{Displacement of the spots on the CCD plane (left) and of the pupil on the LA (right) in function of the tilt applied to the steering mirror. These values have been measured in Zemax moving tilting the steering mirror in the Y direction.}}
  \label{fig:piVScrONccdLA}
\end{figure}

The periscope is composed by two flat mirrors set parallel to each other and inclined by $45^{\circ}$ with respect the light direction. This setup aims to reduce the distance between the 3 LGS beams keeping them parallel to the WFS board plane. The periscope length is $\sim30mm$ so the distance of the 3 beams from the center of the triangle, when the LGS are placed $120arcsec$ off-axis, is reduced from $78.3$ to $49.2mm$.
\\The first periscope mirror is a commercial $12.7mm$ laser-line flat mirror produced by Newport made by $3.1mm$ thick Pyrex substrate polished at $\lambda/10$ precision. The mirror reflectivity has been maximized with a specific $532nm$ dielectric coating that ensures $R>99\%$ at $45^{\circ}$ incidence.
\\The second periscope mirror is a $11mm$ diameter mirror of $2mm$ thickness. It has been produced and coated by Layertec. The surface quality of the mirror is better than $\lambda/10$ and its reflectivity $R>99.9\%$ over the full area of the mirror. This custom made mirror is placed on a plane conjugate to the telescope pupil by the collimating lens. The mirror is glued on a piezo driven tip-tilt platform (Physik Instrumente S334.1SL) capable of a $25mrad$ mechanical deflection (correspondent to $\pm1.4^{\circ}$ of optical deflection) at $1\mu rad$ resolution when it is controlled in closed loop with the Strain Gauges (SG).
\\This stage is needed to compensate for the laser jitter on sky that will introduce an independent tip-tilt error on the 3 different SH sensors. These errors are measured averaging the displacement of the spots on the $15\times15$ subapertures of each SH sensor and are corrected independently by the steering mirror of each WFS arm in open loop at $\sim1kHz$ frequency. Since the telescope pupil reimaged on the steering mirror is $6mm$ in diameter a compression of $7\times10^{-4}$ occurs between the angles on this mirror and on-sky, allowing to compensate for tip-tilt errors in the range of $\pm 3.8 arcsec$ PtV. \\The effects of applying a tilt on the steering mirror is illustrated by the plots in figure \ref{fig:piVScrONccdLA}, obtained in Zemax. The data represent the chief-ray positions measured on the CCD and LA planes in function of a variable tilt applied to the mirror. While the spots position on the CCD is varied tilting the steering mirror, the pupil position on the LA is kept fixed to less than $1/20$ of subaperture in one direction and $1/6$ in the other one (this is asymmetry is due to the fact that tilting the mirror in the positive Y direction more than $0.5^{\circ}$ the blue beam vignette in the Pockels cell). It is visible that $\pm 0.7^{\circ}$ of mechanical deflection of the mirror corresponds to a shift of the spots on the CCD plane of $\pm0.3mm=\pm3.7arcsec$, considering the $82\mu m \; arcsec^{-1}$ plate scale on the CCD.
\\The WFS will be also provided with a control loop able to off-load the tip-tilt correction integrated by the steering mirrors to the ARGOS laser system that will correct at lower frequencies the drift of the LGS position on-sky avoiding to saturate the mirror dynamic range.
\\The controller used to move the steering mirrors is an High Voltage Controller (HVC) produced by Microgate. This controller has been choose because it is integrable on the same bus of the WFS slope computer (Microgate BCU), allowing an higher control speed of the mirror via UDP/IP commands transmitted over the fiber Ethernet interfaces of the devices. Table \ref{tab:HVC} resumes the main HVC characteristics.
\begin{table}
  \centering
  \caption{\footnotesize{Summary of the Microgate HVC characteristics. This controller is directly connected to the ARGOS slope computer, allowing an higher speed of control of the steering mirror.}} \label{tab:HVC}
  \vspace{0.25cm}
\begin{tabular}{|p{4cm}|p{9cm}|}
  \hline
  \textbf{Parameter} & \textbf{Value}\\
  \hline
  \hline
  Number of channels            & 6 independent channels with strain gauge position feedback \\
  \hline
  Output voltage                & $0\div90V$ standard, $0\div125V$ maximum   \\
  \hline
  Output current                & $60mA$ continuous, $120mA$ peak   \\
  \hline
  Output power                  & $7.5W$ continuous, $15W$ peak, each channel   \\
  \hline
  Actuator capacitance          & up to $20\mu F$ \\
  \hline
  SG bandwidth                  & $26.5kHz$ \\
  \hline
  Interfaces                    & one input port summed to the regular actuator command and amplified $12\times$ \\
                                & one output signal replicating the actuator output voltage with gain $0.1V/V$ \\
                                & one low impedance strain gage diagnostic output with gain $400 V/V $ \\
  \hline
\end{tabular}
\end{table}

\section{Gating unit}
\label{sec:pockels}
To gate the required slice of backscattered light ARGOS uses Pockels cell. This choice has been motivated by the fact that having an optical light switch inside the WFS allows a free choice of detector, instead of limit the choice to gated CCDs. The use of Pockels cell to gate laser beacon light has already been demonstrated from LGS systems installed at SOAR telescope \cite{2003_Tokovinin_SOAR_LGS_AO}, Starfire Optical Range (SOR) \cite{1994_Fugate_SOR_LGS_AO} and William Hershel Telescope (WHT) \cite{2008_Martin_GLAS}.
\\The traditional design of Pockels cell consist of an electro-optic active crystal\footnote{F. Pockels in 1893 demonstrated that certain birefringent crystals, such as Lithium Niobate ($LiNb0_3$), can vary their refractive index when an electric field is applied. The so called Pockels effect is distinguished from the Kerr effect by the fact that the induced birefringence is linearly proportional to the electric field while in the Kerr effect it is quadratic. Usually the electric field is created placing the crystal in a parallel plate capacitor. In this way it is possible to modulate the polarization of the light passing through the crystal by varying the potential applied to the capacitor.} located between crossed polarizers. The basic property of electro-optical crystals is the quarter wave voltage: defined as the minimum voltage needed to introduce a phase shift of a quarter wavelength on a wave crossing the crystal.
\\To select the proper crystal it is desirable to have a low quarter wave voltage, a small crystal length and a sufficient large aperture. Other selection criteria are material transparency at the working wavelength and piezoelectric ringing.
\begin{figure}
  \centering
  \includegraphics[width=10cm]{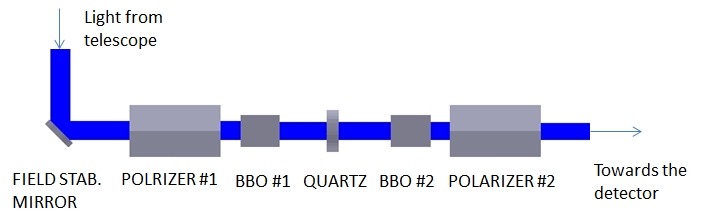}
  \vspace{-0.25cm}
  \caption{\footnotesize{Optical layout of the Pockels cell unit developed by MPE. This unique setup uses 2 crossed polarizers with 2 electro-optical crystal in between. A positive uniaxial crystal (Quartz) allows to compensate for phase delays and to obtain a more uniform suppression over a larger field.}}
  \label{fig:pockels_layout}
\end{figure}
\\Figure \ref{fig:pockels_layout} shows the optical layout of the ARGOS Pockels cell. This unique setup has been entirely developed at MPE to meet the ARGOS requirements in terms of suppression rate and transmitted field. The 2 polarizers are identical, they are designed as Glan-Thompson prisms: made of two right-angled Calcite prisms cemented together\footnote{The Glan-Thompson prism is commonly used as a polarizing beam splitter, it deflects the p-polarized ordinary ray and it transmits the s-polarized extraordinary ray. The Calcite crystals composing the two halves of the prims are aligned parallel to the cement plane and perpendicular to the plane of reflection. With respect to others polarizing beam splitters Glan-Thompson prim has a wider acceptance angle and it can achieve a higher extinction ratio of the ordinary component.}. The polarizers have a clear aperture of $11\times11mm^2$ and they are $27.5mm$ long.
\\The electro-optical crystal chosen is Beta Barium Borate ($\beta Ba B_2 O_4$, BBO). This material ensures both an high suppression rate and a ringing free transmission, that is requisite to avoid the introduction of focus errors on the WFS.
\\To compensate for the relatively small FoV ($<0.8^{\circ}$) over which the suppression rate is uniform the modulator has been splitted in 2 elements of $12mm$ length. Because BBO is a negative uniaxial crystal it has been inserted between them a positive uniaxial crystal to compensate for the phase delays. This compensator is a Quartz ($Si O_2$) crystal of $12mm$ diameter and $3.375mm$ length.
\\The suppression and transmission behaviors obtainable from the ARGOS Pockels cell are plotted in figure \ref{fig:pockels_transmission}. It is visible that a suppression rate of $10^{-3}$ is obtainable over a field of $4^{\circ}$ (third image on the right), reaching $5\times10^{-3}$ over the central $1.6^{\circ}$. At the same time the transmission is clearly ringing free and it reaches values $>97\%$ over the central $1.6^{\circ}$.
\begin{figure}
  \centering
  \includegraphics[width=10cm]{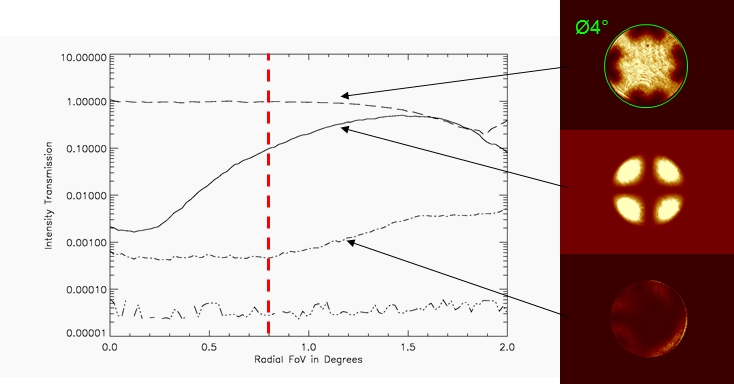}
  \vspace{-0.25cm}
  \caption{\footnotesize{Transmission curve in function of the angular distance from the optical axis of the Pockels cell. The dashed line shows the transmission of a convergent beam when a voltage is applied to the crystal, recorded on a CCD camera (image on the right). The solid line represents the transmission obtainable in the case of a classical Pockels cell, made only of 2 polarizers and a single $12mm$ diameter uncompensated BBO crystal. The dashed-dotted line shows the transmission in the closed state of the ARGOS Pockels cell (layout of figure \ref{fig:pockels_layout}). The lowest line shows the transmission of 2 crossed polarizers alone.}}
  \label{fig:pockels_transmission}
\end{figure}

\section{Refocussing and field lenses}
\label{sec:ref+field_lenses}
\begin{figure}
  \centering
  \includegraphics[width=6.5cm]{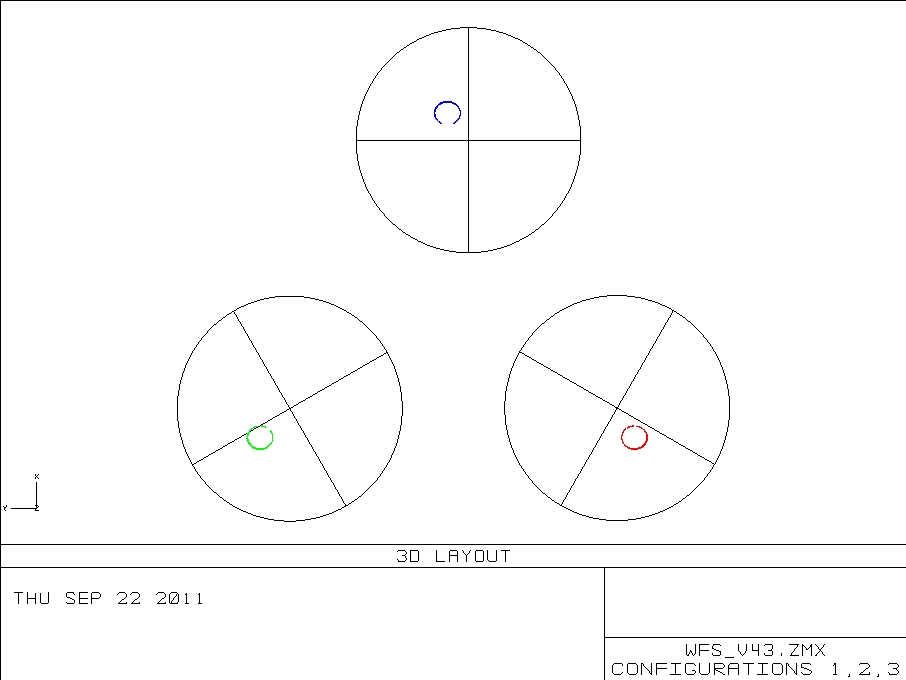}
  \includegraphics[width=6.5cm]{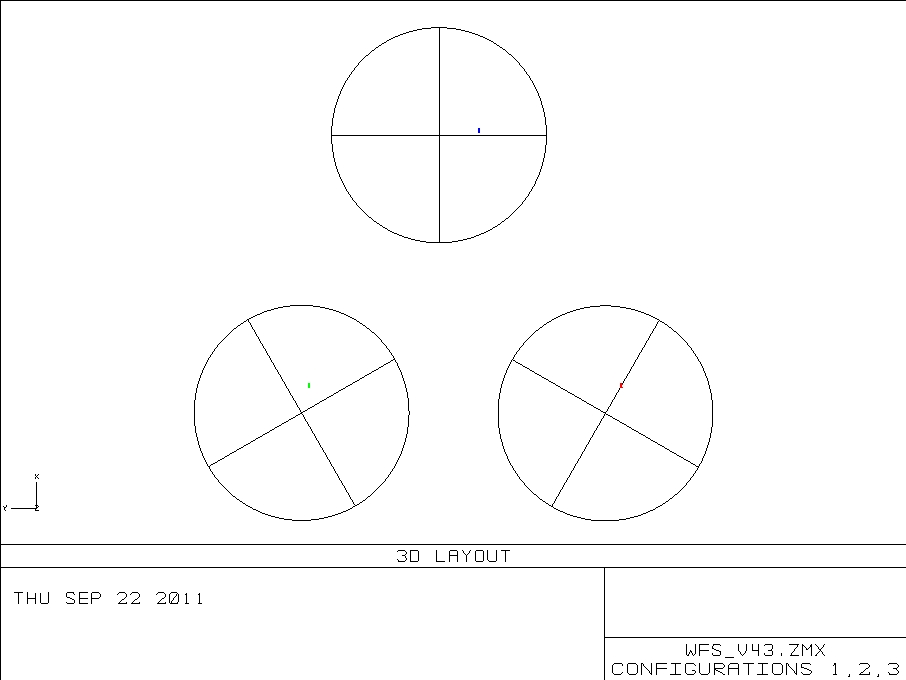}
  \vspace{-0.25cm}
  \caption{\footnotesize{Arrangement of the 3 LGS beams on the RL (left) and FL (right). The diagrams refer to the project of the left WFS. The WFS for the right eye of LBT has the green and red arms exchanged.}}
  \label{fig:RLandFL_layout}
\end{figure}

\begin{figure}
  \centering
  \includegraphics[width=6.5cm]{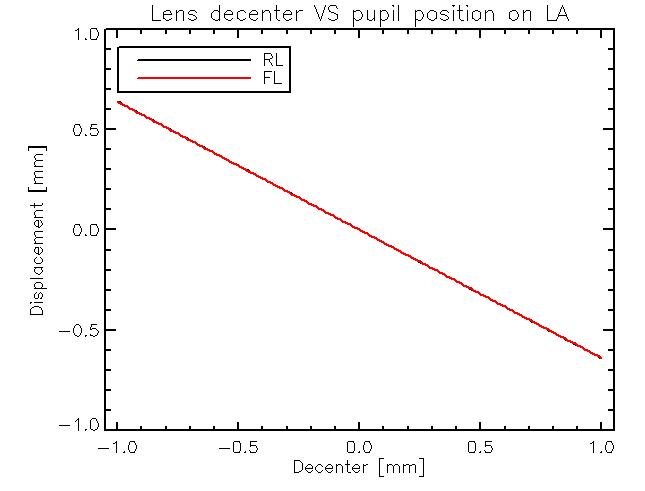}
  \includegraphics[width=6.5cm]{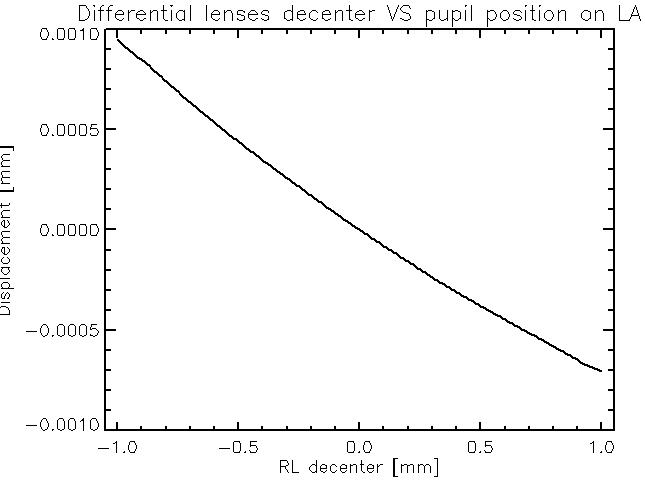}\\
  \includegraphics[width=6.5cm]{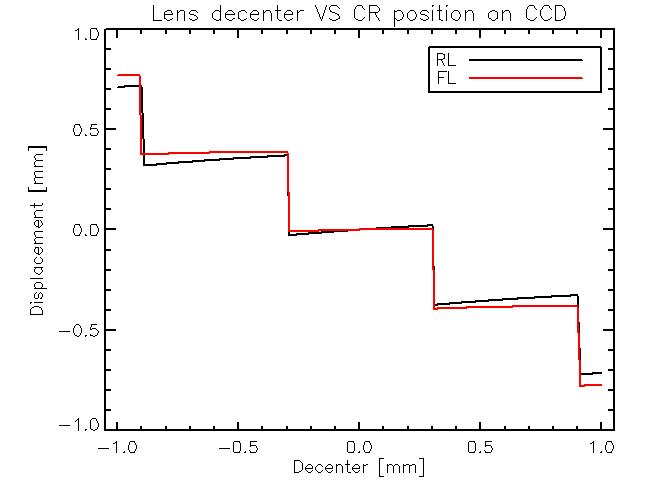}
  \includegraphics[width=6.5cm]{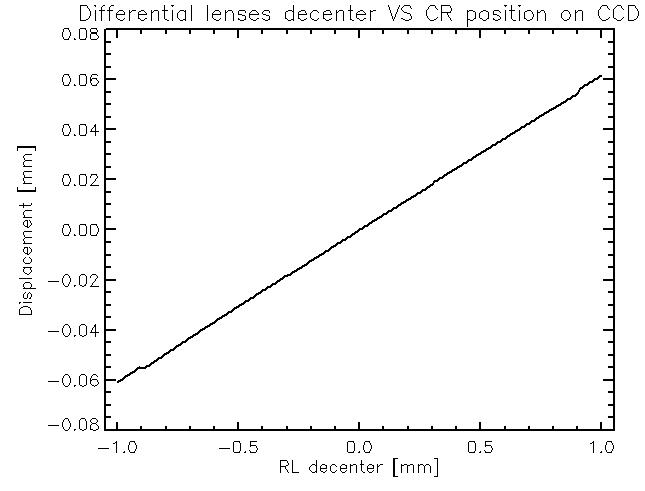}\\
  \vspace{-0.25cm}
  \caption{\footnotesize{Displacement of the pupil position on the LA plane (above) and the central SH spot on the detector (bottom) in function of the RL and FL decenter. The effect of moving independently one lens is shown on the left, while on the right it is shown the effect of decentering both lenses at the same time by the opposite amount.}}
  \label{fig:RLandFLdec}
\end{figure}

The 3 LGS beams are arranged on a triangular pattern inside the laser launch system. On-sky they have a $120arcsec$ distance from the telescope optical axis. This arrangement is maintained in the first part of the WFS, up to the gating units. However this symmetry have to be broken in order to match the 3 beams on the square geometry of the lenslet array and the detector pixel grid.
\\The task of rearranging the triangular symmetric beams on the a square grid is done by a group of 2 lenses working in an off-axis position. The first one is a plano-convex BK7 commercial lens with $200mm$ focal length (Newport KPX199). It re-focuses the collimated light in a $f_{33}$ beam and it makes a virtual image of the pupil at $\sim1m$ distance. The second lens is biconvex singlet with a focal length of $300mm$ (Newport KBX172), it is placed in correspondence of the beam focus acting as a field lens. Figure \ref{fig:RLandFL_layout} shows the LGS footprints on the refocussing lens (RL) and field lens (FL) respectively. On the RL the patches have an average diameter of $6mm$ while on the FL they are less than $1mm$ wide.
\\With this setup it is possible to disentangle the displacement of the pupil on the LA plane and the SH spots position on the focal plane. Figure \ref{fig:RLandFLdec} shows the effects of applying a decenter to the RL and the FL in the WFS Zemax project. In this simulation the lenses of the blue WFS arm have been decentered in the X direction by $\pm1mm$ starting from their nominal position. Considering the 2 plots in figure \ref{fig:RLandFLdec} it is visible that these lenses displaces the pupil by the same amount while only the RL displaces the SH spots on the CCD.

\section{The SH collimator}
\label{sec:SH_coll}
ARGOS WFS is designed to use a single CCD with a single lenslet array to provide 3 SH sensors at the same time. This setup in fact will ease the readout of the 3 sensors and it ensures that the signals of the 3 LGS are affected by the same electronic noise sources.
\begin{figure}
  \centering
  \includegraphics[width=8cm]{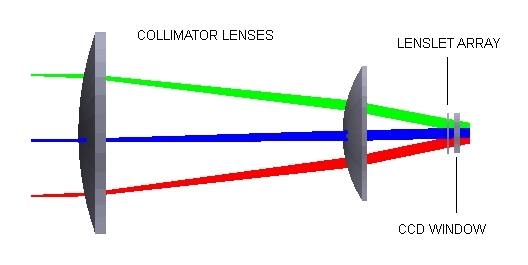}
  \includegraphics[width=5cm]{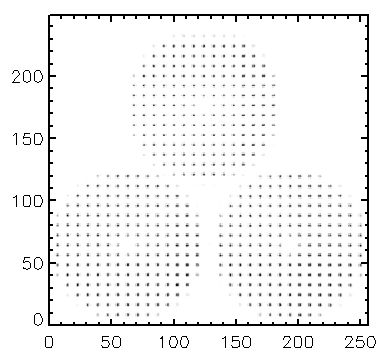}
  \vspace{-0.25cm}
  \caption{\footnotesize{Left: detail of the 3 beams propagation through the collimator up to the SH focal plane. Right: image of the beams arrangements on the detector obtained in Zemax. The image simulates a source with a gaussian profile and $1arcsec$ FWHM placed at a $12km$ distance form the telescope. The central 5 subapertures are vignetted by the pupil central obscuration while the top subaperture row of the blue arm is vignetted by the tertiary mirror.}}
  \label{fig:wfs_final_layout}
\end{figure}
\\Meniscus lenses are widely used for field flattening applications \cite{2006_Handbook_OptSys_Gross}. In the final part of the WFS instead we need to increase the field curvature to fit the 3 beams on a single optical surface of $21mm$ diameter. So we used meniscus lenses in a reversed setup both to collimate and to steer the 3 LGS at the same time. The collimator layout has been optimized in Zemax, substituting the LA with a paraxial lens of equivalent focal length, and minimizing the rms of the 3 LGS wavefronts on the focal plane.
\\From the optimization process we get focal lengths for the 2 lenses of $415$ and $226mm$ respectively, and a spacing of $153mm$ between them, as shown in figure \ref{fig:wfs_final_layout}. Together these lenses acts as a $60mm$ equivalent focal length optic that collimates the 3 beams to a $5.76mm$ diameter. In addition the emerging beams are tilted by $8.6^{\circ}$ so they fit inside a single $21mm$ diameter lenslet array. On the detector each of the $15\times15$ subapertures is imaged on a $8\times8$ pixels area. Figure \ref{fig:wfs_final_layout}, on right, shows a Zemax simulated image of the 3 SH sensors on the detector plane, the arrangement of the pupils and spots pattern is evident.

\subsection{SH collimator assembling}
\begin{figure}
  \centering
  \includegraphics[width=8cm]{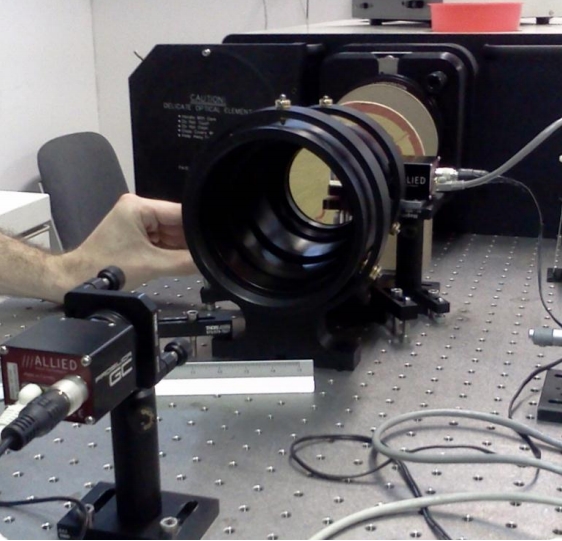}
  \vspace{-0.25cm}
  \caption{\footnotesize{Picture of the setup used to measure the position errors of the 2 SH collimator lenses inside the barrel. The light source is an interferometer at $632nm$ wavelength. Two cameras look at the beams transmitted and reflected by the 2 lenses held in the barrel.}}
  \label{fig:SH_coll_assembly}
\end{figure}

Because of the large dimensions of the 2 collimator lenses and their particular curvature radii they have been custom produced. The selected manufacturer is Optimax Systems that has been commissioned of the production of 3 units of each lens. Table \ref{tab:coll_specs} resumes the specifications for the optical manufacturing and coating of the lenses. All the curvature radii of the surfaces have been optimized in Zemax using Optimax test plates to reduce the production costs.
\begin{table}
  \centering
  \caption{\footnotesize{Manufacturing and optical specifications of the two lenses of the SH collimator.}} \label{tab:coll_specs}
\begin{tabular}{|l|c|c|}
  \hline
  \textbf{Specification} & \textbf{Lens $\mathbf{\sharp1}$} & \textbf{Lens $\mathbf{\sharp2}$}\\
  \hline
  \hline
  Material           &  BK7           & BK7 \\
  Physical dimension &  $(120\pm1)mm$ & $(80\pm1)mm$ \\
  Clear aperture     &  $(108.0\pm0.2)mm$ & $(72.0\pm0.2)mm$ \\
  Thickness          & $(15.0\pm0.2)mm$ & $(10.0\pm0.2)mm$ \\
  Transmitted WFE    & $\lambda/10$ &   $\lambda/10$ \\
  Scratches and digs & 40/20    & 40/20 \\
  Centering tolerance& $5arcmin$ & $5arcmin$ \\
  Coating transmissivity at $532nm$ & $>99\%$ & $>99\%$ \\
  \hline
\end{tabular}
\end{table}
\\These two lenses are held inside a common barrel. We measured the relative decenter and tilt of the lenses inside the barrel using a collimated beam generated by an interferometer and measuring its deflection with 2 cameras, as shown in figure \ref{fig:SH_coll_assembly}. The first camera is placed along on the beam axis after the barrel and it is sensible to the decenter of the lenses. The second one is placed after a beam splitter in correspondence of the lens concave surface radius of curvature. So this second camera looks at the beam reflected by the lens concave surface and it is sensible both to tilts and decenters of the lens. Subtracting the contribution of the decenter measured by the first camera on the second one we measured the lenses tilt. Rotating the barrel around its axis and repeating the measurements we evaluated the maximum decenter and tilt between the 2 lenses and the barrel axis.
\\Table \ref{tab:SH_coll_errors} resumes the position errors of the lenses measured on the 2 collimator units. These errors have a negligible effect on the WFS optical quality and beams geometry.
\begin{table}
  \centering
  \caption{\footnotesize{Decenter and tilt between the collimator lenses and the barrel axis measured on the left and right WFS units.}} \label{tab:SH_coll_errors}
\begin{tabular}{|l|c|c|}
  \hline
  \textbf{Element} & \textbf{Decenter $[\mu m]$} & \textbf{Tilt $[arcmin]$} \\
  \hline
  \hline
  Lens $\sharp1$ SX Collimator & $24.8\pm0.2$ & $4.38\pm0.02$ \\
  Lens $\sharp2$ SX Collimator & $51.1\pm0.2$ & $1.14\pm0.02$ \\
  Lens $\sharp1$ DX Collimator & $33.2\pm0.2$ & $4.20\pm0.02$ \\
  Lens $\sharp2$ DX Collimator & $96.8\pm0.2$ & $2.88\pm0.02$ \\
  \hline
\end{tabular}
\end{table}

\section{Lenslet array}
\label{sec:lenslet}
\begin{figure}
  \centering
  \includegraphics[width=6.5cm]{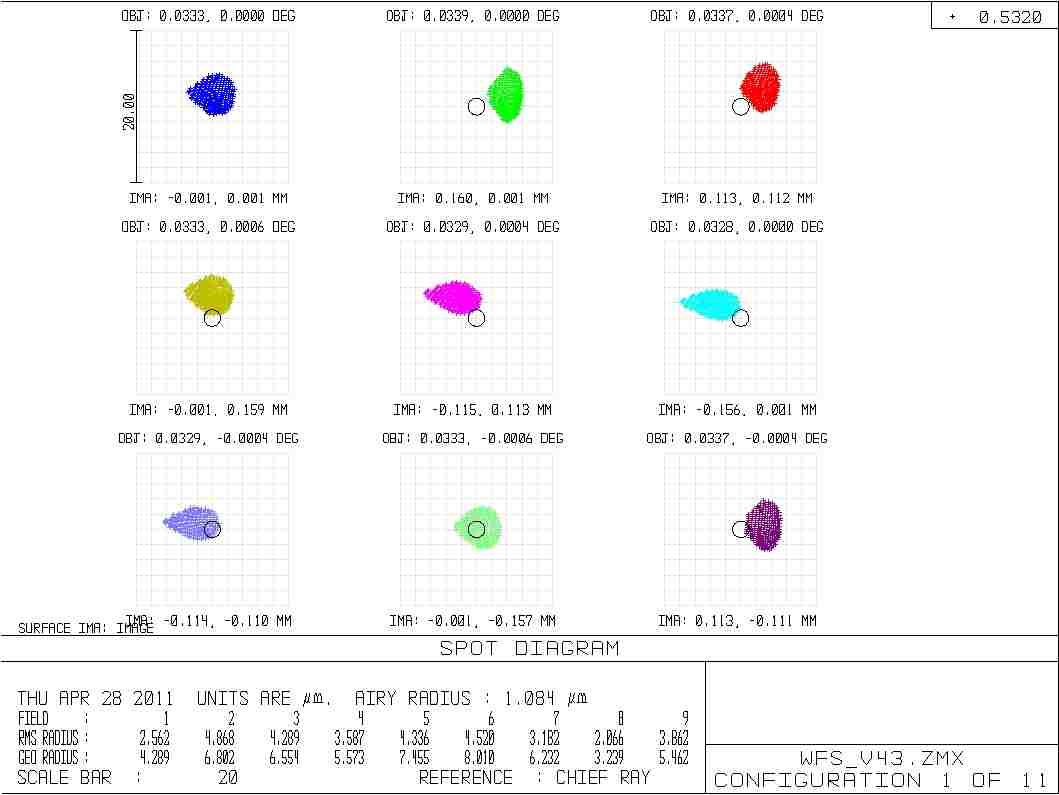}
  \includegraphics[width=6.5cm]{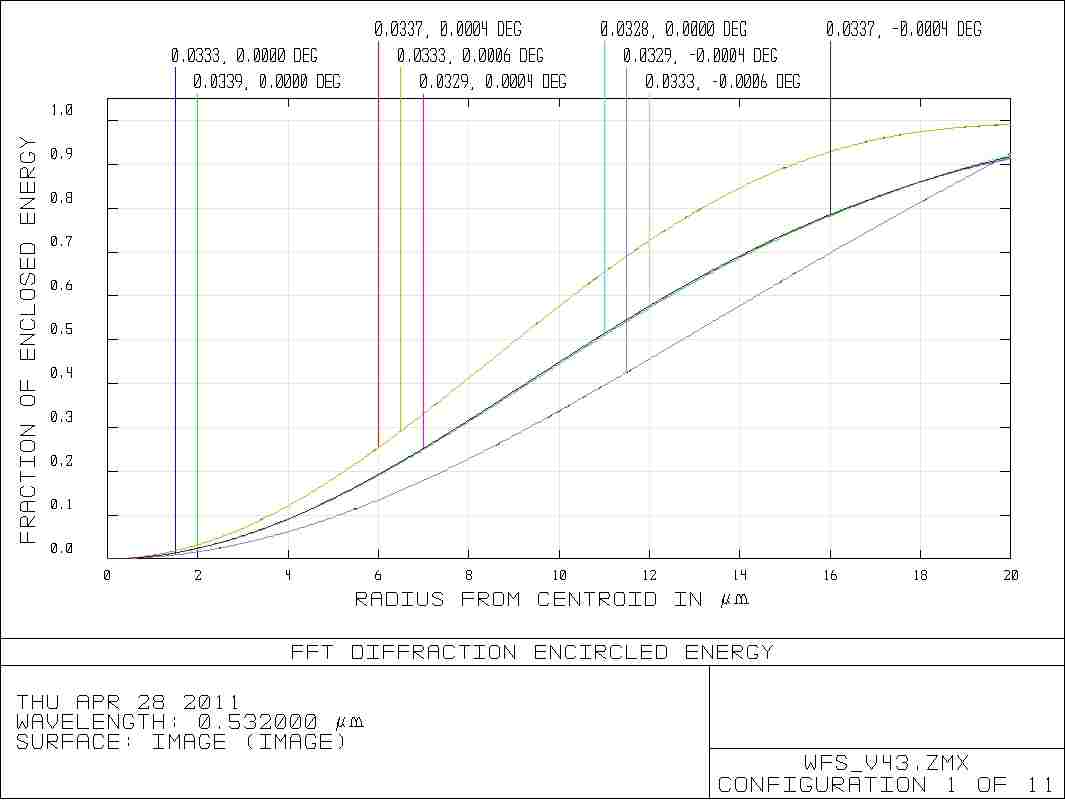}
  \vspace{-0.25cm}
  \caption{\footnotesize{Left: Spot quality on the SH focal plane. The source is point-like and it is positioned in the center of the WFS FoV and on a circle of $2arcsec$ radius. Right: encircled energy as a function of spot radius for the same spot positions.}}
  \label{fig:wfs_spot_quality}
\end{figure}

The WFS lenslet array has plano-convex circular lenses arranged on a square grid and $21mm$ external diameter. The lenslets have an f-number of 32. Figure \ref{fig:wfs_spot_quality} shows the spot quality on the SH focal plane produced by a single lenslet placed at the center of the telescope pupil (just out of the ASM central obscuration). The 9 spots are arranged on a circle of $2arcsec$ radius centered on the LGS nominal position. In this conditions, considering that the plate scale of the telescope is $0.08mm\;arcsec^{-1}$ on the focal plane, a diffraction limited spot results to have a FWHM of $17 \mu m$ at $532nm$. Hence, plots in figure \ref{fig:wfs_spot_quality} shows that by design the WFS yields a spot FWHM of $\sim24\mu m$, equivalent to $0.3arcsec$ on sky.

\subsection{Laboratory test of the SH lenslet array}
\begin{figure}
  \centering
  \includegraphics[width=6.5cm]{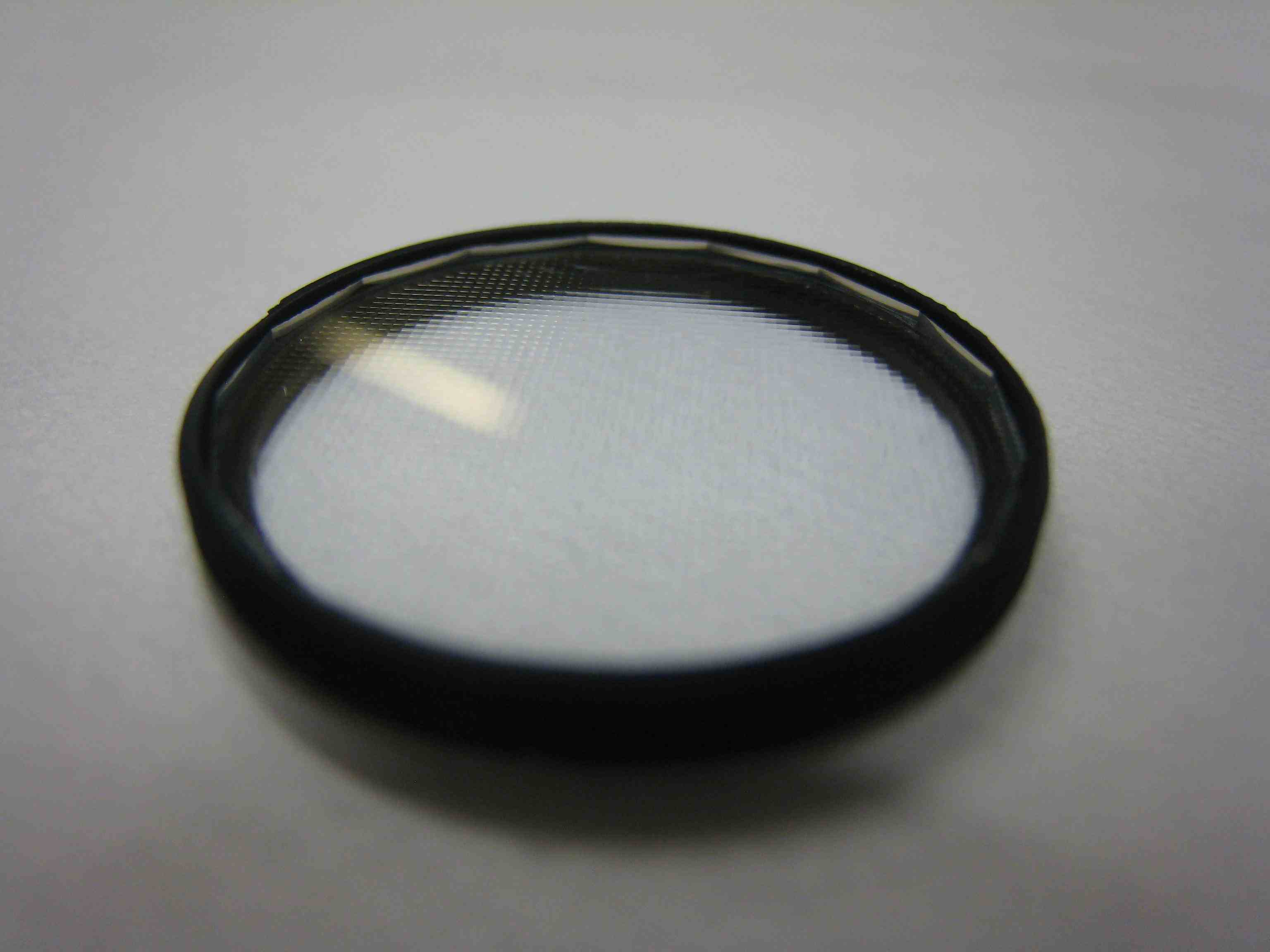}
  \includegraphics[width=6.5cm]{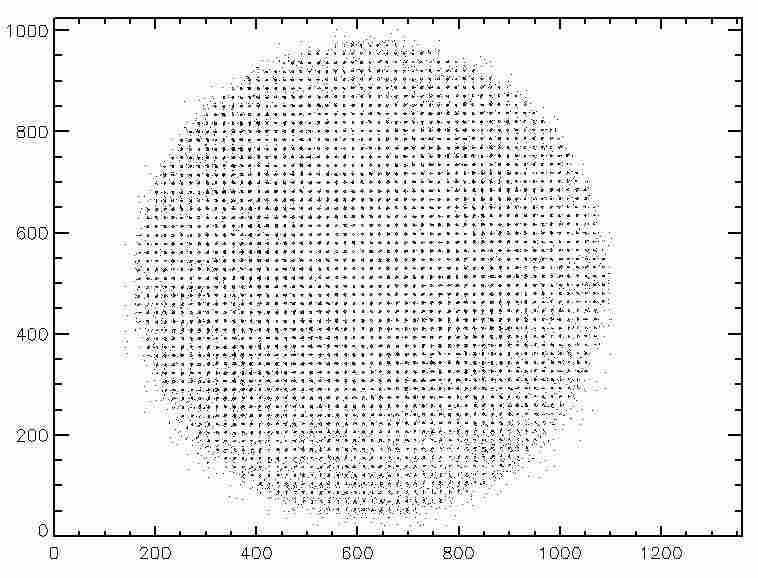}
  \vspace{-0.25cm}
  \caption{\footnotesize{Left: picture of one unit of the 4 lenslet arrays produced by SUSS MicroOptics. Right: image of the SH pattern produced by the full clear aperture of the lenslet array. 55 spots are visible across the pupil diameter.}}
  \label{fig:wfs_la_picture}
\end{figure}

Table \ref{tab:la_data} resumes the specification we produced for the manufacturing of the WFS lenslet array. The production has been assigned to Suss MicroOptics that manufactured and coated 4 units, one of them is shown in figure \ref{fig:wfs_la_picture}. We have tested the main characteristics of the lenslet arrays in laboratory to verify they were compliant to values specified in table \ref{tab:la_data}.
\begin{table}
  \centering
  \caption{\footnotesize{Manufacturing and optical specifications of the lenslet array.}} \label{tab:la_data}
\begin{tabular}{|l|c|}
  \hline
  \textbf{Specification} & \textbf{Value} \\
  \hline
  \hline
  Material          & Fused silica \\
  Lens pitch        & $(384.0\pm0.2)\mu m$ \\
  Lens width        & $(376.0\pm0.2)\mu m$ \\
  Lens RoC          & $6.15mm\pm10\%$ \\
  Number of lenses  & $62\times62$ \\
  Array thickness   & $(1.2\pm0.1)mm$ \\
  Array diameter    & $(24.00\pm0.05)mm$ \\
  Array CA          & $(21.5\pm0.1)mm$ \\
  Surface quality   & $ 5/5 \times 0.04$ \\
  Coating           & AR, both sides, $T>99.8\%$ at $532nm$ \\
  \hline
\end{tabular}
\end{table}
\\The lenslet pitch has been measured using the beam generated by an interferometer (Wyko 4100 RTI) and a CCD camera (AVT/Prosilica GC1350). Assuming that the interferometer beam is perfectly collimated the measured lenslet pitch is $(384.0\pm0.1)\mu m$ in compliance with specification.
\\The number of lenslets within the array clear aperture has been measured re-imaging the SH focal plane through an optical relay having a $0.2$ magnification. The camera frame is shown on right of \ref{fig:wfs_la_picture}, 55 spots are visible across the diameter of the pupil. Considering the measured lenslet pitch the effective clear aperture results $21.12mm$ that is sufficient for the WFS purposes.
\\Using a similar setup we measured the lenslets focal length. Introducing a flat mirror in the optical path it is possible to tilt the collimated beam that is then focussed by the lenslet array on the camera. The tilt applied to the flat mirror ($\theta$) is related to the focal distance $f=d/2\theta$, where $d$ is the differential spot position on the camera before and after tilting the mirror. We measured the focal length of the lenslets to be $(12.0\pm0.3)mm$ that matches the specification of $6.15mm\pm10\%$ on the radius of curvature.

\section{WFS detector}
\label{sec:ccd}
\begin{figure}
  \centering
  \includegraphics[height=5cm]{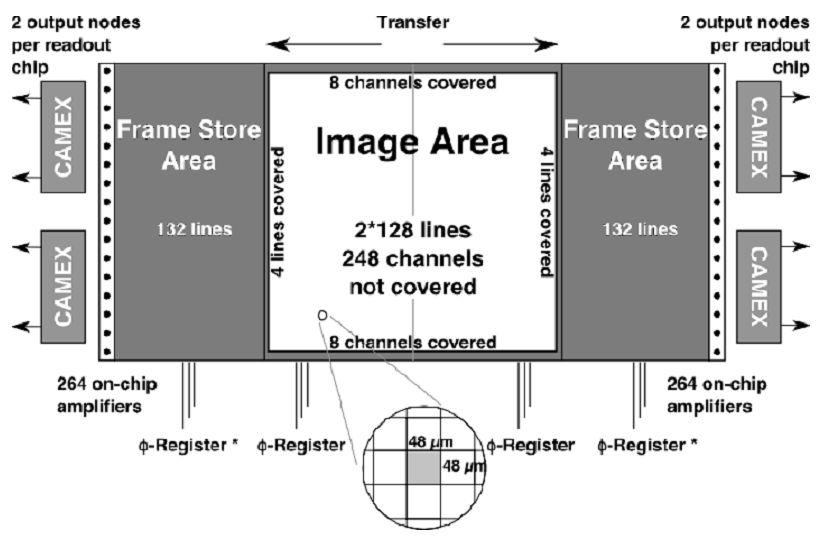}
  \includegraphics[height=5cm]{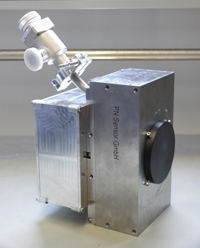}
  \vspace{-0.25cm}
  \caption{\footnotesize{Left: schematic layout of the split frame transfer pnCCD for high speed optical applications. Outside the imaging area there are eight reference columns and four reference lines included on the left and right respectively upper and lower side. For readout of the $2\times264$ CCD channels, two multi channel readout ASICs (CAMEX), each comprising 132 channels, are placed adjacently on each readout side of the detector. Right: picture of the first PnCCD assembled in laboratory. Image courtesy of G. Orban de Xivry, MPE.}}
  \label{fig:ccd_readout}
\end{figure}

The WFS detector is a pnCCD developed by PNSensor in collaboration with MPE as a derivation of the X-ray detector for the XMM/Newton satellite mission of the European Space Agency (ESA) \cite{1996_Meidinger_pnCCD_SPIE}. The design of the pnCCD for high speed optical applications has been adapted to the requirements of a wavefront sensor in a SH system, to allow high speed operations (up to $1kHz$) while maintaining the 2 dimensional imaging capabilities, the pnCCD detector has been designed to operate in a split frame transfer mode. The image area has a size of $11.9\times12.2mm^2$, comprising $248\times256$ sensitive pixel with a size of $48\mu m$. Each half image is transferred to its storage region on opposite sides of the detector within $30\mu s$. Signal readout is accomplished by four readout ASICs having 132 channels each. A schematic of the detector readout is shown in figure \ref{fig:ccd_readout}.
\\While in use the detector chip is cooled at $-35^{\circ}C$ and it is kept under vacuum condition. The main detector characteristics measured in the test campaign performed on the engineering unit are \cite{2010_Xivry_pnCCD_SPIE}:
\begin{itemize}
  \item The average gain factor for the 4 CAMEX is $2.5ADU/e^{-}$ .
  \item Read-out-noise (RON) is $3.6e^{-}$.
  \item The CCD quantum efficiency (QE) in the $500\div800nm$ range is $>90\%$, reaching $QE>98\%$ at $532nm$.
  \item Dark current value is $0.196e^{-}\;px^{-1}\;s^{-1}$ when the CCD is read at $100fps$.
\end{itemize}
Figure \ref{fig:ccd_readout} on right shows a picture of the first PnCCD fully assembled in laboratory.

\section{Internal calibration unit}
\label{sec:cal_unit}
\begin{figure}
  \centering
  \includegraphics[width=4.7cm]{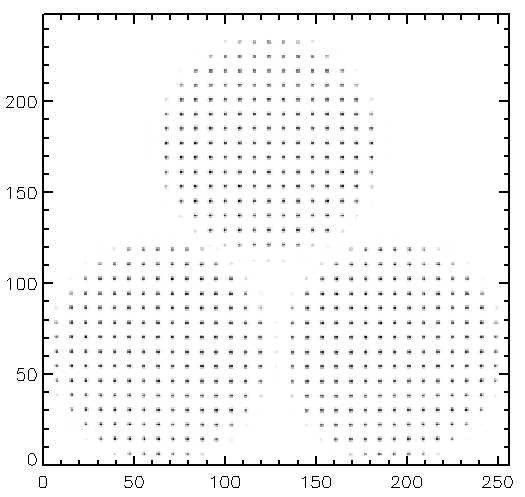}
  \includegraphics[width=7.7cm]{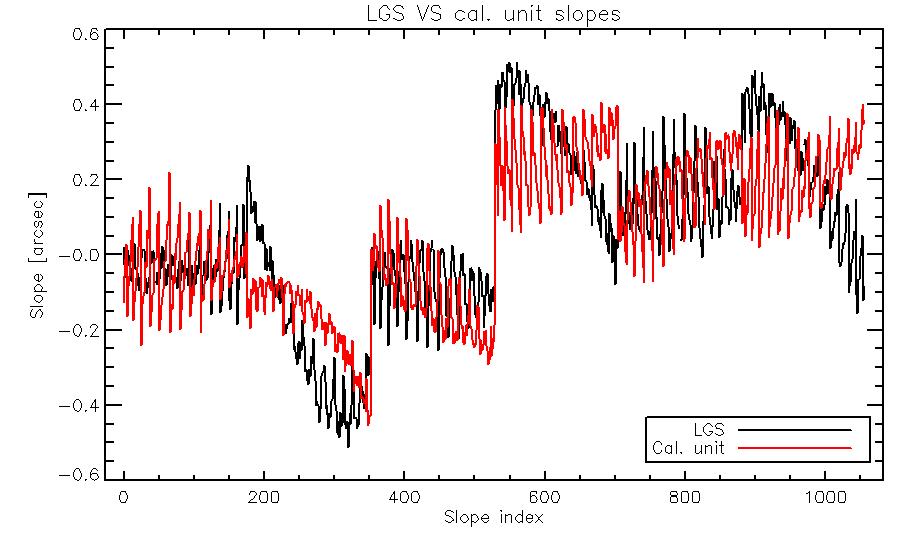}
  \vspace{-0.25cm}
  \caption{\footnotesize{Left: frame simulated with Zemax using the calibration unit as light source and $0.6mm$ FWHM sources with Gaussian profile. Right: comparison of the slopes evaluated on the frame generated with the calibration unit source and the LGS, shown in figure \ref{fig:wfs_final_layout}.}}
  \label{fig:cal_unit_frame}
\end{figure}

The WFS is provided with an internal light source that generates 3 $f_{16.6}$ beams using green LED sources placed inside the WFS electronic rack and remotely controllable. In the rack LED modules the light is first collimated and then coupled into an exchangeable optical fiber. A narrow bandpass filter inserted between the collimator and the fiber coupler allow the LED sources to produce light at $532nm$ with a bandwidth of $3nm$. Three $6m$ long fibers bring light from the rack to the WFS module.
\begin{figure}
  \centering
  \includegraphics[height=5cm]{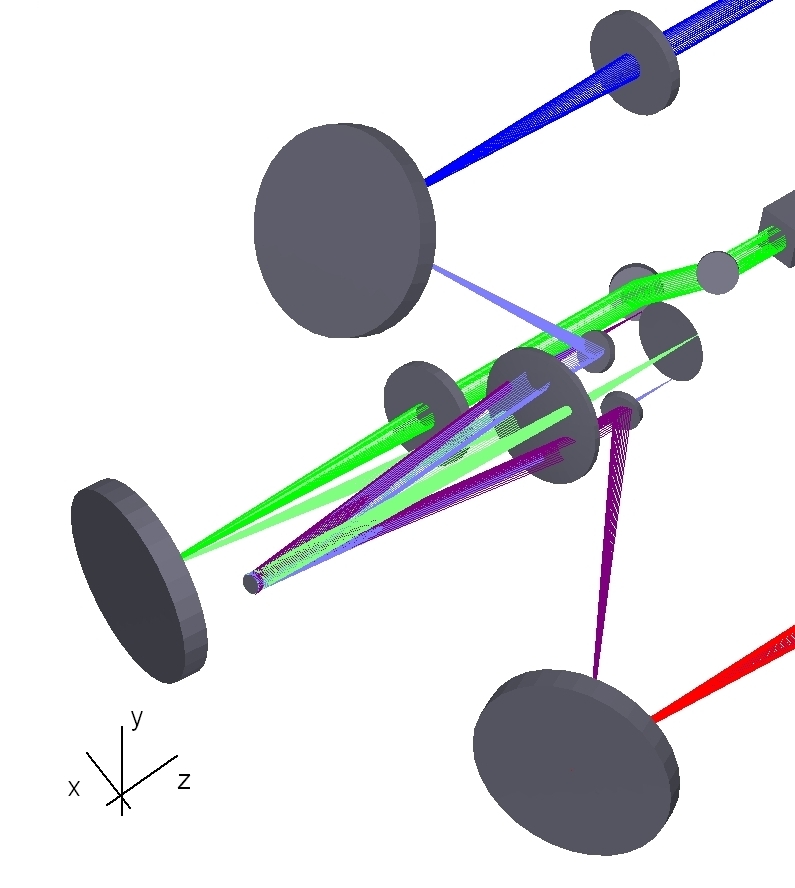}
  \includegraphics[height=5cm]{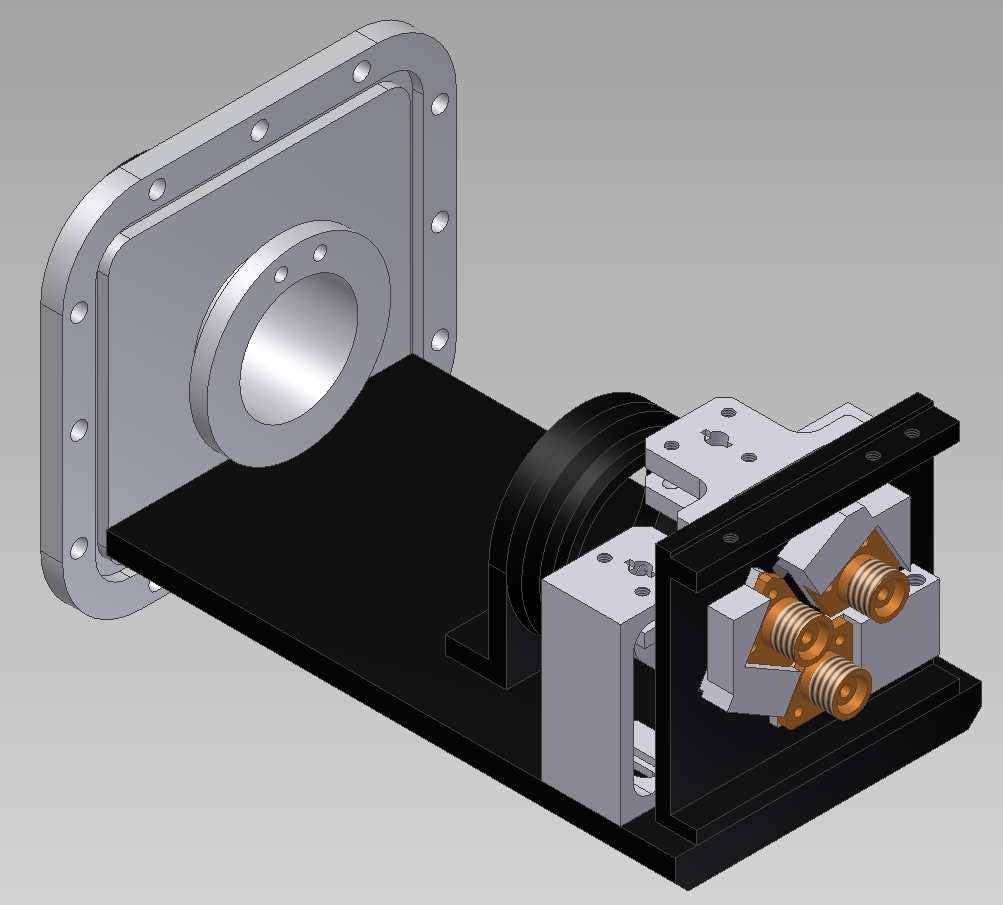}
  \vspace{-0.25cm}
  \caption{\footnotesize{Left: Scheme of the optics and beams arrangement in the WFS internal calibration unit. Light emitted by the 3 fibers at $120^{\circ}$ double passes a biconvex lens being reflected by a flat mirror and then focussed in a $f_{16.6}$ beam and directed toward the EW. Right: 3D model of the calibration unit, the grey flange on the left is the interface with the WFS enclosure and it holds the flat mirror. The fibers and lens support are drawn in black while the 3 small fold mirror mounts are in gray.}}
  \label{fig:cal_unit}
\end{figure}
\\Figure \ref{fig:cal_unit} shows the position of the calibration unit inside the WFS and its opto-mechanical 3D model. The unit is inserted at the center of the 3 WFS arms, between the entrance windows and the periscope assembly. The 3 fibers are attached to a metallic plate and they are arranged at the vertex of a triangle inscribed in a circle of $10mm$ radius, to mimic the geometry of the 3 LGS on-sky. Using a $0.6mm$ core multimode fiber it is possible to simulate a source having $0.85arcsec$ FWHM on-sky. The light emerging from the fibers passes through a $30mm$ diameter lens with $90.7mm$ focal length (OptoSigma 013-2410) placed on the central axis of the calibration unit. So the 3 beams are steered and they are reflected back toward the lens by flat mirror placed at $100mm$ distance. Overlapping a circular stop of $4.8mm$ diameter to the flat mirror it is possible to make $f_{16.6}$ the beams that double pass the lens. The beams are then folded by $10mm$ diameter flat mirrors, inclined by $20.5^{\circ}$, and they are focussed at a $4mm$ distance from on the rear surface of the WFS entrance windows. The rear surface of the EW reflects few percent of the calibration unit light into the WFS optics.
\begin{figure}
  \centering
  \includegraphics[width=8cm]{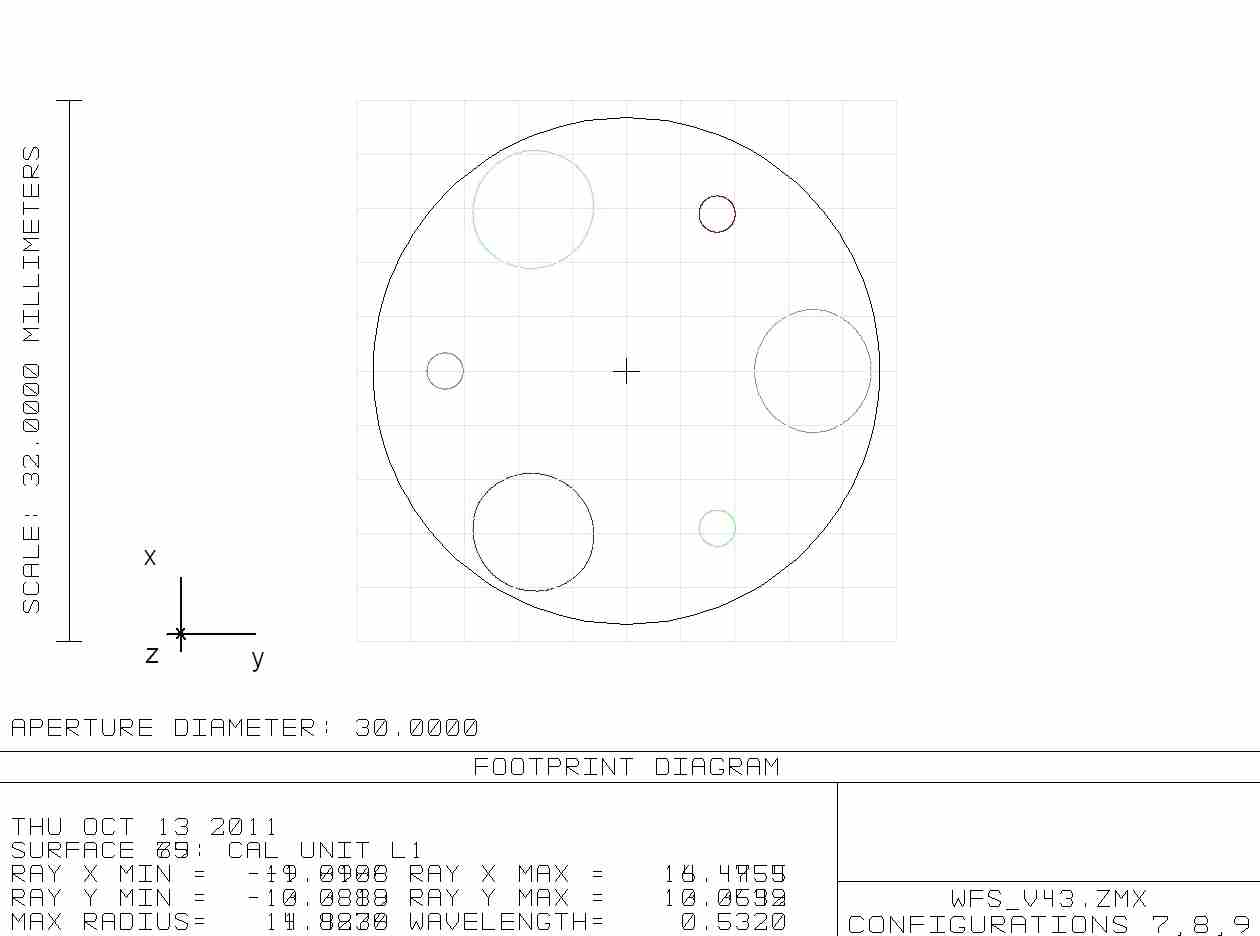}
  \vspace{-0.25cm}
  \caption{\footnotesize{Footprints of the 3 beams on the calibration unit lens. This image is rotated clockwise by $90^{\circ}$ with respect the layout of figure \ref{fig:cal_unit}. The light blue beam lays in the YZ plane.}}
  \label{fig:cal_unit_lens_ftprnt}
\end{figure}
\\The 3 calibration unit beams, that in figure \ref{fig:cal_unit} are drawn in light blue, light green and purple, feed the blue, green and red WFS arms respectively and they are overlapped to the LGS beams coming from the telescope. The image in figure \ref{fig:cal_unit_lens_ftprnt} shows the beams footprint on the surface of the calibration unit lens facing to the fibers outputs.
\\Figure \ref{fig:cal_unit_frame} shows a Zemax simulated frame using the internal calibration unit as light source. On the right we plotted the slopes evaluated from this frame and the one in figure \ref{fig:wfs_final_layout}, that simulates the LGS sources. The PtV of the 2 sets of slopes is within $0.4arcsec$, while the measured rms is $0.23arcsec$ when using the LGS sources and $0.26arcsec$ using the calibration unit source.
\\The main tasks of the internal calibration unit are:
\begin{enumerate}
  \item to provide an optical setup needed to characterize the WFS in laboratory before to ship it to the telescope,
  \item to allow a fast check of the WFS alignment once it will be installed at the telescope.
\end{enumerate}
The first point is accomplished by replacing the flat mirror in the field stop plane with a MEMS deformable mirror, whose details will be given in section \ref{ssec:MEMS}.

\subsection{MEMS deformable mirror}
\label{ssec:MEMS}
\begin{figure}
  \centering
  \includegraphics[width=8cm]{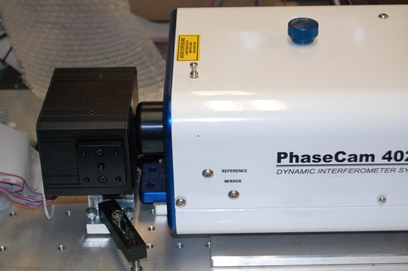}
  \vspace{-0.25cm}
  \caption{\footnotesize{Picture of the Boston Micromachines MULTI-DM installed in front of the 4D Technology interferometer.}}
  \label{fig:MEMS_setup}
\end{figure}

The deformable mirror used in the WFS closed loop tests is a MULTI-DM produced by Boston Micromachines, with $12\times12$ actuators on a $5.4mm$ side (equivalent to $0.45mm$ actuators' pitch). The maximum actuator's stroke is $5.5\mu m$ corresponding to apply $300V$ to the mirror controller.
\\In the calibration unit setup the MEMS will be provided with a circular stop of $4.8mm$ diameter to mask the DM membrane and reproduce a circular pupil. Because the 3 calibration unit beams have a wide angle of incidence ($\sim6^{\circ}$) the mirror membrane the stop must be placed within $0.1mm$ from the DM surface to ensure that the actuator pattern seen by the 3 beams will be shifted less than $1/10$ of subaperture size. To place the stop at this tiny distance from the DM membrane its enclosure has been provided with a removable lid, on which it will be mounted the field stop.
\\The MEMS has been tested in laboratory using a 4D Technology interferometer (model Phase Cam 4020) to check the surface quality obtainable and to calibrate the actuators response. The 4D interferometer produces a collimated beam of $(6.9\pm0.1)mm$ diameter sampled with $990\times998px$ at $632nm$. Figure \ref{fig:MEMS_setup} shows the MEMS installed in front of the 4D interferometer to minimize the aberrations caused by the air flow between the 2 instruments.
\begin{figure}
  \centering
  \includegraphics[width=6.5cm]{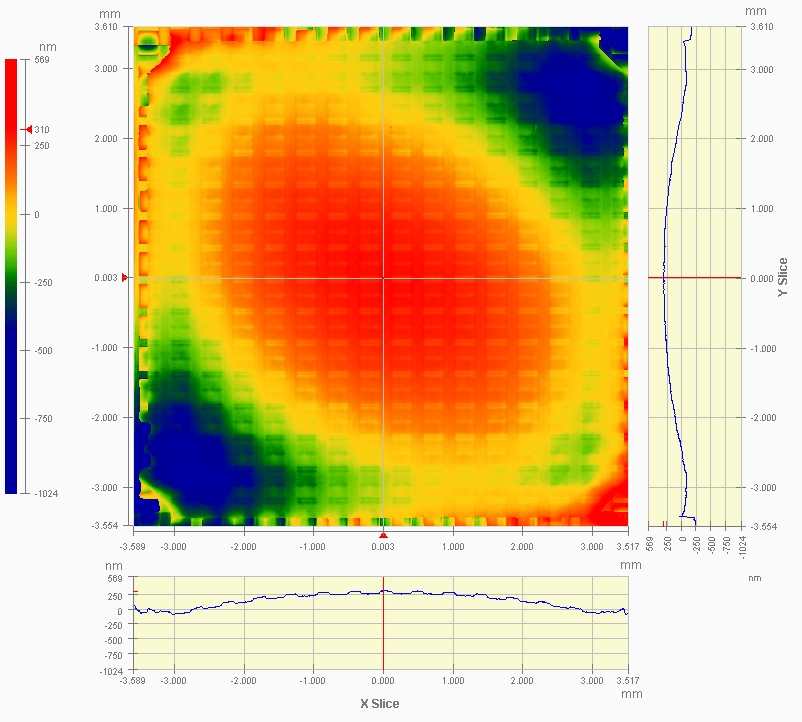}
  \includegraphics[width=6.7cm]{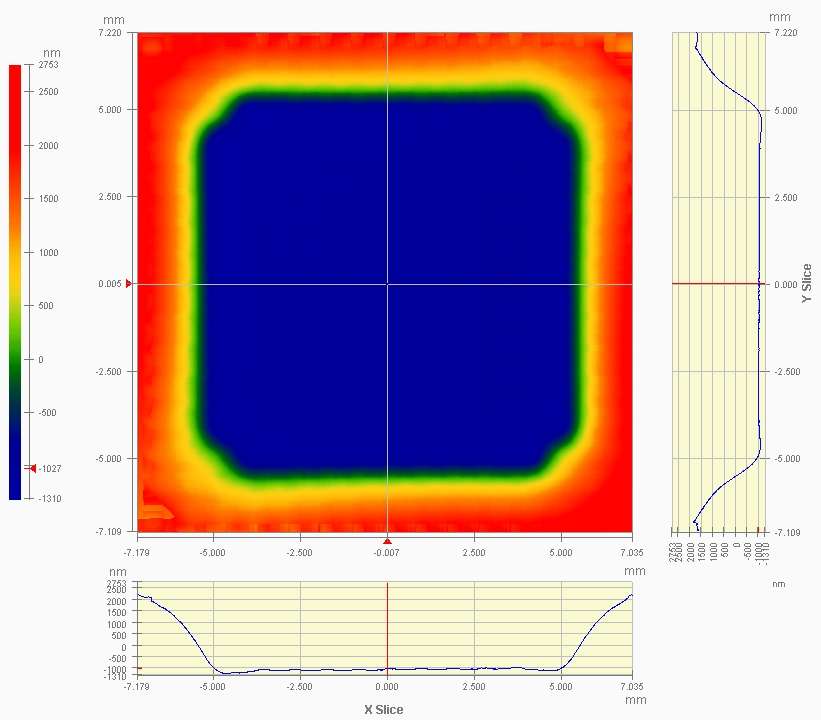}
  \vspace{-0.25cm}
  \caption{\footnotesize{Interferograms of the full aperture MEMS surface. Left: unpowered mirror, right: mirror set at the best flat position provided by Boston Micromachines.}}
  \label{fig:MEMS_surface}
\end{figure}

\subsubsection{Surface quality}
Figure \ref{fig:MEMS_surface} shows the different shape of the mirror when it is unpowered or at its best flat position. Applying a voltage to the actuators the membrane is pulled towards the electrodes substrate. To evaluate the surface quality obtainable we overlapped a circular pupil of variable diameter to the interferometer image. Results of the analysis are summarized in table \ref{tab:MEMS_rms}. The setup designed for the calibration unit, with $11\times11$ actuators across the pupil, ensures that the mirror can be flattened to $32.6nm$ of surface rms. A problem rising when a $4.95mm$ pupil is used is that the actuators on the pupil perimeter are not fully inscribed in pupil, so a larger number of actuators have to be calibrated and controlled.
\begin{table}
  \centering
  \caption{\footnotesize{Summary of the Boston Micromachines MULTI-DM surface error in function of the mask diameter.}} \label{tab:MEMS_rms}
  \vspace{0.25cm}
\begin{tabular}{|c|c|c|c|}
  \hline
  \textbf{Pupil D [mm]} & \textbf{Actuators} & \textbf{Surface PtV [nm]} & \textbf{Surface rms [nm]}\\
  \hline
  \hline
  5.40      &       12      &   1080    &   100.7   \\
  4.95      &       11      &   240     &   32.6    \\
  4.50      &       10      &   180     &   22.6    \\
  \hline
\end{tabular}
\end{table}
\begin{figure}
  \centering
  \includegraphics[width=6.5cm]{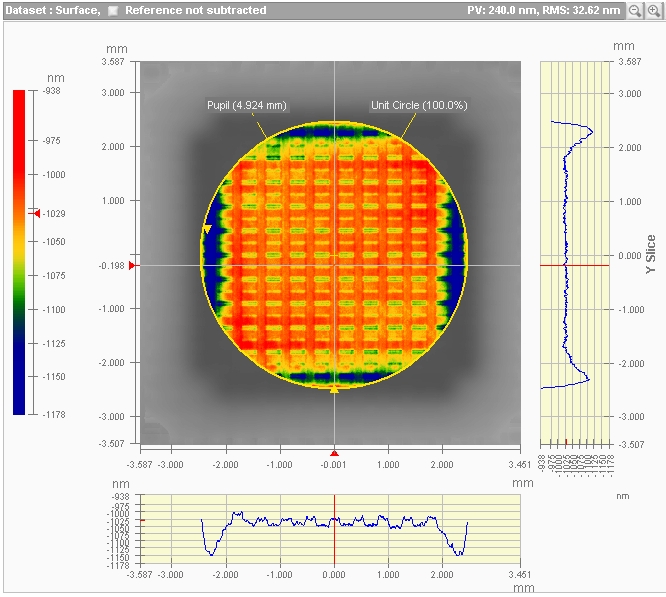}
  \includegraphics[width=6.5cm]{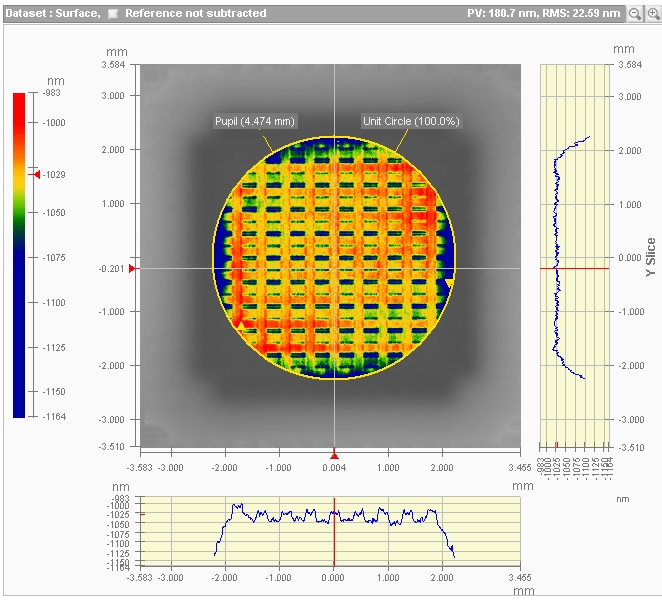}
  \vspace{-0.25cm}
  \caption{\footnotesize{Interferograms of the MEMS surface when a $4.95mm$ (left) or $4.5mm$ diameter pupil are overlapped. The left image shows the half actuators inscribed in the pupil, clearly visible at the top and bottom of the image.}}
  \label{fig:MEMS_surface}
\end{figure}

\subsubsection{Actuators response}
\begin{figure}
  \centering
  \includegraphics[width=6.5cm]{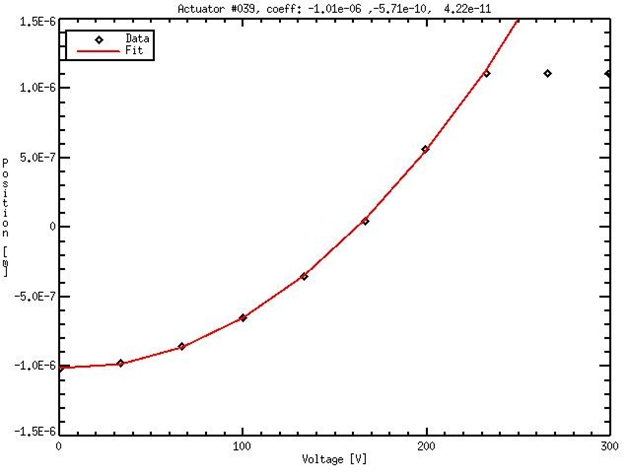}
  \includegraphics[width=6.5cm]{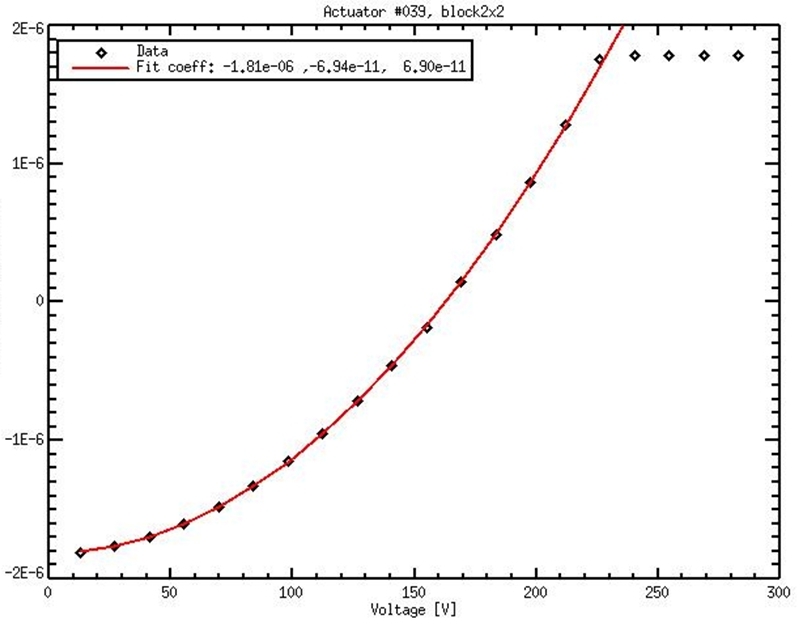}
  \vspace{-0.25cm}
  \caption{\footnotesize{Typical calibration curves obtained for 2 actuators respectively at the center (left) and at the edges (right) of the mirror active area.}}
  \label{fig:MEMS_act_curve}
\end{figure}

To calibrate the actuators deflection vs. voltage curve we applied to each actuator an increasing voltage command in the $0-300V$ range and we measured the membrane shape using the 4D interferometer. From these measurements we could calibrate the response of the actuators to a given voltage fitting the data with a quadratic law, see \cite{1997_Horenstein_electrostatic_MEMS}.
\\Figure \ref{fig:MEMS_act_curve} shows that the effective voltage range usable to calibrate the actuators is limited to the $0-200V$ interval. At greater voltages the \emph{snap through} of actuators occurs. This effect is caused when the two surfaces of the capacitor that moves the mirror membrane become in contact (the short circuit effect is avoided placing a thin nitride layer inside the capacitor).
\\Sweeping a single actuator in the $0-200V$ range introduces a maximum deflection of $2\mu m$ in the mirror membrane. To test the MEMS full stroke a set of adjacent actuators has to be moved simultaneously. We tested this effect using different sets of actuators patterns. Figure \ref{fig:MEMS_act_curve} on right shows that the actuators stroke increases to $\sim4\mu m$ when a block of $2\times2$ actuators is moved together. This effect is due to the reduced membrane stiffness.
\\The actuator response is calibrated fitting the data with the quadratic law:
\begin{equation}\label{eq:MEMS_calib}
    P = A_0 + A_1V + A_2V^2,
\end{equation}
where $P$ is the peak of the membrane deflection with respect the mirror flat position evaluated in correspondence of the actuator and $V$ is the absolute voltage command applied to the actuator. The flat position ($P=0$ in graphs) occurs in the interval of $150-160V$ for the central actuators, while it is necessary to apply $>180V$ at the actuators on the membrane edges. This will reduce the stroke of the actuators.
\\Inverting equation \ref{eq:MEMS_calib} and solving it for positive voltage values we obtain:
\begin{equation}\label{eq:MEMS_V2pos}
    V = -A_1 + \frac{\sqrt{A_1^2 -4A_2(A_0-P)}}{2A_2},
\end{equation}
that gives the voltage required to move the membrane above an actuator by $P$ with respect the best flat position. This formula and the coefficients evaluated in this analysis will be used to drive the deformable mirror in open loop during the WFS laboratory characterization.

%
%

%
%
\chapter{Conclusions}
\label{cap:concl}

The research activities that I conducted during my PhD are part of an international project named ARGOS, aimed to provide the LBT with a ground layer AO system using Rayleigh LGS. ARGOS will supply the LUCI infrared imager and multi-object spectrograph with an uniform AO correction over a FoV as large as $4\times4arcmin$ reaching almost $100\%$ sky coverage.
\\At the beginning of my PhD activity, in 2009, ARGOS just passed through the Preliminary Design Review. The first task in which I was involved was to study the system performance with numerical simulations. I have done this work using CAOS, an open source software developed to solve a large set of AO case studies. I adapted the code to represent most of the subsystems and parameters that characterize the ARGOS system to obtain more realistic results as possible. I evaluated the ARGOS performance considering the gain in terms of PSF FHWM and encircled energy between the seeing limited and the GLAO assisted observations in several directions of the LUCI FoV and under different seeing profiles. The results I obtained showed that ARGOS is able to produce a gain of a factor $1.5-3$ under the different seeing conditions and observing bands. These results agree with the previous ones that have been obtained during the ARGOS design studies.
\\During my PhD I worked also on the optical design of the dichroic window that ARGOS uses to separate the laser light from the scientific one. The criticality of this optic are mainly related to its dimensions, constrained by the LGS beams footprint on the window surface, its working angle and the fact that it has to transmit a convergent $f_{15}$ beam toward LUCI with perfect optical quality and no additional thermal background for the instrument.
\\I took care of the dichroic shape optimization to minimize the aberrations injected by the window on the LUCI focal plane. Designing the rear surface of the window with a $0.56^{\circ}$ wedge and a concave cylinder of $230m$ RoC, with a properly collimating the telescope, the static aberrations on the $f_{15}$ plane are nulled over the full LUCI spectral range.
\\I produced the specifications for the polishing of 2 units of the ARGOS dichroic on stability requirements both in transmission, to keep the objects within the LUCI slits, and reflection, to keep the LGS stable within the WFS subapertures. The WFE injected in transmission and reflection have been kept low enough to avoid non common path aberrations between the systems. The optic measurements done on the units produced by SESO demonstrated that they are within specifications.
\\I followed a similar approach to produce the specifications for the coating of the dichroic units. To avoid losses in the LGS fluxes, and hence in WFS sensitivity, the coating reflectivity as been maximized for s-polarized $532nm$ light. At the same time coating had to maximize the transmissivity at infrared wavelengths both to not reduce the science objects fluxes and to not inject thermal radiation from the dome toward LUCI. I measured the coating produced by Layertec on the dichroic units checking the reflectivity and transmissivity of the windows at sample wavelengths. The measurements confirmed that the coating produced was within specifications.
\\The LGS WFS of ARGOS was designed to comply with several requirements. A single LGS is sensed with $15\times15$ subapertures in a Shack-Hartmann configuration with a $4.7arcsec$ FoV. Pupil stabilizing and vibration compensation systems are provided. Electro-optical shutters allow to reduce the LGS elongation on the detector plane. The 3 LGS are arranged on a single detector, a PnCCD having $248\times256px$ of $48\mu m$ side and $3e^{-}$ RON.
\\Each optical element of the WFS has been independently optimized regarding to the requirements provided. I produced also the final design of the complete system including the WFS, the patrol cameras and the internal calibration unit. I evaluated the tolerances and specifications for the production of the custom optics of the WFS.
\\I took care also to test in laboratory the optics delivered: the folding mirrors, the entrance windows, the 2 meniscus lenses of the SH collimator. I verified that these elements were compliant to specifications and I started the assembling process.  I calibrated the MEMS-DM used for the laboratory characterization of the WFS.

\addcontentsline{toc}{chapter}{{\bf Acknowledgments}}
\include{ark}

\addcontentsline{toc}{chapter}{{\bf Bibliography}}
\bibliographystyle{unsrt}
\bibliography{bibendum}





\end{document}